\documentclass{article}

\usepackage[utf8]{inputenc}
\usepackage[T1]{fontenc}
\usepackage[french]{babel}
\usepackage{amsmath}
\usepackage{amssymb}
\usepackage{graphicx}
\usepackage{tabularx}
\usepackage{textcomp}
\usepackage{fancybox}
\usepackage[hidelinks]{hyperref}
\usepackage[margin=2cm]{geometry}
\usepackage{lmodern}
\usepackage{siunitx}
\usepackage{cite}
\usepackage{color}
\usepackage{tikz}
\usetikzlibrary{plotmarks}
\usetikzlibrary{external}
\usetikzlibrary{patterns}
\usepackage{epsf} 
\usepackage{pgfplots}
\pgfplotsset{compat=newest}
\usepackage{subfigure}
\usepackage{chemfig}
\usepackage{enumitem}

\title{Capillarité}
\author{P. Lidon}

\makeindex

\begin{document}

\maketitle

\section*{Préambule}

Ce document contient et complète le cours de quatre heures donné sur la capillarité en février 2015 à la préparation à l'agrégation de physique de l'ENS de Lyon. En cours, nous aborderons en priorité les notions essentielles pour la leçon et le montage faisant explicitement référence à la capillarité. Le reste des notions présentées dans ce document pourra vous servir dans d'autres leçons ou à préparer les écrits.

Pour vous entraîner, vous pouvez essayer de traiter les thèmes que nous n'aurons pas eu le temps de voir en cours. Vous pouvez en outre regarder les sujets
\begin{itemize}
\item épreuve C 2012, dans son intégralité,
\item épreuve A 2004, partie B-4,
\item épreuve A 1999, partie B,
\item physique 2 PC du concours d'entrée à l'ENS de 2008.
\end{itemize}

La capillarité peut généralement être traitée d'un point de vue mécanique ou d'un point de vue thermodynamique. J'ai volontairement choisi de présenter principalement l'aspect thermodynamique, car il permet un traitement généralement plus rigoureux, et il est moins traité dans les ouvrages. En contrepartie, de nombreux passages ne peuvent pas être directement récupérés depuis la bibliographie : à vous de vous les approprier, ou de consulter la bibliographie si vous souhaitez une présentation différente et aisée à retrouver.

\textbf{Prérequis :}
\begin{itemize}
\item Thermodynamique, en particulier les potentiels thermodynamiques et le potentiel chimique.
\item Hydrodynamique : écoulements parfaits, viscosité, équation de Navier-Stokes, nombre de Reynolds.
\item Notions de physique des ondes.
\item Notions de physique statistique : ensemble microcanonique, champ moyen.
\end{itemize}

\textbf{Remerciements et précaution :}

Ce polycopié a d'ores et déjà bénéficié des relectures scrupuleuses de Michel Fruchart, Robin Guichardaz et \'{E}tienne Thibierge, qu'ils en soient remerciés. Il n'en contient pas moins vraisemblablement encore de nombreuses erreurs à de multiples points de vue.

\newpage

\tableofcontents

\newpage

\section{Généralités sur la tension superficielle}
\label{sec:gen}
La capillarité est l'étude des phénomènes impliquant des interfaces entre plusieurs phases. Nous définissons ici la tension de surface, grandeur caractéristique des effets interfaciaux, et en étudions quelques premières conséquences.

\subsection{Définition}
\label{subsec:gen_definition}

\begin{figure}[htb]
\centering
\subfigure[Sur un cadre contenant une tige mobile, on crée un film de savon. Quand on perce le film d'un côté du rail, ce dernier se déplace vers le film restant.]{\label{fig:experience_cadrerail}
\includegraphics{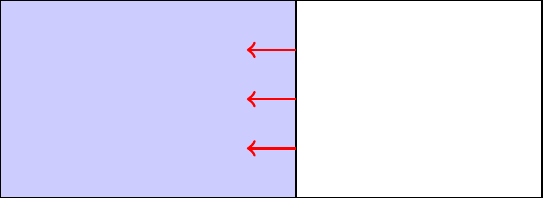}
} \hspace{0.5cm}
\subfigure[Sur un cadre contenant un fil formant une boucle, on réalise un film de savon. Quand on perce le film au centre de la boucle, celle-ci s'étire et prend une forme circulaire.]{\label{fig:experience_cadreboucle}
\includegraphics{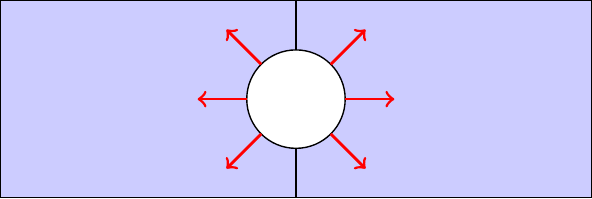}
}
\caption{Expériences illustrant la tension de surface. \label{fig:experience}}
\end{figure}

Commençons par décrire deux expériences simples, que vous avez certainement déjà vu réaliser, représentées figure \ref{fig:experience}.

Dans le premier cas (figure \ref{fig:experience_cadrerail}), on utilise un cadre à l'intérieur duquel se trouve un rail mobile, et on réalise un film de savon sur ce cadre, séparé en deux par le rail. Si l'on perce l'un des films, on constate que le rail se déplace dans la direction du film restant.

Dans le second cas (figure \ref{fig:experience_cadreboucle}, un fil formant une boucle est accroché au cadre. Lorsque l'on forme un film de savon, la boucle a une forme quelconque, mais dès lors que l'on perce le film à l'intérieur de la boucle, celle-ci s'étend pour prendre une forme circulaire.

Ces deux expériences simples montrent que pour un système multiphasique, la présence d'interfaces représente un coût en énergie qu'il convient de minimiser. Plus une interface est étendue, plus elle est défavorable énergétiquement. En l'absence d'autre force, cela explique la forme adoptée par une interface entre deux phases : par exemple, une goutte d'eau dans de l'air prend une forme sphérique pour minimiser l'aire de contact.

On donne alors une définition thermodynamique de la tension de surface $\gamma$. Pour augmenter infinitésimalement de $\mathrm{d}A$ l'aire de l'interface du système constitué de l'ensemble des deux phases en présence, il faut lui fournir de manière réversible un travail $\delta W^\text{rév} = \gamma \mathrm{d}A$. Ainsi, pour un système avec une interface, l'aire $A$ de l'interface devient une nouvelle variable interne\footnote{Si l'on n'impose pas une forme à l'interface, aire et volume sont deux paramètres indépendants.} et la première identité thermodynamique s'écrit $\mathrm{d}U = T\mathrm{d}S - P\mathrm{d}V + \mu \mathrm{d}N + \gamma \mathrm{d}A$. On définit alors la tension de surface comme la dérivée de la fonction d'état caractéristique $U(S,V,N,A)$ par rapport à l'aire de l'interface :
\begin{equation}
\gamma = \left. \frac{\partial U}{\partial A} \right| _{S,V,N}.
\end{equation}

Cette formulation est cependant peu commode : expérimentalement, il est rare que l'on contrôle l'entropie du système ! Pour changer les paramètres d'état utilisés pour décrire le système sans perdre d'information, il faut changer la fonction d'état employée via une transformation de Legendre. La température est généralement un paramètre d'état plus approprié : dans ce cas, on emploiera l'énergie libre $F(T,V,N,A) = U(S,V,N,A) - TS$ comme fonction d'état. Il est également parfois plus intéressant de prendre la pression comme paramètre d'état plutôt que le volume : dans ce cas, on utilisera l'enthalpie libre $G(T,P,N,A) = U(S,V,N,A) + PV - TS$. On obtient alors des définitions équivalentes de la tension de surface, plus commodes en pratique :
\begin{equation}
\gamma = \left. \frac{\partial F}{\partial A} \right| _{T,V,N} = \left. \frac{\partial G}{\partial A} \right| _{T,P,N}.
\end{equation}

La tension de surface est donc une énergie surfacique, ou de façon équivalente, une force linéique. Elle est usuellement exprimée en $\SI{}{\newton\per\meter}$. On verra par la suite que ces deux visions permettent d'appréhender de façons un peu différentes les phénomènes, avec des avantages et des inconvénients.

\textbf{Références :} 
\cite{gouttes} - Chapitre 1 - 1.2. ; \cite{diu_thermo} - Chapitre 3 - Complément A - 2.

\subsection{Origine microscopique}
\label{subsec:gen_origmicro}

Si la thermodynamique nous permet de quantifier le coût énergétique de la formation d'une interface, elle ne nous renseigne aucunement sur l'origine de ce coût.

\begin{figure}[!htb]
\centering
\includegraphics{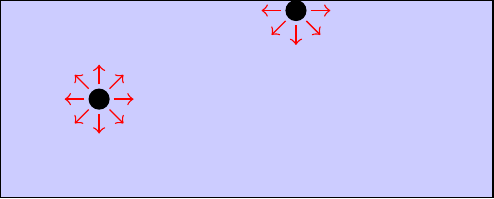}
\caption{Origine microscopique de la tension de surface : une molécule à l'interface interagira plus faiblement avec le reste du fluide qu'une particule dans le volume.\label{fig:origine_micro}}

\end{figure}

Au sein d'une phase, de multiples interactions existent entre les différents constituants (liaisons covalentes, interactions de Van der Waals, liaisons hydrogènes, etc.) et assurent sa cohésion. Dans deux phases différentes, la nature et l'intensité de ces interactions changent. Dès lors, comme illustré figure \ref{fig:origine_micro}, une molécule placée à l'interface entre deux phases n'aura pas les mêmes interactions qu'une molécule située dans le volume : cette distinction entre volume et surface est à l'origine de la tension de surface.

On peut alors obtenir un ordre de grandeur de la tension de surface pour une interface liquide-gaz. Notons $a$ la taille d'une molécule, $U$ l'énergie de cohésion par particule dans le liquide et considérons que la cohésion du gaz est négligeable. Amener une molécule à l'interface augmentera l'aire de cette dernière de $a^2$ et fera perdre une énergie de cohésion de l'ordre de $U$ : dès lors, $\gamma \sim U/a^2$. Pour estimer $U$, on peut considérer qu'à l'ébullition, l'agitation thermique compense la cohésion $U \sim k_\text{B} T_\text{éb}$. On peut ainsi estimer $\gamma \sim k_\text{B} T_\text{éb} / a^2$. Dans le cas de l'eau, on trouve $\gamma \sim \SI{0.1}{\newton\per\meter}$ qui se compare de façon acceptable à la valeur tabulée à $\SI{20}{\celsius}$ : $\gamma_\text{tab} = \SI{72.75}{\milli\newton\per\meter}$.

La tension de surface dépend :
\begin{itemize}
\item de la nature des phases en présence (dans le cas d'une interface avec un gaz, la nature du gaz n'a que peu d'influence à des pressions raisonnables),
\item de la présence d'impuretés dans l'une des phases : elle abaisse généralement la tension de surface mais ce n'est pas systématique\footnote{Par exemple, la présence de sel augmente légèrement la tension de surface de l'eau.},
\item de la température : la tension de surface diminue généralement quand la température augmente.
\end{itemize}

Nous ne nous intéresserons dans ce cours qu'au cas d'interfaces avec un fluide isotrope. Pour des solides ou des cristaux liquides, par exemple, la tension de surface dépend de l'orientation de l'interface par rapport aux axes des phases considérées.

Des explications plus détaillées sur l'origine microscopique de la tension de surface sont fournies dans l'article \cite{Marchand_2011}.

\textbf{Références :} 
\cite{gouttes} - Chapitre 1 - 1.1. ; \cite{GHP} - Chapitre 1 - 1.4.1. ; \cite{handbook} - Section Fluids

\subsection{Forme d'une goutte sur une fil}
\label{subsec:gen_forme}

\begin{figure}[htb]

\centering
\includegraphics{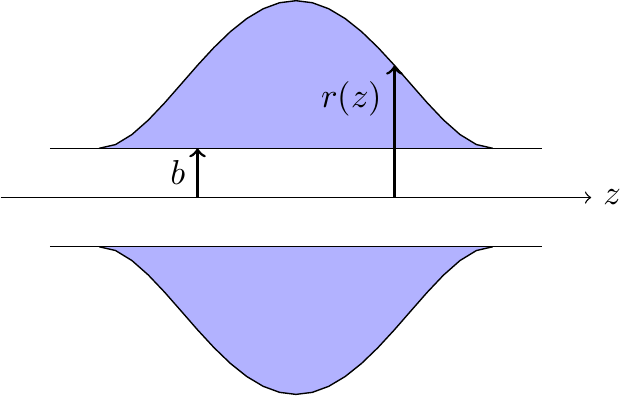}
\caption{Une goutte d'un liquide est posé sur un cylindre de rayon $b$. En l'absence de pesanteur, la goutte adopte une forme dite ondoïdale.\label{fig:goutte_onduloidale}}

\end{figure}

En guise d'exemple, cherchons à déterminer la forme adoptée par une goutte, initialement sphérique de rayon $R$, déposée sur un fil cylindrique de rayon $b$\footnote{On peut se poser ce genre de problème dans des géométries diverses. Comme nous le verrons plus tard, il s'agit de minimiser la courbure moyenne de l'interface. Par exemple, la forme adoptée par un film de savon tendu entre deux cadres circulaires est une surface appelée caténoïde, dont la courbure moyenne est nulle en tout point.}. On supposera la goutte assez petite pour pouvoir négliger les effets de la pesanteur\footnote{On verra plus loin que cela nécessite $R \ll \ell_\text{c}$, où $\ell_\text{c}$ est la longueur capillaire.}. On suppose enfin le problème à géométrie cylindrique, et la forme de la goutte est alors décrite par la fonction $r(z)$ où $\vec{e}_z$ est l'axe du fil. La situation est représentée figure \ref{fig:goutte_onduloidale}.

L'aire d'une tranche de goutte d'exprime dans une telle géométrie selon $\mathrm{d} S = 2\pi r \mathrm{d}s$ où $s$ est l'abscisse curviligne le long de la goutte. Le théorème de Pythagore nous donne $\mathrm{d}s^2 = \mathrm{d}r^2 + \mathrm{d}z^2 = \mathrm{d}z^2 (1+r'^2)$ où $r' = \mathrm{d}r/\mathrm{d}z$. Dès lors, l'aire de la goutte s'écrit
\begin{equation}
S = \int 2\pi r \sqrt{1+r'^2} \mathrm{d}z.
\end{equation}

En outre, le volume $V$ de la goutte est imposé, égal à sa valeur initiale $V_0=4\pi R^3/3$. Il s'exprime selon
\begin{equation}
V = \int \pi (r^2(z) - b^2) \mathrm{d}z.
\end{equation}

La forme de la goutte peut donc être obtenue en minimisant sa surface sous la contrainte d'un volume constant, soit en minimisant la fonctionnelle 
\begin{equation}
F[z(x)] = \gamma S - \lambda V = \int \left[2\pi\gamma r \sqrt{1+r'^2} - \lambda \pi (r^2-b^2)\right] \mathrm{d}z = \int f(r,r') \mathrm{d}z
\end{equation}
\noindent où $\lambda$ est un multiplicateur de Lagrange associé à la contrainte en volume\footnote{Plus précisément, il faudrait écrire $\lambda (V-V_0)$ mais $V_0$ ne dépendant pas de $z$, il disparaîtra ensuite.}.

On doit donc résoudre l'équation d'Euler-Lagrange associée
\begin{flalign}
&& &\frac{\mathrm{d}}{\mathrm{d}z} \left( \frac{\partial f}{\partial r'} \right) = \frac{\partial f}{\partial r} && \\
\text{soit : } && \gamma &\left[-\frac{r''}{(1+r')^{3/2}} + \frac{1}{r\sqrt{1+r'^2}} \right] = \lambda &&
\end{flalign}
\noindent Cette équation détermine le multiplicateur de Lagrange\footnote{Et si vous êtes habitués à la géométrie, ce qui n'est pas mon cas, vous reconnaîtrez aisément l'expression des deux courbures principales de la courbe, soit donc la loi de Laplace $\gamma (C_1 + C_2) = \lambda$ que nous verrons au paragraphe \ref{subsec:Laplace}. $\lambda$ apparaît comme la différence de pression entre l'intérieur et l'extérieur de la goutte. Ce n'est pas surprenant : une vision un peu formelle de la thermodynamique définit la pression comme le multiplicateur de Lagrange associé à la contrainte de volume !}.

L'équation d'Euler-Lagrange admet une intégrale première
\begin{flalign}
&& r' \frac{\partial f}{\partial r'} - f &= \text{cste} &&\\
\text{soit : } && -\lambda \frac{z^2}{2} + \gamma\frac{z}{\sqrt{1+z'^2}} &= \text{cste}. &&
\end{flalign}

La constante peut être imposée en choisissant un raccordement lisse sur la surface du fil\footnote{C'est-à-dire un angle de contact nul comme on le verra plus tard.} : $r'(r=b) = 0$. Alors,
\begin{equation}
-\lambda \frac{r^2 - b^2}{2} + \gamma\left(\frac{r}{\sqrt{1+r'^2}} - b \right) = 0.
\end{equation}

Cette équation décrit une forme de goutte dite ondoïdale : pousser plus loin la résolution nécessite une intégration numérique. Le rayon maximal de la goutte est obtenu pour $r'=0$ soit
\begin{equation}
r_\text{max} = \frac{2\gamma}{\lambda} - b.
\end{equation}

\textbf{Références :} 
\cite{gouttes} Chapitre 1 - 1.5.2. et annexe

\subsection{Importance de la tension de surface}
\label{subsec:gen_importance}

La tension de surface se manifeste de façon importante quand les propriétés interfaciales dominent le comportement du système :
\begin{itemize}
\item si les termes surfaciques sont prépondérants devant les termes volumiques : cela demande de se placer à des petites échelles de longueur, c'est le cas pour les écoulements dans des milieux poreux, dans des films minces, etc.
\item si le système comporte un nombre important d'interfaces : c'est le cas pour les émulsions (dispersion de goutelettes d'un liquide dans un autre liquide) ou les mousses (dispersion de bulles de gaz dans un liquide)\footnote{Pour ces deux derniers cas, la comparaison de la tension de surface avec les forces induisant un écoulement permet de déterminer dans quel mesure les objets inclus dans le fluide se déforment.}.
\end{itemize}

\section{Interfaces entre deux fluides}
\label{sec:interfaces}

La tension de surface caractérise le coût en énergie pour modifier l'aire d'une interface. Or, donner une courbure à une interface modifie sa surface : c'est ce que l'on va étudier dans ce paragraphe.

\subsection{Loi de Laplace}
\label{subsec:Laplace}

\subsubsection{Bulle de gaz en équilibre avec le liquide}
\label{subsubsec:Laplace_bulle}

\begin{figure}[htb]
\centering
\includegraphics{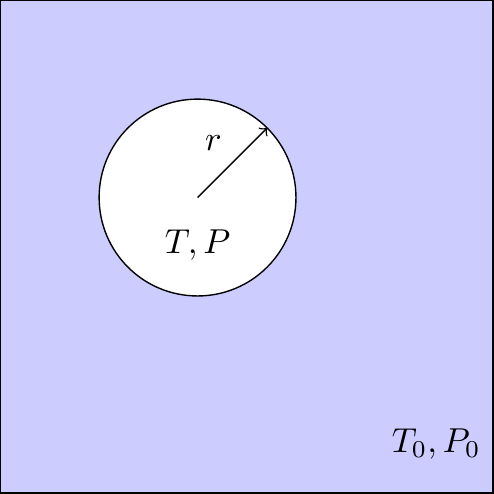}
\caption{On considère une bulle de gaz sphérique en équilibre avec la phase liquide, considérée comme un barostat. La pression à l'intérieur de la bulle est supérieure à celle du liquide du fait de la tension de surface. \label{fig:laplace_demo}}
\end{figure}

Considérons la situation représentée figure \ref{fig:laplace_demo}. On s'intéresse à un système composé d'une bulle de gaz sphérique, de température $T$, de pression $P$ et de rayon $r$, entourée de liquide. Le liquide se comporte comme un thermostat de température $T_0$ et de pression $P_0$\footnote{On peut se compliquer un peu la vie en considérant le liquide comme étant en fait en équilibre avec un thermostat et un barostat. On obtient qu'à l'équilibre, sa pression et sa température sont celles du réservoir, et les conclusions pour la bulle sont les mêmes.}. Le système est considéré comme fermé : cela suppose soit que l'interface ne laisse pas passer les molécules, hypothèse absurde dans notre situation mais raisonnable pour le cas d'une bulle de savon (on n'oubliera alors pas qu'une bulle est constituée de deux interfaces liquide/gaz), soit que l'échelle de temps sur laquelle les équilibres mécaniques et thermiques s'établissent est petite devant celle de l'équilibre osmotique, ce qui est raisonnable\footnote{L'hypothèse d'équilibre mécanique est bien vérifiée, car les équilibres mécaniques ne s'établissent pas par des processus diffusifs. L'hypothèse d'équilibre thermique rapide est acceptable, mais tout juste, le coefficient de diffusion thermique de l'eau ayant à peine un ordre de grandeur de plus que le coefficient d'autodiffusion. On atteint là un problème assez récurrent : nous cherchons à décrire des phénomènes hors équilibre avec une approche thermodynamique, adaptée aux situations d'équilibre.}.

Le choix d'une bulle sphérique n'est pas anodin : on sait qu'il s'agit de la forme qui minimise l'aire à volume donné, il s'agit donc de la forme d'équilibre de la goutte. On cherche alors quelle est la pression dans le gaz à rayon~$r$ donné.

On travaille en contact avec un thermostat à $T_0$ et un barostat à $P_0$ : les paramètres d'état sont la température $T$ de la bulle, sa pression $P$, et son rayon $r$\footnote{Ayant imposé la forme sphérique, le rayon décrit à la fois le volume (qui n'est pas ici un paramètre d'état) et l'aire de la bulle : $V=4\pi r^3 / 3$ et $A=4\pi r^2$.}, et le potentiel thermodynamique adapté à la situation est $G^*(U,r;T_0,P_0) = U + P_0 V - T_0 S$. Sa différentielle s'écrit
\begin{flalign}
&& \mathrm{d}G^* &= \mathrm{d}U + P_0 \mathrm{d} V - T_0 \mathrm{d} S = (T-T_0) \mathrm{d} S + (P_0-P) \mathrm{d} V + \gamma \mathrm{d} A \nonumber && \\
\text{soit :} && \quad \mathrm{d}G^* &= (T-T_0)\mathrm{d}S + 4\pi r^2 \left(P_0 - P + \frac{2 \gamma}{r} \right) \mathrm{d}r.
\end{flalign}

À l'équilibre, le potentiel est extrémal : $\mathrm{d}G^* = 0$ donc $T=T_0$ et on trouve la loi de Laplace pour une sphère
\begin{equation}
P = P_0  +\frac{2 \gamma}{r}
\end{equation}
\noindent qui traduit une surpression à l'intérieur de la bulle, induite par la courbure de l'interface.

\textbf{Références :} 
\cite{gouttes} Chapitre 1 - 1.4. ; \cite{GHP} Chapitre 1 - 1.4.2. ; \cite{diu_thermo} - Chapitre 5 - Complément A ; \cite{fluides} - Partie 3, 1.

\subsubsection{Généralisation à une interface de forme quelconque} 
\label{subsubsec:Laplace_general}

\paragraph{Loi de Laplace :}

\begin{figure}[htb]
\centering
\includegraphics{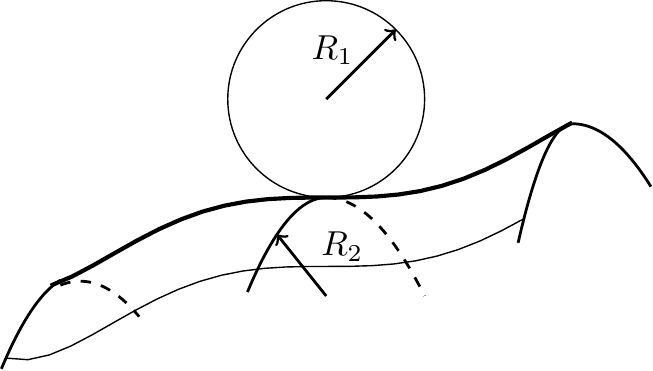}
\caption{Les rayons de courbure $R_1$ et $R_2$ en un point d'une surface bidimensionnelle dans un espace tridimensionnel sont les rayons des cercles osculateurs à ce point dans deux plans orthogonaux.\label{fig:courbures}}
\end{figure}

Définissons tout d'abord les rayons de courbure en un point M d'une surface. On prend l'intersection de la surface avec deux plans $\Pi_1$ et $\Pi_2$ orthogonaux passant par M. On obtient alors deux courbes : on appelle rayons de courbure les rayons $R_1$ et $R_2$ des cercles tangents en M à ces courbes, tracés figure \ref{fig:courbures}. Les valeurs de $R_1$ et $R_2$ dépendent du choix des plans, mais on peut trouver une combinaison indépendante de ce choix : la courbure moyenne $C = 1/R_1 + 1/R_2$.

De façon similaire à ce que l'on a fait pour la bulle sphérique au paragraphe \ref{subsubsec:Laplace_bulle}, pour une interface de géométrie quelconque, on montre que la discontinuité de pression quand on traverse une interface courbée est donnée par :
\begin{equation}
P_\text{int} = P_\text{ext} + \gamma \left(\frac{1}{R_1} + \frac{1}{R_2}\right) = P_\text{ext} + \gamma C.
\end{equation}

Notons que les rayons de courbures sont des grandeurs algébriques : la surpression se situe du côté concave de l'interface, c'est-à-dire vers l'intérieur de la courbure. En outre, on n'oubliera pas que dans le cas général, ce sont des grandeurs locales, qui varient avec le point de la surface considéré. L'équilibre mécanique de la goutte implique que la courbure soit homogène de façon à ce qu'il n'y ait pas de gradient de pression en son sein.

Considérons enfin deux cas particuliers. Pour une interface plane, $R_1,R_2 \rightarrow \infty$ donc il n'y a pas de discontinuité de pression au passage de l'interface. Pour une sphère de rayon $R$, $R_1=R_2=R$ et l'on retrouve la loi du paragraphe précédent.

\textbf{Références :} 
\cite{gouttes} Chapitre 1 - 1.4. ; \cite{barrat_hansen} Chapitre III - 6.4.

\subsubsection{Applications} 
\label{subsubsec:Laplace_applications}

\paragraph{Mûrissement d'Ostwald :}

Une expérience assez aisée à réaliser consiste à relier deux bulles de savon de rayons différents avec un tuyau muni d'un robinet en son centre. Lorsque le robinet est ouvert, on observe que la petite bulle se vide dans la grosse. En effet, la petite bulle étant plus courbée, la surpression qui y règne est plus importante. Lors de la mise en contact, l'air quitte la petite bulle.

Un phénomène similaire, appelé mûrissement d'Ostwald, contribue au vieillissement des mousses et des émulsions : on observe une augmentation de la taille moyenne des bulles ou des goutelettes au cours du temps. Bloquer, ou du moins limiter, le mûrissement est un enjeu dans les industries pharmaceutique, cosmétique, agroalimentaire, etc. qui emploient de tels systèmes.

\paragraph{Relation de Kelvin :}

La courbure d'une interface liquide-gaz modifie la pression de vapeur saturante. En effet, considérons une goutte sphérique de rayon $r$ de liquide (de pression $P$, de température $T$ et de potentiel chimique $\mu_\ell(T,P)$) plongée dans sa phase gazeuse (de pression $P_0$, de température $T_0$ et de potentiel chimique $\mu_\text{g}(T_0,P_0)$). À l'équilibre thermodynamique, $T=T_0$ et $P$ est donné par la relation de Laplace. Il y a en outre égalité des potentiels chimiques des deux phases : $\mu_\ell(T_0,P)=\mu_\text{g}(T_0,P_0)$. 

Si le gaz est supposé parfait, on a :
\begin{equation}
\mu_\text{g}(T_0,P_0)=\mu_\text{g}(T_0,P_\text{sat}(T_0)) + k_\text{B} T_0 \ln{\frac{P_0}{P_\text{sat}(T_0)}}
\end{equation}
\noindent où $P_\text{sat}(T_0)$ est la pression de vapeur saturante considérée.

Développons en outre le potentiel chimique du liquide autour de $P_\text{sat}(T_0)$ :
\begin{equation}
\mu_\ell(T_0,P) \simeq \mu_\ell(T_0,P_\text{sat}(T_0)) + (P-P_\text{sat}(T_0)) \left.\frac{\partial \mu_\ell}{\partial P}\right|_T = \mu_\ell(T_0,P_\text{sat}(T_0)) + \frac{P-P_\text{sat}(T_0)}{\rho}
\end{equation}
\noindent où $\rho$ désigne la densité particulaire du liquide.

Or, par définition de la pression de vapeur saturante, $\mu_\ell(T_0,P_\text{sat}(T_0))=\mu_\text{g}(T_0,P_\text{sat}(T_0))$ donc, à l'équilibre, en utilisant la loi de Laplace, l'égalité des potentiels chimiques fournit la relation de Kelvin\footnote{En pratique, le terme $P_0 - P_\text{sat}(T_0)$ est négligeable devant le premier membre de l'équation et n'est pas pris en compte.}
\begin{equation}
k_\text{B} T_0 \rho \ln{\frac{P_0}{P_\text{sat}(T_0)}} = P_0 - P_\text{sat}(T_0) + \frac{2\gamma}{r}.
\label{eq:Kelvin}
\end{equation}

La relation de Kelvin montre que la pression de vapeur saturante pour une interface courbée est modifiée. Le cas limite d'une interface plane $r \rightarrow \infty$ redonne $P=P_\text{sat}(T_0)$.

\textbf{Références :} 
\cite{gouttes} Chapitre 1 - 1.4. ; \cite{barrat_hansen} Chapitre III - 6.4. ; \cite{fluides} - Partie 3, 2.

\subsection{Métastabilité et nucléation homogène}
\label{subsec:nuclhomo}

\subsubsection{Phénoménologie}
\label{subsubsec:nuclhomo_phenom}

Dans de nombreux cas de transitions de phase du premier ordre, il est possible d'observer des phases métastables. Dans ce cas, on peut observer sur des durées conséquentes une phase dont la thermodynamique prédit qu'elle n'est pas la plus stable. Dans de multiples cas, la tension de surface est responsable de ces retards à la transition.

C'est ce que l'on observe pour la transition liquide-gaz. La conservation d'une phase liquide dans le domaine de stabilité du gaz est appelé surchauffe. On peut par exemple conserver de l'eau liquide à pression atmosphérique à des températures supérieures à $\SI{100}{\celsius}$\footnote{C'est l'origine de quelques ébouillantements avec de l'eau sortant du micro-onde : l'augmentation de température s'y faisant de façon homogène, il n'y a que peu de convection, et pour peu que le récipient soit suffisamment lisse pour ne pas générer de nucléation hétérogène, l'eau peut être mise en surchauffe. Dès qu'elle reçoit suffisamment d'énergie, par un choc du récipient par exemple, elle se met à bouillir avec les projections agréables que l'on peut imaginer.}. En laboratoire, dans des conditions expérimentales très soigneuses, il a été possible de conserver de l'eau liquide à des températures atteignant $\SI{300}{\celsius}$. C'est le principe sur lequel reposent les chambres à bulles : on place un liquide métastable dans une enceinte, et le passage de particules suffit à fournir l'énergie nécessaire à la nucléation. Le phénomène inverse, la conservation d'une phase gazeuse métastable dans le domaine de stabilité du liquide, est appelé surcondensation : c'est ce que l'on utilise dans les chambres à brouillard, dont le principe est symétrique de celui des chambres à bulles.

Pour la transition solide-liquide, il est possible de conserver une phase liquide dans le domaine de stabilité du solide : il s'agit de surfusion. C'est par exemple ce que l'on observe lors de pluies verglaçantes, où de l'eau reste liquide à pression atmosphérique malgré une température inférieure à $\SI{0}{\celsius}$ : on parle alors de surfusion. Il est en revanche difficile d'obtenir une phase solide métastable dans le domaine du liquide, pour des raisons que nous verrons par la suite.

En TP, il est assez aisé d'observer des surfusions. En refroidissant lentement un tube à essai propre et sans aspérité contenant de l'eau distillée\footnote{Toutes ces précautions sont prises pour éviter une nucléation hétérogène.} dans un bain de glace, on peut obtenir de l'eau liquide à température négative. Il est également possible d'observer la surfusion de la benzophénone, ou encore de l'étain.

\subsubsection{Nucléation homogène}
\label{subsubsec:nuclhomo_demo}

\'Etudions maintenant plus en détail le processus de nucléation et considérons un gaz\footnote{Pas nécessairement métastable, nous verrons dans la suite où cette condition intervient.} homogène. On étudie le processus de nucléation homogène, c'est-à-dire que la nucléation peut se faire équiprobablement n'importe où dans le volume\footnote{En pratique, et nous le reverrons ensuite, il est difficile d'observer la nucléation homogène car un autre processus, la nucléation hétérogène, se produit généralement bien avant. Il faut travailler avec des fluides très purs et dans des récipients très lisses, de façon à ce qu'il n'y ait pas de germes autour desquels initier la transition.}. On étudie la formation d'un germe sphérique de rayon $r$ de liquide.

Qualitativement, la variation d'énergie provoquée par la formation du germe contient deux termes :
\begin{itemize}
\item un terme volumique, qui traduit la différence d'énergie de cohésion entre le liquide et le gaz, et qui sera favorable à la nucléation si l'on se place dans le domaine de stabilité du liquide,
\item un terme surfacique, qui traduit le coût de création d'une interface, toujours défavorable.
\end{itemize}
\noindent On voit dès lors que l'énergie ne sera abaissée que si l'on se trouve dans le domaine de stabilité du liquide, et si le germe est suffisamment gros, de façon à ce que le terme de volumique  domine le terme surfacique.

On va considérer le gaz comme un thermostat (de température $T_0$), un barostat (de pression $P_0$), et un réservoir de particules (de potentiel chimique $\mu_\text{g}=\mu_0$)\footnote{Encore une fois, on peut le considérer comme étant en contact avec de tels réservoirs, mais cela ne fait que compliquer le traitement.} : c'est la différence avec ce que l'on a fait pour démontrer la loi de Laplace, le système considéré est ouvert. Dès lors, les paramètres d'états pertinents sont la température $T$ de la goutte, sa pression $P$, son potentiel chimique $\mu_\ell$ et son rayon $r$. Le potentiel thermodynamique adapté est le grand potentiel $\Omega^*(S,r,N;T_0,P_0,\mu_0) = U + P_0 V - T_0 S - \mu_0 N$ où $N$ est le nombre de molécules dans la goutte.

Plutôt que de seulement chercher le sens d'évolution du système, on va aussi essayer d'estimer la barrière énergétique à franchir pour provoquer la transition : au lieu de considérer la différentielle du potentiel (qu'il faudrait ensuite intégrer pour avoir la barrière), on va calculer sa variation $\Delta \Omega^*$ entre un état de référence sans bulle ($r=0$) et un état où la bulle a un rayon $r$ quelconque. On a
\begin{align}
\Delta \Omega^* &= \Delta U + P_0 \Delta V - T_0 \Delta S - \mu_0 \Delta N \nonumber \\
			    &= (T-T_0) \Delta S + (\mu_\ell - \mu_0) \Delta N + (P_0 - P)\Delta V + \gamma \Delta A \nonumber \\
				&= (T-T_0) \Delta S + \left[\rho(\mu_\ell - \mu_0) + P_0 - P \right] \cdot \frac{4}{3} \pi r^3 + \gamma \cdot 4\pi r^2
\label{eq:grpot}
\end{align}
\noindent où $\rho$ désigne la densité particulaire du liquide. Nous ne pouvons pas encore conclure, car il reste des dépendances en $r$ dans la pression $P$ du liquide et dans son potentiel chimique $\mu_\ell$.

Supposons maintenant une séparation des échelles de temps et considérons que les équilibres thermique et mécanique s'établissent rapidement par rapport à l'équilibre osmotique. Dès lors, on peut considérer que $T=T_0$ et $P = P_0 + 2\gamma / r$. En outre, on considère le cas de fluides simples pour lesquels le potentiel chimique est imposé par la pression et la température, via l'équation d'état : pour le gaz, $\mu_0 = \mu_0 (T_0,P_0)$ et pour le liquide, $\mu_\ell = \mu_\ell(T,P) = \mu_\ell(T_0,P_0 + 2 \gamma / r)$. On considère le cas de germes de taille assez grande pour pouvoir effectuer un développement limité de $\mu_\ell$ au premier ordre autour de $P_0$ :
\begin{equation}
\mu_\ell\left(T_0,P_0 + \frac{2 \gamma}{r}\right) \simeq \mu_\ell(T_0,P_0) + \frac{2\gamma}{r} \left.\frac{\partial \mu_\ell}{\partial P}\right|_T = \mu_\ell(T_0,P_0) + \frac{2\gamma}{\rho r}.
\label{eq:DLpotentiel}
\end{equation}

Vérifions à quelle condition ce développement peut être tronqué au premier ordre. Si on le pousse à l'ordre suivant,
\begin{equation}
\mu_\ell\left(T_0,P_0 + \frac{2 \gamma}{r}\right) \simeq \mu_\ell(T_0,P_0) + \frac{2\gamma}{r} \left.\frac{\partial \mu_\ell}{\partial P}\right|_T + \left(\frac{2\gamma}{r}\right)^2 \left.\frac{\partial^2 \mu_\ell}{\partial P^2}\right|_T = \mu_\ell(T_0,P_0) + \frac{2\gamma}{\rho r} - \left(\frac{2\gamma}{r}\right)^2 \frac{\chi_T}{\rho}.
\end{equation}
\noindent où $\chi_T \simeq \SI{4e-10}{\per\pascal}$ est la compressibilité isotherme du liquide. Il est légitime de négliger le terme du second ordre tant que $r \gg 2\gamma \chi_T = \SI{3e-11}{\meter}$ dans le cas de l'eau, ce qui paraît assez légitime, la thermodynamique n'étant plus valable depuis longtemps à de si petites échelles !

Sous ces hypothèses, l'équation \eqref{eq:grpot} devient alors 
\begin{equation}
\Delta \Omega^* = \rho [\mu_\ell(T_0,P_0) - \mu_0(T_0,P_0)] \cdot \frac{4 \pi r^3}{3} + 4\pi r^2 \gamma.
\label{eq:potentiel_homogene}
\end{equation}

\begin{figure}[htb]
\centering
\includegraphics{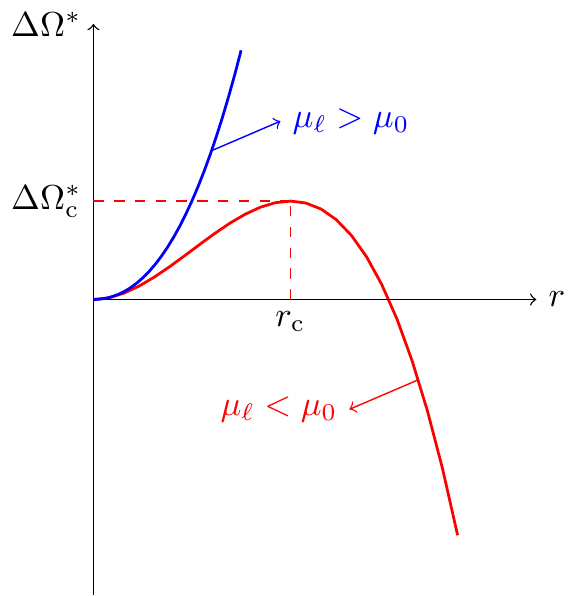}
\caption{Variation du potentiel thermodynamique $\Omega^*$ lors de la formation d'une goutte de rayon $r$. Lorsque la phase liquide est plus stable, une goutte ne se développe que si sa taille dépasse un rayon critique $r_\text{c}$, ce qui nécessite de franchir une barrière énergétique $\Delta \Omega ^*_\text{c}$. \label{fig:courbe_nucl_hom}}
\end{figure}

L'évolution de $\Delta \Omega^*$ avec $r$ est tracée figure \ref{fig:courbe_nucl_hom}. On peut distinguer deux cas. Si $\mu_0(T_0,P_0)<\mu_\ell(T_0,P_0)$, alors $\Delta \Omega^*$ est une fonction strictement croissante de $r$ : quelle que soit la taille de la goutte, elle s'évapore. Au contraire, si $\mu_0(T_0,P_0)>\mu_\ell(T_0,P_0)$, alors $\Delta \Omega^*$ passe par un maximum en 
\begin{equation}
r_\text{c} = \frac{2\gamma}{\rho [\mu_0(T_0,P_0) - \mu_\ell(T_0,P_0)]} > 0
\end{equation}
\noindent et vaut alors
\begin{equation}
\Delta \Omega^*_\text{c} = \frac{16 \pi \gamma^3}{3\rho^2 [\mu_0(T_0,P_0) - \mu_\ell(T_0,P_0)]^2}.
\end{equation}
\noindent Dans ce cas, puisque le potentiel $\Omega^*$ décroît au cours de l'évolution du système, si $r<r_\text{c}$, la goutte s'évapore, mais si $r>r_\text{c}$, elle croît au contraire.

Il nous reste à évaluer l'écart entre les potentiels chimiques pour obtenir des ordres de grandeur du rayon critique et de la barrière énergétique. Pour cela, on suppose que la pression $P_0$ du gaz est proche de la pression de vapeur saturante $P_\text{s} (T_0)$ de façon à développer le potentiel chimique autour de celle-ci :
\begin{equation}
\left\{\begin{array}{l}
\mu_\ell(T_0,P_0) \simeq \mu_\ell(T_0,P_\text{s}(T_0)) + \frac{P_0-P_\text{s}(T_0)}{\rho} \\
\mu_0(T_0,P_0) \simeq \mu_0(T_0,P_\text{s}(T_0)) + \frac{P_0-P_\text{s}(T_0)}{\rho_\text{gaz}(T_0)}
\end{array} \right.
\end{equation}
\noindent puisque $(\partial \mu_\ell / \partial P)_T = 1/\rho $. Or, par définition de la pression de vapeur saturante, $\mu_\ell(T_0,P_s(T_0))=\mu_0(T_0,P_s(T_0))$, et la densité $\rho_\text{gaz}(T_0)$ est négligeable devant celle du liquide $\rho$ si l'on se place suffisamment loin du point critique, ce que l'on va supposer. Dès lors, en considérant en outre le gaz comme parfait, on peut réécrire la différence de potentiel chimique selon
\begin{equation}
\mu_0(T_0,P_0)-\mu_\ell(T_0,P_0) \simeq \frac{P_0 - P_s(T_0)}{P_s(T_0)} \cdot k_\text{B} T_0.
\end{equation}

Ainsi, on obtient 
\begin{equation}
r_\text{c} = \frac{2 \gamma}{\rho k_\text{B} T_0} \frac{P_\text{s}(T_0)}{P_0 - P_\text{s}(T_0)} \qquad \text{et} \qquad \Delta \Omega^*_\text{c} = \frac{16 \pi \gamma^3}{3} \left[\frac{1}{\rho k_\text{B} T_0} \frac{P_\text{s}(T_0)}{P_0 - P_\text{s}(T_0)}\right]^2.
\end{equation}

Cherchons alors un ordre de grandeur pour une pression supérieure de $1\%$ à la pression de vapeur saturante à $T_0=\SI{100}{\celsius}$. Pour de l'eau à cette température, $\gamma = \SI{59}{\milli\newton\per\meter}$ et $\rho = \rho_\text{m} N_\text{A} / M = \SI{3.2e28}{\per\meter\cubed}$ et l'on obtient $r_\text{c} \simeq \SI{7e-8}{\meter}$ et $\Delta \Omega^*_\text{c} \simeq \SI{1e-15}{\joule}$\footnote{N'oublions pas qu'il s'agit d'une barrière énergétique par molécule ! Elle correspond à une énergie molaire $\Delta \omega ^*_\text{c} \simeq \SI{7e8}{\joule\per\mole}$.}.

On peut donc proposer un scénario pour la nucléation homogène. La transition ne peut se faire que si une fluctuation de densité mène à un germe de taille supérieure à $r_\text{c}$, mais une telle fluctuation correspond à une barrière énergétique $\Delta \Omega^*_\text{c}$. Dès lors, plus on s'éloigne de la ligne de coexistence, plus la taille critique est faible, et plus la barrière énergétique est ténue. En considérant la nucléation comme un processus de franchissement de barrière thermiquement activé, le taux de nucléation (nombre de nucléations par unité de volume et par unité de temps) est donné par une relation d'Arrhénius :
\begin{equation}
\Gamma = \Gamma_{\! 0} \exp{\left(-\frac{\Delta \Omega^*_\text{c}(T_0)}{k_\text{B} T_0}\right)}
\end{equation}
\noindent où $\Gamma_{\! 0}$ correspond à une fréquence d'essai, qui doit être déterminée à partir de théorie cinétique.

Notons $\mathbb{P}(t)$ la probabilité que le système soit encore en phase gazeuse au temps $t$, si l'on applique une rampe de température $T(t) = T_\text{coex} - \alpha t$ où $T_\text{coex}$ est la température de coexistence entre liquide et gaz, à la pression considérée. On a $\mathbb{P}(t+\mathrm{d}t)=\mathbb{P}(t)(1-\Gamma_{\! 0} V \mathrm{d}t)$. On peut alors intégrer cette équation, sachant que $\mathbb{P}(t=0)=1$ puisque l'on part de la coexistence. En remplaçant la variable temporelle par la température, on obtient
\begin{equation}
\ln{(\mathbb{P}(T))} = - \int_{T_\text{coex}}^T \frac{\Gamma_{\! 0} V}{\alpha} \exp{\left(-\frac{\Delta \Omega^*_\text{c}(\theta)}{k_\text{B} \theta}\right)} \mathrm{d}\theta.
\end{equation}
\noindent Il s'agit là de la théorie classique de nucléation, donnant des résultats qualitatifs corrects, mais dont le principal problème réside en l'estimation du préfacteur $\Gamma_{\! 0}$.

\textbf{Références :} 
\cite{diu_thermo} Supplément G - 2.

\subsection{Ondes gravito-capillaires}
\label{subsec:ondesgc}

On s'intéresse dans ce paragraphe aux ondes à la surface d'un fluide au repos, couplant la déformation de la surface au champ de vitesse du fluide. Ces ondes sont aisément observables : vagues, rides engendrées par le vent ou par la chute d'un objet, etc.

\begin{figure}[htb]
\centering
\includegraphics[height=5cm]{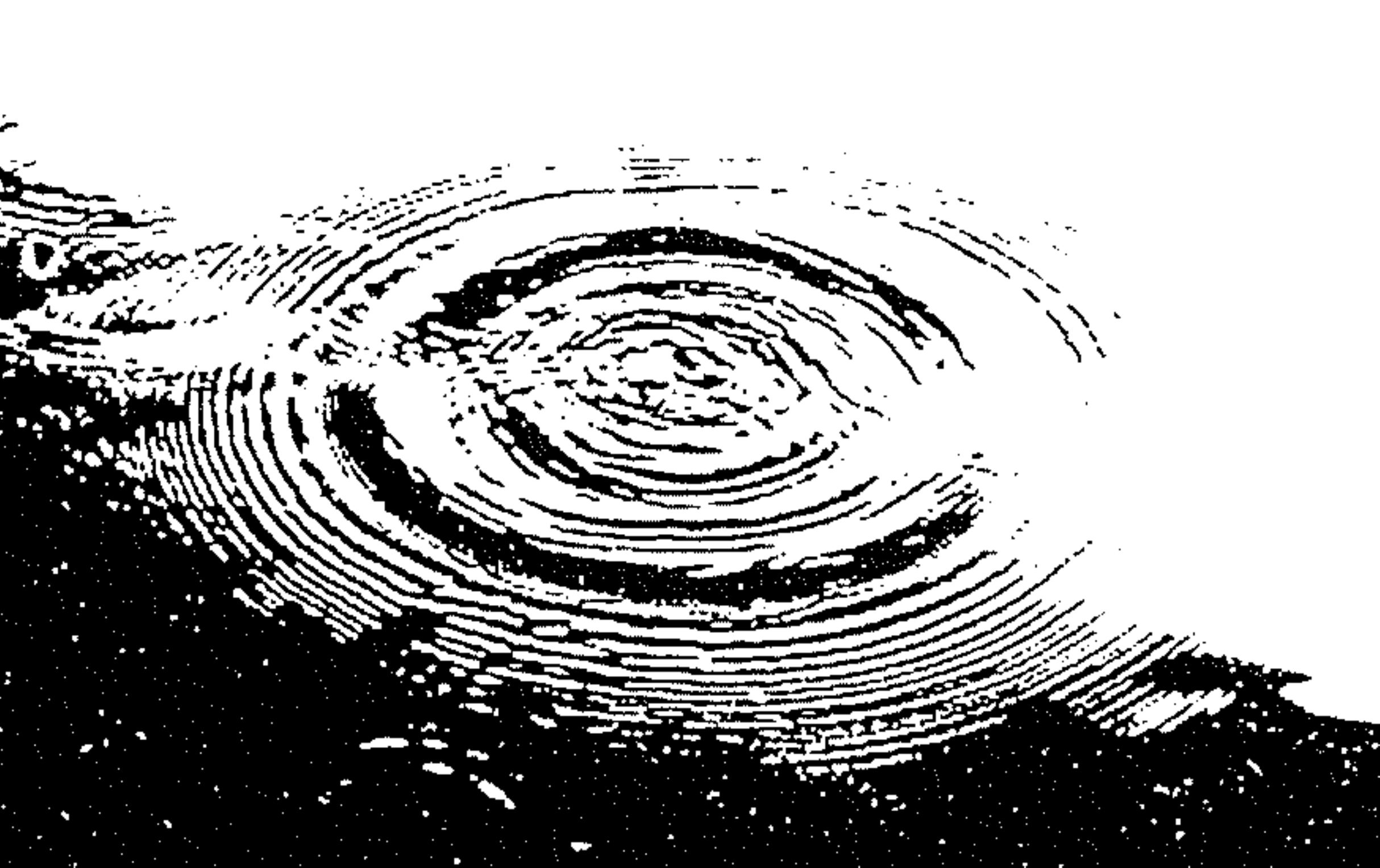}
\caption{Photographie de la surface d'une étendue d'eau environ $\SI{1}{\second}$ après la chute d'un caillou. On observe deux séries d'ondes, de vitesses et de longueurs d'onde distinctes. Source : B. Lahaye, \og Propagation des ondes. Vitesse de phase, vitesse de groupe. \fg{}, BUP 649 (1982). \label{fig:chute_caillou}}

\end{figure}

Considérons l'exemple d'un caillou tombant dans l'eau : comme on le voit sur la photographie \ref{fig:chute_caillou}, on observe après quelques instants l'existence de deux séries d'ondes, l'une étant plus rapide et à des longueurs d'ondes plus courtes. Nous allons tenter d'expliquer ces observations.

\subsubsection{Traitement hydrodynamique}
\label{subsubsec:ondesgc_demo}

\begin{figure}[htb]
\centering
\includegraphics{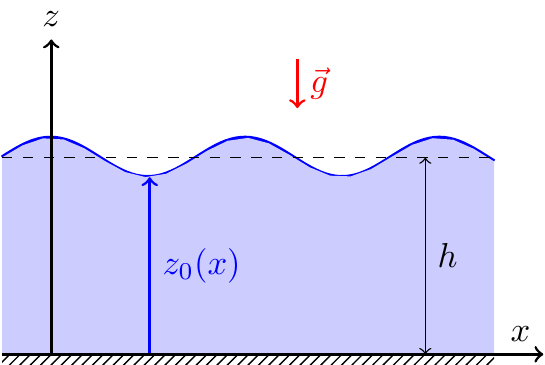}
\caption{La surface d'une couche de fluide d'épaisseur $h$ est perturbée par des vaguelettes. On note $z_0(x,t)$ l'altitude de la surface libre.\label{fig:ondes_surface_notations}}
\end{figure}

\paragraph{Position du problème :} Considérons un fluide, infini dans les directions horizontales $\vec{e}_x$ et $\vec{e}_y$, reposant sur un fond plat, de normale $\vec{e}_z$ et qui fixe l'origine des altitudes. On note $h$ l'épaisseur du fluide au repos et $\rho$ sa masse volumique. On s'intéresse au problème bidimensionnel dans le plan $(xz)$. On note $\vec{v}(x,z,t)$ le champ de vitesse de l'écoulement, $P(x,z,t)$ le champ de pression et $\vec{g}=-g\vec{e}_z$ le champ de pesanteur. Notons enfin $z_0(x,t)$ l'altitude de la surface libre. Les notations sont résumées dans la figure \ref{fig:ondes_surface_notations}.

On va étudier la propagation d'ondes à la surface du fluide, problème couplant l'écoulement en volume du fluide et la déformation de sa surface. Qualitativement, trois phénomènes sont en compétition : l'inertie du fluide tend à le faire persévérer dans son mouvement, alors que la pesanteur et la tension superficielle tendent respectivement à abaisser le centre de gravité du fluide et à en lisser la surface.

\paragraph{\'Equations hydrodynamiques :} On suppose l'écoulement parfait : nous pouvons dès lors négliger la dissipation visqueuse dans l'écoulement, ce dont nous rediscuterons par la suite. L'écoulement du fluide est alors décrit par l'équation d'Euler :
\begin{equation}
\rho \left[ \frac{\partial \vec{v}}{\partial t} + (\vec{v} \cdot \overrightarrow{\nabla}) \vec{v} \right] = \rho \vec{g} - \overrightarrow{\nabla} P .
\end{equation}

L'écoulement étant parfait, la vorticité est conservée, et puisque le fluide est initialement au repos, elle est nulle. Le champ de vitesses découle alors d'un potentiel $\phi (x,z,t)$ tel que $\vec{v} = \overrightarrow{\nabla} \phi$.

On suppose en outre l'écoulement incompressible ce qui implique la conservation locale du débit $\overrightarrow{\nabla} \cdot \vec{v} = 0$. Le potentiel des vitesses satisfait donc l'équation de Laplace sans source
\begin{equation}
\Delta \phi = 0.
\label{eq:Laplace}
\end{equation}

En utilisant le fait que $(\overrightarrow{\nabla} \times \vec{v})\times \vec{v} = (\vec{v} \cdot \overrightarrow{\nabla}) \vec{v} - \overrightarrow{\nabla} (v^2/2)$, on peut écrire l'équation d'Euler pour un écoulement irrotationnel :
\begin{equation}
\overrightarrow{\nabla}\left( \rho \frac{\partial \phi}{\partial t} + \frac{\rho v^2}{2} + \rho g z + P\right) = \vec{0}.
\end{equation}

Nous considérons enfin que l'onde est de faible amplitude, c'est-à-dire que la vitesse de l'écoulement et le déplacement du fluide sont d'ordre 1 : nous préciserons par la suite les grandeurs auxquelles il faut les comparer. On peut négliger le terme quadratique et obtenir ainsi la relation de Bernoulli pour un écoulement lent, parfait, incompressible, irrotationnel et instationnaire\footnote{Pour mémoire, l'équation de Bernoulli traduit la conservation de l'énergie.} :
\begin{equation}
\frac{\partial \phi}{\partial t} +  g z + \frac{P}{\rho} = \, K
\label{eq:Bernoulli}
\end{equation}
\noindent où $K$ désigne une constante\footnote{À strictement parler, $K$ peut dépendre du temps. Cependant il s'agira alors d'un terme non propagatif, qui ajoutera une contribution indépendante de l'espace au potentiel : cette contribution disparaîtra donc dans la vitesse, qui est la quantité physique intéressante. On peut donc librement choisir $K$ comme indépendante du temps.}.

\paragraph{Conditions aux limites :}

Nous disposons de trois conditions aux limites.
\begin{itemize}
\item La discontinuité de pression à la traversée de la surface libre est donnée par la loi de Laplace. En se souvenant que la courbure est donnée par $\partial^2_{x,x} z_0$, on obtient
\begin{equation}
P(x,z=z_0(x,t),t)=P_0 - \gamma \frac{\partial ^2 z_0}{\partial x^2}.
\label{eq:CL_Laplace}
\end{equation} 
\item La composante verticale de la vitesse est nulle au fond du récipient
\begin{equation}
v_z(z=0)=\left( \frac{\partial \phi}{\partial z} \right) (x,z=0,t) = 0.
\label{eq:CL_fond}
\end{equation}
\item La composante normale de la vitesse de l'interface est égale à la composante normale de la vitesse du fluide au niveau de la surface libre. Pour des déformations de faible amplitude, la normale à l'interface s'identifie à la verticale et on obtient
\begin{equation}
v_z(x,z=z_0(x,t),t)=\left( \frac{\partial \phi}{\partial z} \right) (x,z=z_0(x,t),t) = \frac{\partial z_0}{\partial t}.
\label{eq:CL_surface}
\end{equation}
\end{itemize}

Dans l'hypothèse d'ondes de faible amplitude, les relations \eqref{eq:CL_Laplace} et \eqref{eq:CL_surface} peuvent être prises en~$z=h$ plutôt qu'en~$z=z_0(x,t)$, les corrections induites étant d'ordre supérieur.

Les équations \eqref{eq:Bernoulli} et \eqref{eq:CL_surface} couplent l'écoulement, caractérisé par son potentiel $\phi$, et la déformation de l'interface, décrite par~$z_0$. Leur combinaison permet d'obtenir l'équation satisfaite par le potentiel des vitesses à la surface~$z=z_0 \simeq h$ 
\begin{equation}
\left[ \frac{\partial^2 \phi}{\partial t^2} + g \frac{\partial \phi}{\partial z} - \frac{\gamma}{\rho} \frac{\partial^3 \phi}{\partial x^2 \partial z} \right] (x,z=h,t) = 0.
\label{eq:CL_finale}
\end{equation}

\paragraph{Résolution du problème :}

Nous cherchons une onde de surface plane progressive se propageant selon $\vec{e}_x$ et inhomogène dans la profondeur. Nous allons donc chercher une solution en séparant les variables~$z$ et~$u=x-ct$, où $c$ reste à déterminer :
\begin{equation}
\phi(x,z,t) = \chi(u) \psi(z).
\end{equation}

\noindent En réinjectant cette forme de solution dans l'équation de Laplace \eqref{eq:Laplace}, on obtient
\begin{equation}
\chi '' (u) \psi(z) + \chi (u) \psi''(z) = 0 \quad \text{soit :} \quad  \left(\frac{\chi''}{\chi}\right)(u)=-\left(\frac{\psi''}{\psi}\right)(z)
\end{equation}

\noindent Les deux membres de cette dernière égalité étant fonction de variables indépendantes, ils sont tous deux simultanément constants, égaux à une constante notée $-k^2$\footnote{On choisit une constante négative de façon à avoir propagation selon $x$. Prendre une constante positive donnerait une solution amortie qui ne nous intéresse pas ici.}. Dès lors,
\begin{equation}
\left\{\begin{array}{l}
\chi(u) = A \text{e}^{\mathrm{i} ku} + B \text{e}^{-\mathrm{i} ku} \\
\psi(z) = C \cosh{(kz)} + D \sinh{(kz)}.
\end{array} \right. 
\end{equation}

La condition aux limites \eqref{eq:CL_fond} impose que $D=0$ et on choisit par commodité de se restreindre aux solutions se propageant dans le sens des $x$ croissants : le potentiel des vitesses s'écrit alors 
\begin{equation}
\phi(x,z,t) = A \text{e}^{\mathrm{i} (kx - \omega t)} \cosh{(kz)}
\label{eq:potentiel_solution}
\end{equation}
\noindent où $A$ est une constante caractérisant l'amplitude de l'onde et où l'on a posé $\omega = c k$.

Il nous reste à obtenir la relation de dispersion. Pour cela, on injecte la solution $\eqref{eq:potentiel_solution}$ dans l'équation de propagation~\eqref{eq:CL_finale} et l'on trouve la relation de dispersion, ainsi que l'expression de la vitesse de phase
\begin{equation}
\omega^2 = \left(gk + \frac{\gamma k^3}{\rho} \right) \tanh{(kh)} \quad \text{et} \quad c = \frac{\omega}{k} =\sqrt{ \left(\frac{g}{k} + \frac{\gamma k}{\rho} \right) \tanh{(kh)}}.
\end{equation}

Nous pouvons enfin introduire une longueur caractéristique
\begin{equation}
\ell_\text{c} = \sqrt{\frac{\gamma}{\rho g}}
\end{equation}
\noindent appelée longueur capillaire\footnote{Donnons en tout de suite un ordre de grandeur : pour de l'eau pure à température ambiante, $\ell_\text{c} \simeq \SI{2.7}{\milli\meter}$.} et dont nous discuterons plus en détail dans un paragraphe ultérieur. La relation de dispersion et la vitesse de phase se réécrivent alors
\begin{equation}
\omega^2 = gk(1+k^2 \ell_\text{c}^2) \tanh{(kh)} \quad \text{et} \quad c = \frac{\omega}{k} =\sqrt{\frac{g}{k} (1+k^2\ell_\text{c}^2) \tanh{(kh)}}.
\end{equation}

\paragraph{Discussion des hypothèses :}

Nous avons effectué diverses hypothèses au cours de la démonstration qu'il convient maintenant de discuter.

L'hypothèse de petite perturbation recouvre en fait trois conditions.
\begin{itemize}
\item L'amplitude $v_0$ de la vitesse doit être suffisamment faible pour que l'on puisse négliger le terme d'accélération convective dans l'équation d'Euler (ou de façon équivalente, le terme d'énergie cinétique dans l'équation de Bernoulli) devant le terme d'instationnarité. Il faut donc avoir $v_0 \ll c = \omega / k$.
\item L'amplitude de la déformation de l'interface doit être suffisamment lisse pour que l'on puisse assimiler la normale à la surface libre à la verticale : il faut pour cela $\partial_x z_0 \ll 1$.
\item L'amplitude de la déformation de l'interface doit être suffisamment faible pour que l'on puisse considérer que les conditions aux limites sont prises en $z=h$ : il faut donc $|z_0-h| \ll h$.
\end{itemize}

Nous avons en outre considéré l'écoulement comme parfait, c'est-à-dire que nous avons négligé la dissipation visqueuse\footnote{Même si l'hypothèse d'écoulement parfait néglige tous les phénomènes diffusifs, c'est la viscosité qui entre en compte dans ce problème.}. Cette hypothèse peut être prise en défaut quand l'épaisseur de fluide devient faible, nous en discuterons plus en détails dans le paragraphe sur l'instabilité de Rayleigh-Taylor. Dans la limite d'eaux profondes, la dissipation visqueuse intervient dans l'écoulement en volume du fluide. On peut obtenir un temps caractéristique d'atténuation~$\tau$ en équilibrant l'inertie (qui est le moteur du mouvement) et la dissipation visqueuse : $\rho v_0 / \tau \sim \eta v_0 / \lambda^2$ soit
\begin{equation}
\tau \sim \frac{\rho \lambda^2}{\eta} = \frac{\lambda^2}{\nu}.
\end{equation}
\noindent et la distance caractéristique d'atténuation est $L^* \sim c \tau$. On retrouve des lois d'échelles courantes de phénomènes de diffusion.

\textbf{Références :} 
\cite{GHP} Chapitre 6 - 4.1. ; \cite{gouttes} Chapitre 5 - 5.

\subsubsection{Analyse dimensionnelle}
\label{subsubsec:ondesgc_analysedim}

De façon générale, les divers paramètres intervenant dans le problème sont la masse volumique $\rho$, l'intensité de la pesanteur $g$, la tension de surface $\gamma$, l'épaisseur de fluide $h$, la pulsation de l'onde $\omega$, son nombre d'onde $k$, et l'amplitude de la perturbation de la surface libre $\zeta_0$. Nous disposons donc de sept paramètres pour trois dimensions indépendantes. Le théorème de Buckingham implique alors que le problème est déterminé par quatre nombres sans dimension que nous choisissons comme
\begin{equation}
\tilde{K} = k \ell_\text{c}, \quad \tilde{H}=kh, \quad \tilde{Z}=k\zeta_0 \quad \text{et} \quad \tilde{\Omega} = \omega /\sqrt{k g}.
\end{equation}

\noindent Nous pouvons donc seulement affirmer que la relation de dispersion se met sous la forme
\begin{equation}
\tilde{\Omega}=f(\tilde{K},\tilde{H},\tilde{Z}).
\end{equation}

Considèrons alors le cas d'eaux profondes, soit $\tilde{H} \gg 1$, et d'une onde de faible amplitude, soit $\tilde{Z} \ll 1$. Supposons ainsi que dans ce régime, $f$ ne dépend plus que de $\tilde{K}$\footnote{Cela correspond au cas d'une auto-similitude de première espèce, que nous supposons adaptée. Si cette hypothèse impliquait des résultats en désaccord avec l'expérience, il faudrait rechercher $f(\tilde{K},\tilde{H},\tilde{Z})$ sous la forme $\tilde{H}^\alpha \tilde{Z}^\beta \psi(\tilde{K}\tilde{H}^{-\gamma} \tilde{Z}^{-\delta})$. Notons néanmoins que $f$ doit être indépendante de $\tilde{Z}$ puisque l'on effectue une analyse linéaire, où l'amplitude de l'onde est infinitésimale.}.

Tout d'abord, notons qu'il ne reste pour décrire le mouvement du fluide qu'une échelle de temps $1/\omega$ et une échelle de longueur $1/k$ pertinentes, la longueur capillaire étant une caractéristique statique du fluide. Dès lors, la vitesse d'une particule de fluide a pour ordre de grandeur $v_0 = \omega / k$, la vitesse de phase de l'onde de surface. Nous pouvons alors obtenir les comportements asymptotiques de $f$ avec $\tilde{K}$\footnote{C'est là que tout peut sembler un peu fumeux, dans la mesure où les forces que l'on tente d'équilibrer agissent dans des directions orthogonales. Mais ce que l'on prend pour de la magie n'est que la puissance de l'analyse dimensionnelle : essayer de distinguer deux directions reviendrait à ajouter une longueur dans le problème.}.
\begin{itemize}
\item Cas purement gravitationnel : Si l'on considère la situation $\tilde{K} \ll 1$, c'est la gravité qui domine la capillarité. On doit donc équilibrer l'inertie du fluide avec la pesanteur : $\partial_t \vec{v} \sim \vec{g}$. En ordre de grandeur on obtient $\omega^2 \sim k g$.
\item Cas purement capillaire : Si l'on considère la situation $\tilde{K} \gg 1$, c'est la capillarité qui domine la gravité. On doit donc équilibrer l'inertie du fluide avec le gradient de pression capillaire à l'interface : $\partial_t \vec{v} \sim \overrightarrow{\nabla} P / \rho$. En ordre de grandeur, $\|\overrightarrow{\nabla} P \| \sim k \, \Delta P \sim k \times (\gamma k)$ donc $\omega^2 \sim \gamma k^3 / \rho$.
\end{itemize}

On obtient ainsi les deux comportements limites de la fonction $f$ :
\begin{equation}
 \text{pour}\; \tilde{K} \ll 1, \; \tilde{\Omega}=f(\tilde{K}) \sim 1 \; \text{et pour} \; \tilde{K} \gg 1, \; \tilde{\Omega}=f(\tilde{K}) \sim \tilde{K}. 
\end{equation}

\noindent L'analyse dimensionnelle ne permet pas de prédire le comportement général de $f$, il se trouve que c'est la somme des deux cas limites, ce qui n'est pas aberrant puisque les effets liés à la gravité et ceux liés à la capillarité apparaissent comme une somme dans l'équation de propagation \eqref{eq:CL_finale}.

\subsubsection{Discussion des résultats}
\label{subsubsec:ondesgc_discussion}

\begin{figure}[htb]
\centering
\includegraphics{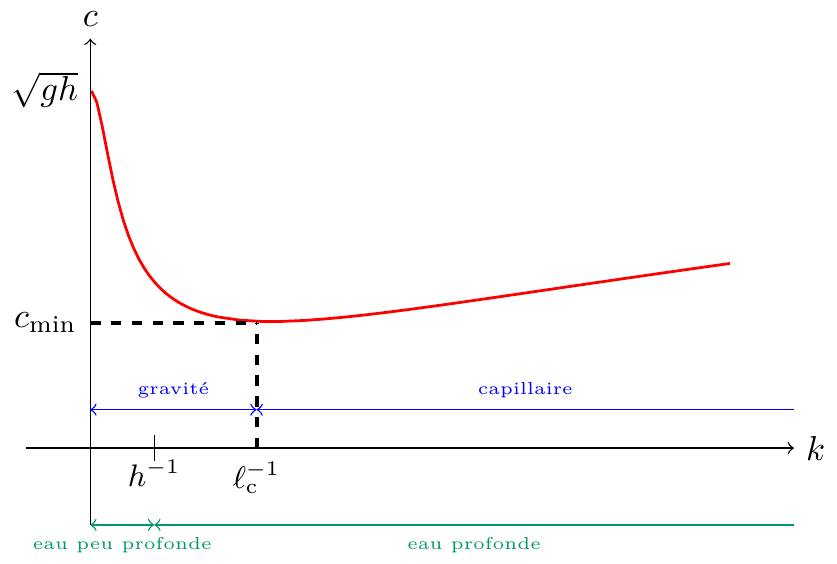}
\caption{Relation de dispersion des ondes gravito-capillaires à la surface d'un fluide.\label{fig:rel_disp_surface}}
\end{figure}

\paragraph{Relation de dispersion :} Commençons par discuter de la relation de dispersion, tracée figure \ref{fig:rel_disp_surface}. Divers régimes peuvent être distingués :
\begin{itemize}
\item selon la valeur de $kh$ : si $\lambda \gg h$, on est dans le régime d'eaux peu profondes et $\tanh{(kh)} \sim kh$, alors que si $\lambda \ll h$, on est dans le régime d'eaux profondes et $\tanh{(kh)} \sim 1$. La tangente hyperbolique variant cependant assez lentement loin de $0$ \footnote{On a $\tanh{x} = 0.9$ pour $x \simeq 1.5$.}, l'hypothèse d'eaux profondes est assez peu restrictive et reste valable dans de nombreux cas. Nous nous y plaçons pour la suite.
\item selon la valeur de $k\ell_\text{c}$ : si $\lambda \ll \ell_\text{c}$, la capillarité domine et on a $\omega^2 \simeq gk^3\ell_\text{c}^2$ et $c \simeq \sqrt{gk}\ell_\text{c}$ alors que si $\lambda \gg \ell_\text{c}$, la gravité domine et on a $\omega^2 \simeq gk$ et $c\simeq \sqrt{g/k}$ \footnote{On remarque, chose attendue, que la capillarité est influente à petite échelle.}.
\end{itemize}

Dans l'expérience du caillou tombant dans un lac, la série d'ondes rapides de courte longueur correspond aux ondes capillaires, alors que la série d'onde lentes et de grande longueur d'onde correspond aux ondes gravitaires.

On remarque en outre que la vitesse des ondes passe par un minimum non nul $c_\text{min}$ pour $k\ell_\text{c} = 1$\footnote{Dans l'eau à température ambiante, $c_\text{min} \simeq \SI{23}{\centi\meter\per\second}$.}. Il s'agit de la vitesse relative minimale à laquelle un obstacle doit  se déplacer par rapport au fluide environnant pour laisser un sillage stationnaire, en forme de V comme celui que l'on peut observer derrière un canard ou un bateau. Si l'obstacle se déplace à une vitesse inférieure à $c_\text{min}$ la perturbation restera localisée autour de lui. L'émission d'un sillage dissipe de l'énergie, intervenant dans les problèmes de résistance à la progression en surface d'un fluide (on parle de résistance de vague)\footnote{Ce type de critère se retrouve dans de multiples domaines, dès qu'une relation de dispersion présente une vitesse minimale non-nulle : critère de Landau pour la superfluidité, amortissement Landau dans les plasmas, rayonnement Cherenkov, etc.}.

\paragraph{Retour sur la dissipation visqueuse :} Munis des vitesses de phase approximatives, on peut évaluer les distances typiques d'atténuations dans les deux régimes. Dans le régime capillaire, on trouve $L^* \sim (\rho \gamma \lambda^3)^{1/2}/\eta$. Pour $\lambda=\SI{1}{\milli\meter}$ par exemple, on obtient $L^* \simeq \SI{30}{\centi\meter}$. Dans le régime gravitaire, $L^* \sim \rho g^{1/2} \lambda^{5/2} / \eta$ et les longueurs d'atténuation deviennent rapidement très élevées (quelques kilomètres pour $\lambda \sim \SI{1}{\meter}$). Les vagues en mer correspondent à ce second régime.

\paragraph{\'Ecoulement induit par l'onde :} Munis du potentiel des vitesses \eqref{eq:potentiel_solution}, il est aisé d'obtenir le profil d'écoulement
\begin{equation}
\vec{v}(x,z,t) = -Ak\sin{(kx - \omega t)}\cosh{(kz)} \vec{e}_x + Ak\cos{(kx - \omega t)}\sinh{(kz)} \vec{e}_z.
\end{equation}

\noindent L'intégration du profil de vitesse fournit alors une équation paramétrique des trajectoires :
\begin{equation}
\Delta x (t) = A \frac{k}{\omega}\cos{(kx - \omega t)}\cosh{(kz)} \quad \text{et} \quad \Delta z (t) = A \frac{k}{\omega}\sin{(kx - \omega t)}\sinh{(kz)}.
\end{equation}

\noindent En éliminant le temps, on trouve
\begin{equation}
\left(\frac{\Delta x}{\cosh{kz}}\right)^2 + \left(\frac{\Delta z}{\sinh{kz}}\right)^2 = \frac{A^2k^2}{\omega^2}.
\end{equation}
\noindent Les trajectoires sont donc des ellipses. 

Dans le cas des eaux profondes, et loin du fond, $kz \gg 1$ donc $\cosh{(kz)} \simeq \sinh{(kz)}$ et les trajectoires deviennent circulaires. On\footnote{C'est peut être un peu audacieux j'en conviens. Mais si la puissance de l'analyse dimensionnelle a priori est énorme, celle de l'analyse dimensionnelle a posteriori est infinie.} aurait pu le prédire par analyse dimensionnelle, puisque nous avons vu que dans le régime d'eaux profondes, les directions verticale et horizontale doivent être traitées de façon équivalente.

\textbf{Références :} 
\cite{GHP} Chapitre 6 - 4.1. et 4.2. ; \cite{gouttes} Chapitre 5 - 5.3. ; \cite{portelli} - Thème 21

\section{Ligne triple et mouillage}
\label{sec:ligne}
Très souvent, on s'intéresse aux propriétés d'une goutte d'un liquide déposée sur un solide, faisant intervenir une ligne de contact à trois phases, entre solide liquide et gaz. C'est aux phénomènes propres à ce type de systèmes que nous nous consacrons dans ce paragraphe.

\subsection{Généralités sur la ligne triple et le mouillage}
\label{subsec:gentriple}

\subsubsection{Paramètre d'étalement}
\label{subsubsec:gentriple_parametreetalement}

\begin{figure}[htb]
\centering
\includegraphics{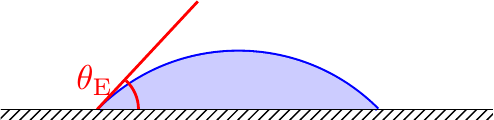}
\caption{Une goutte de liquide posée sur la surface d'un solide s'étale plus ou moins selon les affinités relatives des phases en présence. Si la goutte ne s'étale pas complètement, elle forme une calotte sphérique dont l'angle à la base est appelé angle de contact $\theta_\text{E}$.\label{fig:angle_contact_def}}
\end{figure}

Le mouillage est l'étude de l'étalement d'un liquide déposé sur un substrat, solide ou liquide. Cet étalement donne lieu à une ligne de contact entre trois phases : le liquide déposé, le substrat, et le gaz environnant. Cette ligne est appelée ligne triple. L'angle $\theta_\text{E}$ entre les interfaces liquide-gaz et solide-liquide au niveau de la ligne triple, comme représenté figure \ref{fig:angle_contact_def} est appelé angle de contact.

Le mouillage résulte de la compétition entre les affinités relatives des trois phases les unes pour les autres, décrites par les tensions de surface entre solide et gaz $\gamma_\text{SG}$, entre solide et liquide $\gamma_\text{SL}$ et entre liquide et gaz, $\gamma$. On appelle paramètre d'étalement $S$ la différence d'énergie surfacique entre le substrat sec et le substrat mouillé :
\begin{equation}
S  =E_\text{sec} - E_\text{mouillé} = \gamma_\text{SG} - (\gamma_\text{SL} + \gamma)
\end{equation}

Si $S>0$, l'énergie surfacique est abaissée en recouvrant le substrat avec le liquide : le liquide s'étale complètement en un film, dont l'épaisseur résulte de la compétition entre la capillarité et les forces d'interaction à l'échelle moléculaire. On parle de mouillage total.

Si $S<0$, le liquide ne s'étale pas entièrement mais forme une calotte sphérique, faisant un angle de contact $\theta_\text{E}$ avec le substrat\footnote{Le cas $\theta_\text{E} = 0$ correspond en fait à la situation de mouillage total, comme nous le verrons par la suite.}. On est dans une situation de mouillage partiel. On distingue alors un liquide plutôt mouillant, pour lequel $\theta_\text{E} \leq \pi/2$, et un liquide plutôt non mouillant, pour lequel $\theta_\text{E} \geq \pi/2$\footnote{On comprendra bien qu'il s'agit d'un abus de language : un liquide est plutôt mouillant ou non pour un substrat donné.}.

\textbf{Références :} 
\cite{gouttes} Chapitre 1 - 2.1. ; \cite{GHP} Chapitre 1 - 4.3.

\subsubsection{Loi de Young-Dupré}
\label{subsubsec:gentriple_youngdupre}

\paragraph{Démonstration :}

Jusque là, nous avons considéré la tension de surface comme une énergie surfacique. On peut aussi la considérer comme une force linéique s'exerçant sur la ligne triple.

\begin{figure}[htb]
\centering
\includegraphics{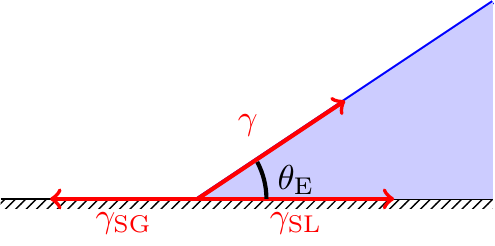}
\caption{Au niveau de la ligne triple, les tensions de surfaces des différentes interfaces sont en compétition.\label{fig:young_demo}}
\end{figure}

À l'équilibre, les forces de traction sur la ligne triple dues aux trois tensions de surface s'équilibrent (voir figure~\ref{fig:young_demo}), ce qui donne en projection dans le plan du substrat :
\begin{equation}
\gamma \cos{\theta_\text{E}} = \gamma_\text{SG} - \gamma_\text{SL}.
\end{equation}
\noindent Il s'agit de la loi de Young-Dupré. Elle montre que l'angle de contact résulte de la compétition entre les affinités des différentes phases en présence. Expérimentalement, cette relation permet, ayant mesuré deux des tensions superficielles et l'angle de contact, d'obtenir la troisième tension de surface. C'est une méthode commode pour obtenir les tensions de surface avec les phases solides car elles sont souvent très faibles et par conséquent difficiles à mesurer directement.

On peut alors réécrire le paramètre d'étalement 
\begin{equation}
S = \gamma (\cos{\theta_\text{E}} -1).
\end{equation}
\noindent On constate que l'angle de contact ne peut être défini que dans le cas d'un mouillage partiel, pour lequel $S<0$.

\paragraph{Limitation :}

Lorsque l'on pose une goutte sur une surface quelconque, l'angle de contact $\theta_\text{E}$ observé est souvent différent de celui prévu par la loi de Young-Dupré. En effet, celle-ci n'est valable que sur une surface idéale, sans impureté (défauts chimiques) ni rugosité (défauts physiques). Nous verrons plus tard des modèles permettant de prendre ces défauts en compte.

Sur une surface non-idéale, l'angle de contact n'est pas unique. Si l'on injecte du liquide dans la goutte, l'angle de contact $\theta$ augmente progressivement, et la ligne triple reste immobile jusqu'à ce que $\theta$ atteigne l'angle d'avancée~$\theta_\text{A}$, supérieur à l'angle $\theta_\text{E}$ prédit par la relation de Young-Dupré. A contrario, si l'on aspire du liquide, la ligne triple ne commence à reculer que quand $\theta$ atteint l'angle de reculée $\theta_\text{R}$, inférieur à $\theta_\text{E}$. On parle d'hystérésis de la ligne triple.

Cette hystérèse est due à l'ancrage de la ligne triple sur les défauts du substrat. Lorsque la ligne triple rencontre un défaut, elle se déforme (pour s'y accrocher si le défaut est une zone très mouillable, ou pour l'éviter si c'est une zone peu mouillable) puis finit par s'en arracher, ce qui dissipe de l'énergie.

Une bonne surface pour l'étude du mouillage donnera un faible hystérésis, c'est-à-dire que la différence entre les angles d'avancée et de reculée $\theta_\text{A}- \theta_\text{R}$ sera faible. Obtenir de telles surfaces nécessite un traitement particulier et de grandes précautions de manipulation.

\paragraph{La tension de surface vue comme une force linéique :}

Le point de vue adopté ici pour démontrer la loi de Young-Dupré a ses forces et ses faiblesses. Notons déjà qu'il est tout à fait possible de démontrer cette loi par des considérations énergétiques. Cette vision en termes de force linéique est assez intuitive, mais peut par là même se révéler trompeuse et soulever des interrogations légitimes de certaines personnes\footnote{Vous voyez de qui je veux parler, je pense...}. Pour une explication rigoureuse de ce point de vue, en regard avec les considérations thermodynamiques, on consultera l'article \cite{Marchand_2011}.

\begin{figure}[htb]
\centering
\includegraphics{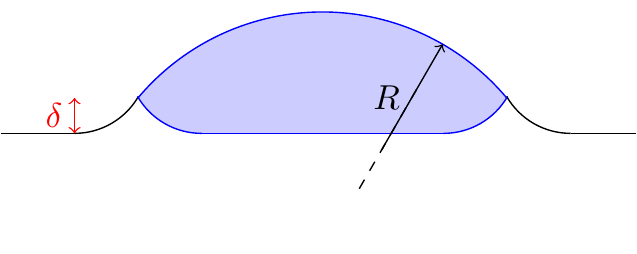}
\caption{La résultante verticale des forces de tension de surface sur la ligne triple déforme la surface solide jusqu'à être équilibrée par les forces de rappel élastique engendrées.\label{fig:longueur_elastocap}}
\end{figure}

Une première question saute aux yeux : nous avons équilibré les différentes forces dans le plan du substrat, mais \textit{quid} de la résultante dans la direction orthogonale ? Cette force existe, et déforme le substrat jusqu'à être compensée par les forces élastiques engendrées par la déformation. Comme représenté figure \ref{fig:longueur_elastocap}, notons $R$ le rayon de la goutte, $\delta$ la déformation du substrat et $E$ son module élastique : la déformation du substrat coûte une énergie $E \delta^2 R$ mais fait gagner une énergie de surface $\gamma \delta R$. \'Equilibrer les deux termes mène à une déformation $\delta \sim \gamma / E = \ell_\text{ec}$ : $\ell_\text{ec}$ est appelée longueur élasto-capillaire\footnote{Pour plus de détails sur les effets elastocapillaires, le lecteur consultera avec émerveillement les travaux de José Bico : \url{http://www.pmmh.espci.fr/~jbico/Research_fr.html}.}.

Le principal problème de cette approche est que l'on a tendance à oublier rapidement l'origine physique des phénomènes. En effet, on cherche souvent à équilibrer les forces dues aux tensions de surface liquide/solide et solide gaz : les résultats dépendent alors de la différence $\gamma_\text{SL}-\gamma_\text{SG}$, que l'on peut exprimer alors en fonction de la seule tension de surface liquide/gaz. Il ne faut cependant pas oublier que le phénomène sous-jacent est le mouillage, et donc la compétition entre les différentes tensions superficielles ! Le point de vue mécanique occulte généralement le rôle du substrat. C'est pour cette raison que je préfère généralement employer l'approche thermodynamique, bien qu'elle soit parfois plus lourde. Une autre vision de la tension de surface est de raisonner en termes de pressions, via la loi de Laplace : ce point de vue a les mêmes avantages et inconvénients que l'approche en termes de forces.

\textbf{Références :} 
\cite{gouttes} Chapitre 1 - 2.1. ; \cite{GHP} Chapitre 1 - 4.3. ; \cite{israelachvili} Chapitre 17 - 6.

\subsubsection{Nucléation hétérogène}
\label{subsubsec:gentriple_nuclhetero}

On va envisager un second scénario de nucléation, qui se produit généralement en l'absence de précaution particulière : on considère à nouveau la liquéfaction d'un gaz, mais cette fois-ci en présence d'impuretés (poussières, défauts sur les parois du récipient, etc.). On étudie maintenant un germe en forme de calotte sphérique\footnote{Là encore, nous considérons une calotte sphérique car nous savons qu'il s'agit de la forme minimisant l'énergie.}, de rayon~$r$ et d'angle de contact~$\theta$, se formant sur la surface solide d'une impureté. 

On conserve les notations du paragraphe \ref{subsubsec:nuclhomo_demo} : $\mu_\ell$ représente le potentiel chimique du liquide et $\mu_0$, celui du gaz qui tient lieu de réservoir. Le calcul de la contribution volumique au grand potentiel $\Omega^*$ reste valable mais il nous faut changer la contribution surfacique, puisque l'on a maintenant l'intervention d'une phase solide
\begin{equation}
\Delta \Omega ^* = \rho [\mu_\ell(T_0,P_0)-\mu_0(T_0,P_0)] V_\text{germe} + \Delta \Omega^*_\text{surf}.
\label{eq:potentiel_heterogene}
\end{equation}

Notons $A_\text{LG}$ la surface du germe en contact avec la vapeur, et $A_\text{P}$, celle en contact avec la paroi du substrat. Dans l'état de référence, le solide est entouré par la phase gazeuse, alors qu'en présence du germe, il est en partie couvert, donc
\begin{equation}
\Delta \Omega^*_\text{surf} = (\gamma A_\text{LG} + \gamma_\text{SL} A_\text{P}) - \gamma_\text{SG} A_\text{P} = \gamma A_\text{LG} + (\gamma_\text{SL}-\gamma_\text{SG})A_\text{P}. 
\end{equation}

\noindent Considérons une situation de mouillage partiel : la loi de Young-Dupré nous permet de réécrire
\begin{equation}
\Delta \Omega^*_\text{surf} = \gamma (A_\text{LG} - \cos{\theta} A_\text{P}).
\end{equation} 

\noindent Il nous reste à calculer les différents volumes et surfaces intervenant dans ces expressions. La surface de solide recouverte par le germe est un disque de rayon $r \sin{\theta}$ donc $A_\text{P} = \pi r^2 (\sin{\theta})^2$. La surface d'une calotte sphérique de rayon $r$ et de hauteur $h$ vaut $A_\text{LG} = 2\pi r h = 2\pi r^2(1-\cos{\theta})$. Enfin, son volume vaut $V_\text{germe} = \pi h^2 (3r-h)/3 = \pi r^3 (1-\cos{\theta})^2(2+\cos{\theta})/3$. Dès lors, l'équation \eqref{eq:potentiel_heterogene} se réécrit
\begin{equation}
\Delta \Omega ^* = \rho [\mu_\ell(T_0,P_0)-\mu_0(T_0,P_0)] \cdot \frac{\pi r^3}{3} (1-\cos{\theta})^2(2+\cos{\theta}) + \gamma [ 2\pi r^2(1-\cos{\theta}) - \pi r^2 \cos{\theta}(\sin{\theta})^2].
\end{equation}
\noindent Or, $2(1-\cos{\theta})-\cos{\theta}(\sin{\theta})^2 = (1-\cos{\theta})^2(2+\cos{\theta})$, donc, en utilisant \eqref{eq:potentiel_homogene}, on obtient :
\begin{equation}
\Delta \Omega ^* = \frac{(1-\cos{\theta})^2(2+\cos{\theta})}{4} \left[\rho [\mu_\ell(T_0,P_0) - \mu_0(T_0,P_0)] \cdot \frac{4 \pi r^3}{3} + 4\pi r^2 \gamma\right] = \frac{(1-\cos{\theta})^2(2+\cos{\theta})}{4} \left.\Delta \Omega^* \right|_\text{hom}.
\end{equation}

\begin{figure}[htb]
\centering
\includegraphics{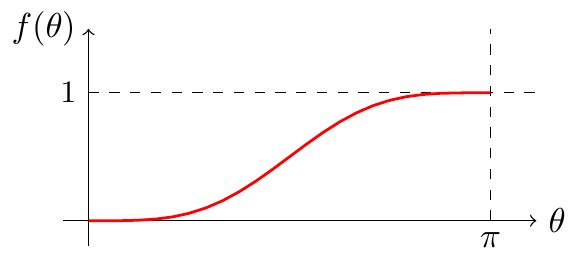}
\caption{Dans le cas de la nucléation hétérogène, la variation d'énergie associée à la formation d'une goutte est corrigée par un facteur $f(\theta)$, dépendant de l'angle de contact.\label{fig:courbe_nucleation_heterogene}}
\end{figure}

Le préfacteur $f(\theta) = (1-\cos{\theta})^2(2+\cos{\theta}) / 4$, tracé figure \ref{fig:courbe_nucleation_heterogene}, est une fonction croissante de $\theta$ sur $[0,\pi]$ et reste toujours inférieur à 1. Dès lors, les variations de $\Delta \Omega^*$ restent les mêmes que dans le cas homogène, et le scénario de transition reste similaire : la transition ne peut se faire que si un germe suffisamment gros apparaît lors d'une fluctuation, mais du fait du mouillage, la barrière énergétique associée à une telle fluctuation est notablement abaissée, d'autant plus que le liquide mouille la paroi. La transition se fera donc préférentiellement à partir d'une impureté.

De façon alternative, on peut considérer que le mouillage induit un abaissement effectif de la tension de surface liquide/gaz. Pour cela réécrivons $\Delta \Omega^*$ et $\left.\Delta \Omega^* \right|_\text{hom}$ en fonction du volume du germe :
\begin{equation}
\left\{ \begin{array}{l}
\Delta \Omega^* = \rho V_\text{germe} [\mu_\ell(T_0,P_0)-\mu_0(T_0,P_0)] + (36\pi)^{1/3} \gamma V_\text{germe}^{2/3} \\
\left.\Delta \Omega^* \right|_\text{hom} = \rho V_\text{germe} [\mu_\ell(T_0,P_0)-\mu_0(T_0,P_0)] + (36\pi)^{1/3} \gamma \left(\frac{1-\cos{\theta}}{2}\right)^{2/3} (2+\cos{\theta})^{1/3} V_\text{germe}^{2/3}.
\end{array}\right. 
\end{equation}

\noindent Le cas hétérogène est donc similaire au cas homogène, mais avec une tension de surface effective abaissée $\gamma_\text{eff} = \gamma [f(\theta)]^{1/3} \leq \gamma$.

Commentons enfin le cas du mouillage total, où l'on ne peut pas garder de contact à trois phases à l'équilibre. Si $\gamma_\text{SL} > \gamma_\text{SG}$, la valeur limite de l'angle de contact vaut $\pi$ et la goutte forme une sphère parfaite sur la paroi. Dans ce cas, $f(\pi)=1$ et il n'y a aucun changement par rapport à la nucléation homogène. Dans le cas contraire, l'angle de contact tend vers $0$ et le liquide recouvre entièrement le solide. Dans ce cas, $f(0)=0$ et la barrière énergétique s'annule : il n'y a plus d'état métastable. C'est ce qu'il se produit souvent dans le cas de la fusion, le liquide mouillant généralement bien sa propre phase solide, il n'y a pas de retard à la transition.

\textbf{Références :} 
\cite{diu_thermo} Complément G

\subsection{Tensiométrie par arrachement}
\label{subsec:arrachement}

Lorsque l'on tente de retirer un objet d'un fluide qui le mouille, le fluide se déforme pour accompagner l'objet et exerce une force sur celui-ci, liée à la tension de surface. La mesure de tension de surface à partir de la force d'arrachement est appelée méthode de Wilhelmy.

Pour mesurer cette force, on attache un objet (une plaque carrée pour la balance d'arrachement, un anneau toroïdal pour la méthode de du Nouÿ) à un dynamomètre, puis on le plonge dans le liquide à étudier. En abaissant lentement et sans à-coup\footnote{Lentement pour éviter d'ajouter une contribution dynamique à la force ; sans à-coup pour améliorer la précision de la mesure, puisqu'une la force capillaire ne peut être mesurée qu'avant l'arrachement.} le récipient contenant le liquide, on observe que la force mesurée par le dynamomètre augmente, passe par un maximum puis diminue après l'arrachement de l'objet.

\begin{figure}[htb]
\centering
\includegraphics{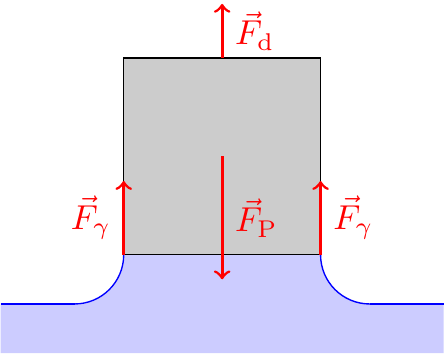}
\caption{Une plaque en surface d'un liquide est soumis à son propre poids et aux forces capillaires, verticales à la limite d'arrachement. \label{fig:arrachement}}
\end{figure}

Effectuons un bilan des forces appliquées à l'obstacle plongé dans le fluide, compensées par le rappel élastique $F_\text{d}$ exercé par le dynamomètre. L'objet est soumis à son poids $F_\text{P}$, à la poussée d'Archimède $F_\text{A}$, et à la traction exercée par la ligne triple $F_\gamma = p \gamma \cos{\theta}$\footnote{Plus précisément, la résultante verticale des tractions exercées par les trois tensions de surface en jeu.} où $p$ est le périmètre de la ligne triple\footnote{On n'oubliera pas qu'il y a parfois deux lignes triples à prendre en compte, de part et d'autre de l'obstacle.} et $\theta$ l'angle de contact.

Lorsque la force capillaire est alignée avec la verticale, situation représentée figure \ref{fig:arrachement}, la force mesurée par le dynamomètre est maximale\footnote{Pour être exact, la force passe par un maximum légèrement avant l'arrachement. En effet, l'épaisseur du film de liquide étant non nulle, il peut encore se recourber légèrement vers l'intérieur avant l'arrachement. La force qu'il convient de mesurer est bien la force maximale.} : à ce moment là, l'obstacle ne plonge plus dans le liquide, donc la poussée d'Archimède est nulle, et l'angle de contact est nul également. On a donc
\begin{equation}
F_\text{max} = p \gamma + F_\text{P}.
\end{equation}
\noindent Connaissant le poids de l'obstacle seul et ses caractéristiques géométriques, la mesure de $F_\text{max}$ permet d'obtenir une valeur de la tension de surface.

Cette méthode est relativement simple à mettre en \oe{}uvre et fournit des résultats d'une précision acceptable. Un biais est introduit par la contribution à la force mesurée du poids de fluide soulevé par l'obstacle : il est possible d'ajouter des corrections à la formule proposée, dépendant de la géométrie de l'objet. Néanmoins, en TP, votre principale source d'écart aux valeurs tabulées\footnote{Mise à part votre envisageable maladresse bien entendu.} proviendra de la pollution des fluides par des impuretés.

\textbf{Références :} 
\cite{gouttes} Chapitre 2 - 6.3. ; \cite{GHP} Chapitre 1 - 4.5.

\subsection{Milieux poreux}
\label{subsec:poreux}

\subsubsection{Condensation capillaire}
\label{subsubsec:poreux_condcap}

Considérons le cas d'un milieu poreux entouré d'une phase gazeuse à une pression inférieure à la pression de vapeur saturante, c'est-à-dire dans des conditions défavorables à la liquéfaction. Il est néanmoins possible de former une phase liquide à l'intérieur des pores si le liquide est plutôt mouillant : en effet, il sera alors plus favorable de recouvrir les parois solides des pores avec du liquide plutôt que de les laisser en contact avec le gaz.

\begin{figure}[htb]
\centering
\includegraphics{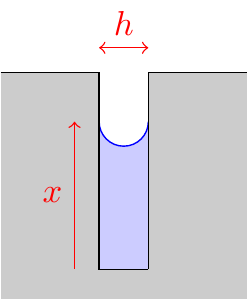}
\caption{Du fait de la tension de surface, il est possible de condenser une phase gazeuse dans des pores à des pressions inférieures à la pression de vapeur saturante.\label{fig:condensation_capillaire}}
\end{figure}

Pour rendre compte plus quantitativement de ce phénomène, considérons une fente, représentée figure \ref{fig:condensation_capillaire}, de largeur $h$ et de longueur $L$ dans un solide, en contact avec une phase gazeuse, considérée comme un réservoir de température $T_0$, de pression $P_0$ et de potentiel chimique $\mu_0(T_0,P_0)$\footnote{On suppose à nouveau avoir affaire à un fluide simple.}. On s'intéresse à l'évolution d'une phase liquide confinée dans la fente\footnote{On s'intéresse donc au problème de la stabilité d'une phase liquide déjà condensée. Le problème de la condensation se traite de façon similaire mais la géométrie est différente : le liquide commence par se condenser aux parois, formant une sorte de gaine intérieure, jusqu'à ce que les films de part et d'autre de la paroi se rejoignent (modèle de Cohan). Le phénomène est qualitativement le même, mais la géométrie modifie un peu les résultats, ce qui induit une hystérèse.}. Le potentiel thermodynamique adapté au problème est à nouveau $\Omega^*$.

Nous supposons en outre une séparation des échelles de temps : les équilibres mécaniques (forme, courbure, pression) au niveau de l'interface et l'équilibre thermique sont considérés comme très rapides par rapport aux équilibres osmotiques. Dès lors, le ménisque conserve à toute instant la forme d'une portion de cylindre, les lois de Young-Dupré ($\gamma \cos{\theta} = \gamma_\text{SG} - \gamma_\text{SL}$) et de Laplace ($P=P_0 - 2\gamma /r$\footnote{Attention au signe, le liquide est en dépression par rapport au gaz.}) sont valables, et la température $T$ du liquide égale celle du gaz. On note $r$ et $\theta$ le rayon de courbure et l'angle de contact de l'interface : on a $h = r \sin{\theta}$.

En notant $x$ la position du ménisque dans la fente, et en négligeant le volume occupé par le ménisque, on a\footnote{On n'oubliera pas qu'il y a deux interfaces avec la paroi, de part et d'autre de la fente !}
\begin{equation}
\mathrm{d} \Omega^* = \frac{2 \gamma}{r} hL\mathrm{d}x + \left[\mu_\ell\left(T_0,P_0-\frac{2 \gamma}{r}\right) - \mu_0(T_0,P_0)\right] \rho hL \mathrm{d}x + 2 (\gamma_\text{SL} - \gamma_\text{SG}) L \mathrm{d}x. 
\end{equation}

Considérons le gaz comme étant parfait : en notant $P_\text{sat}(T_0)$ la pression de vapeur saturante, on a
\begin{equation}\mu_0(T_0,P_0) = \mu_0(T_0,P_\text{sat}(T_0)) - k_\text{B} T_0 \ln{(P_\text{sat}(T_0)/P_0)}.
\end{equation}

\noindent Développons par ailleurs au premier ordre le potentiel chimique du liquide autour de la pression du gaz $P_0$ :
\begin{equation}
\mu_\ell\left(T_0,P_0 - \frac{2\gamma}{r}\right) \simeq \mu_\ell(T_0,P_0) - \frac{2 \gamma}{\rho r}.
\end{equation}
\noindent En supposant la pression $P_0$ proche de la pression de vapeur saturante, on peut développer le potentiel chimique $\mu(T_0,P_0)$ du liquide à interface plane autour de la pression de vapeur saturante : 
\begin{equation}
\mu_\ell(T_0,P_0) \simeq \mu_\ell(T_0,P_\text{sat}(T_0)) + (P_0 - P_\text{sat}(T_0))/\rho.
\end{equation} 

En outre, par définition de la pression de vapeur saturante, $\mu_\ell(T_0,P_\text{sat}(T_0)) = \mu_0(T_0,P_\text{sat}(T_0))$. On a donc
\begin{equation}
\mu_\ell\left(T_0,P_0 - \frac{2\gamma}{r}\right) \simeq \mu_0(T_0,P_0) + k_\text{B} T_0 \ln{\frac{P_\text{sat}(T_0)}{P_0}} + \frac{P_0 - P_\text{sat}(T_0)}{\rho} - \frac{2 \gamma}{\rho r}.
\end{equation}

\noindent Enfin, $P_0 - P_\text{sat}(T_0) \leq P_0 = k_\text{B}T_0 \rho_\text{gaz}$ : puisque $\rho_\text{gaz} \ll \rho$, on va négliger le terme $(P_0 - P_\text{sat}(T_0))/\rho$ par rapport à $k_\text{B} T_0 \ln{(P_0/P_\text{sat}(T_0))}$ donc
\begin{equation}
\mu_\ell\left(T_0,P_0 - \frac{2\gamma}{r}\right) - \mu_0(T_0,P_0) \simeq k_\text{B} T_0 \ln{\frac{P_\text{sat}(T_0)}{P_0}} - \frac{2 \gamma}{\rho r}.
\end{equation}

\noindent La différentielle du potentiel se réécrit dès lors
\begin{equation}
\mathrm{d} \Omega^* = \left[ 2(\gamma_\text{SL} - \gamma_\text{SG}) + k_\text{B} T_0 h \rho \ln{\frac{P_\text{sat}(T_0)}{P_0}} \right] L \mathrm{d}x.
\end{equation} 

Le potentiel étant décroissant au cours de l'évolution, la fente ne se remplit d'eau que pour $\mathrm{d}x > 0$, c'est-à-dire $\partial_x \Omega^* <0$, soit encore\footnote{Cette équation fait penser à celle que l'on a trouvé avec la relation de Kelvin \eqref{eq:Kelvin} : c'est normal, il s'agit plus ou moins du même problème, mais vu sous un angle différent. Ici, la courbure de l'interface est imposée par le mouillage, ce qui modifie la pression de coexistence liquide-gaz.}
\begin{equation}
h < h_\text{c} = \frac{2(\gamma_\text{SG} - \gamma_\text{SL})}{k_\text{B} T_0 \rho \ln{(P_\text{sat}(T_0)/P_0)}}.
\end{equation}

Ainsi, il est possible d'avoir condensation à des pressions inférieures à la pression de vapeur saturante si l'on a un liquide mouillant est favorable, c'est-à-dire $\gamma_\text{SG} > \gamma_\text{SL}$ : dans ce cas, la fente se remplit d'eau.

La grandeur $P_0/P_\text{sat}(T_0)$ définit l'humidité relative de la phase gazeuse. Pour une humidité de $50\%$, en considérant un angle de contact faible, la taille critique de pore pour avoir condensation capillaire de l'eau à température ambiante vaut $h_\text{c} \simeq \SI{1.5}{\nano\meter}$.

\textbf{Références :} 
\cite{gouttes} Chapitre 1 - 2.1. ; \cite{GHP} Chapitre 1 - 4.3. ; \cite{barrat_hansen} Chapitre 6 - 4.

\subsubsection{Succion capillaire}
\label{subsubsec:poreux_succion}

La condensation capillaire explique la propension des matériaux granulaires à prendre l'humidité. En effet, de nombreux\footnote{S'il y en a 99, on peut en faire 100. \og Gouverner, c'est prévoir. \fg{} \'E. de Girardin.} pores nanométriques existent, et permettent le stockage d'eau liquide. Les ponts capillaires ainsi formés à l'intérieur des pores exercent des forces attractives entre grains, responsables de la cohésion des matériaux granulaires humides.

\begin{figure}[htb]
\centering
\includegraphics{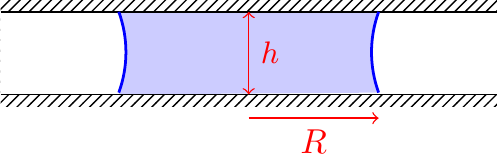}
\caption{Un pont capillaire entre deux surfaces exerce une force de succion entre celles-ci.\label{fig:succion}}
\end{figure}

Considérons la situation de la figure \ref{fig:succion} : deux plans distants de $h$ et reliés par un pont capillaire liquide de base circulaire de rayon $R$\footnote{Nous supposons donc implicitement avoir un liquide mouillant : $\gamma_\text{SG} > \gamma_\text{SL}$.}, et calculons le travail $\delta W$ nécessaire pour les éloigner de $\mathrm{d} h$.

En négligeant le volume des ménisques\footnote{Ce qui revient à supposer $h \ll R$, la courbure du ménisque étant imposée par la relation de Young-Dupré.}, le pont capillaire a un volume $V=\pi R^2 h$. La surface de contact entre le pont et les plaques est $S=2 \cdot (\pi R^2)$. \'Eloigner les plans de $\mathrm{d} h$ diminue la surface mouillée de $\mathrm{d}S = 4 \pi R \mathrm{d} R$ donc le travail à fournir au système est
\begin{equation}
\delta W = (\gamma_\text{SG} - \gamma_\text{SL}) \cdot 4 \pi R \mathrm{d} R.
\end{equation}

\noindent Si l'on suppose en outre que la transformation se fait à volume constant, $2(\mathrm{d}R / R) + (\mathrm{d}h)/h =0$, on trouve :
\begin{equation}
\delta W = - (\gamma_\text{SG} - \gamma_\text{SL}) \cdot \frac{2 \pi R^2}{h} \mathrm{d}h = - \frac{2 \pi R^2 \gamma \cos{\theta}}{h} \mathrm{d}h
\label{eq:travail_succion}
\end{equation}
\noindent où l'on a employé la loi de Young-Dupré.

On peut dès lors en tirer la force de traction exercée par le pont capillaire sur les plaques
\begin{equation}
F = \frac{2\pi R^2 \gamma \cos{\theta}}{h}.
\end{equation}
\noindent Au vu du signe dans l'équation \eqref{eq:travail_succion}, cette force s'oppose à l'écartement des plaques. Il est facile de se convaincre de son existence en essayant d'éloigner deux plaques de plexiglas séparées par un fin film d'eau. Une fois de plus, j'ai choisi un point de vue thermodynamique, mais la démonstration peut se faire aisément en utilisant directement la loi de Laplace.

Cette force permet aussi d'expliquer la cohésion des matériaux granulaires humides. La force exercée par le pont capillaire entre deux grains est de plusieurs ordres de grandeurs supérieure au poids des grains\footnote{La formule que nous avons établie ici ne peut être directement utilisée, la géométrie différant quelque peu. On consultera \cite{andreotti} pour plus de détails sur la cohésion des milieux granulaires, et leur physique plus généralement.}.

\textbf{Références :} 
\cite{gouttes} Chapitre 1 - 1.4. ; \cite{israelachvili} Chapitre 17 - 11.

\subsubsection{Imprégnation et loi de Washburn}
\label{subsubsec:poreux_washburn}

\paragraph{Position du problème :}

Un liquide a tendance à envahir spontanément une matrice poreuse qu'il mouille. C'est ce que l'on observe par exemple en trempant un sucre dans du café.

On va ici considérer le problème de l'imprégnation entre deux plaques de largeur $w$, distantes de $e \ll w$, placées au contact d'un réservoir de liquide à la pression atmosphérique. Nous supposerons les plaques horizontales, de façon à ne pas considérer l'influence de la pesanteur\footnote{Le résultat obtenu est également valable aux premiers instants de la montée dans un dispositif vertical, tant que le poids du fluide soulevé reste négligeable devant les forces capillaires.}. 

Nous allons ici adopter une description hydrodynamique du problème : le mouillage impose un angle de contact, donné par la loi de Young-Dupré, et donc une courbure de l'interface. Le liquide au niveau de l'interface est alors en dépression par rapport au réservoir et un écoulement s'établit. Trois phénomènes entrent en compétition : la dépression due à la capillarité, moteur de l'écoulement d'une part, et l'inertie et les frottements visqueux, qui s'y opposent d'autre part.

\begin{figure}[htb]
\centering
\includegraphics{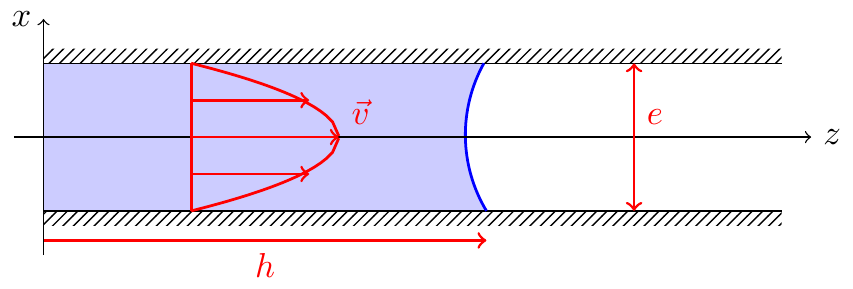}
\caption{Infiltration d'un liquide entre deux plans horizontaux distants d'une épaisseur $e$. La zone $z<0$ est occupée par un réservoir de liquide à pression atmosphérique. \label{fig:washburn}}
\end{figure}

Plus précisément, nous allons résoudre l'équation de Navier-Stokes. Soit $\vec{e}_z$ l'axe de l'écoulement et $\vec{e}_x$ l'axe orthogonal aux plaques, on prendra l'origine à l'entrée des plaques, au milieu du canal, comme présenté figure \ref{fig:washburn}. Loin de l'interface, le profil de vitesses prend alors la forme $\vec{v} = v(x) \vec{e}_z$.

\paragraph{Régime stationnaire :}

On suppose tout d'abord être en régime stationnaire\footnote{Cela peut sembler étrange puisque $h$ dépend du temps. Plus précisément, nous négligeons le terme d'instationnarité $\rho \partial_t \vec{v}$ par rapport au terme visqueux $\eta \partial^2_{x,x} \vec{v}$ dans l'équation de Navier-Stokes. Cela nécessite que l'échelle de temps de l'écoulement soit grande devant le temps~$e^2/\nu$ de diffusion de la quantité de mouvement sur l'épaisseur $e$ : il s'agit de l'hypothèse que nous reverrons avec l'approximation de lubrification.} et à bas nombre de Reynolds\footnote{Hypothèse peu restrictive puisque l'on est dans un milieu très confiné : pour de l'eau dans un (gros) capillaire de $\SI{1}{\milli\meter}$, le nombre de Reynolds vaut $1$ pour des écoulements à $\SI{1}{\meter\per\second}$, vitesse assez considérable à ces échelles !}. L'écoulement est alors décrit par l'équation de Stokes stationnaire
\begin{equation}
\eta \frac{\partial^2 v}{\partial x^2} = \frac{\partial P}{\partial z}.
\label{eq:Stokes}
\end{equation}

\noindent La vitesse ne dépendant pas de $z$, le gradient de pression dans la direction de l'écoulement est constant et vaut, d'après la loi de Laplace, 
\begin{equation}
\frac{\partial P}{\partial z} = \frac{P(h(t)) - P(0)}{h(t)} = - \frac{2\gamma \cos{\theta}}{e h(t)} < 0
\end{equation}
\noindent où $\theta$ est l'angle de contact à la ligne triple et $h(t)$ la position du ménisque.

\noindent Dès lors, en imposant le non-glissement du fluide à la paroi $v(\pm e/2)=0$, l'intégration de l'équation de Stokes \eqref{eq:Stokes} fournit
\begin{equation}
v(x) = -\frac{1}{2\eta} \left[ \left(\frac{e}{2}\right)^2 - x^2 \right] \frac{\partial P}{\partial z} = \frac{\gamma \cos{\theta}}{\eta e h(t)} \left[ \left(\frac{e}{2}\right)^2 - x^2 \right].
\end{equation}

Sans grande surprise\footnote{On a un écoulement visqueux généré par un gradient de pression.}, on obtient un profil d'écoulement de Poiseuille. On peut alors en calculer le débit volumique
\begin{equation}
Q = w \int_{-e/2}^{e/2} v(x) \mathrm{d}x = \frac{w e^2 \gamma \cos{\theta}}{6 \eta h}.
\end{equation}

\noindent La conservation du débit\footnote{Nous considérions un écoulement incompressible, vous vous en serez doutés.} fournit alors $Q = \dot{h} e w$ soit
\begin{equation}
\dot{h} = \frac{e\gamma \cos{\theta}}{6 \eta h}
\end{equation}
\noindent qui s'intègre, en prenant $h(0)=0$, en
\begin{equation}
h^2(t)=\frac{e\gamma \cos{\theta}}{3 \eta h} t.
\end{equation}
\noindent Il s'agit de la loi de Washburn ($h(t) \sim \sqrt{t}$) qui décrit la dynamique d'imprégnation d'un capillaire.

En présence de pesanteur, cette relation reste valable tant que le poids de la colonne d'eau entraînée est négligeable devant la force capillaire, c'est-à-dire tant que $\rho g \ll \partial_z P$ soit encore, en utilisant la solution obtenue, tant que
\begin{equation}
t \ll \tau = \frac{\eta \gamma \cos{\theta}}{e^3 (\rho g)^2}.
\end{equation}

Notons enfin que nous avons négligé les contraintes pouvant s'exercer sur la ligne triple du fait de son élasticité.

\textbf{Références :} 
\cite{gouttes} Chapitre 5 - 4.1. et 4.2. ; \cite{GHP} Chapitre 1 - 4.3.

\subsubsection{Quelques mots sur les écoulements dans les milieux poreux}
\label{subsubsec:poreux_rqs}

\paragraph{Loi de Darcy :}

Comme nous venons de le voir, pour un écoulement de Poiseuille, le débit volumique est proportionnel au gradient de pression, ce correspond à une loi de transport linéaire. On peut effectuer une analogie électrocinétique, où le débit équivaut à l'intensité et le gradient de pression à la tension. Dans le cas de l'écoulement de Poiseuille plan, on a 
\begin{equation}
\| \nabla P \| = \frac{24 \eta}{e^3 w} Q = R_\text{h} Q
\end{equation}
\noindent où $R_\text{h}$ est appelée résistance hydrodynamique et dépend des propriétés géométriques du canal considéré ainsi que de la viscosité du fluide, à laquelle elle est toujours proportionnelle. Dans le cas de l'écoulement de Poiseuille cylindrique, $R_\text{h} = \pi r^4 / (8\eta)$.

Dans un milieu poreux, les écoulements se font en parallèle dans de nombreux canaux micrométriques (voire nanométriques). En appliquant une loi de composition de résistances hydrodynamiques en parallèle, on obtient la loi de Darcy pour un milieu poreux soumis à un gradient de pression
\begin{equation}
Q = \left( \sum_\text{$i$ canaux} \frac{1}{R_\text{h}^i} \right) \|\nabla P \|.
\end{equation}

\noindent En supposant les écoulements homogènes, on en tire une loi locale entre la vitesse et le gradient de pression
\begin{equation}
\vec{v} = - \frac{k}{\eta} \overrightarrow{\nabla} P
\end{equation}
\noindent où~$k$ désigne la perméabilité du milieu, homogène à une surface, et qui décrit l'organisation microscopique des canaux d'écoulement. Divers modèles théoriques\footnote{Un des modèles les plus utilisés est celui de Carman-Kozeny : pour un milieu poreux formé de particules de rayon~$a$ à une compacité~$\phi$, on a~$k=(1-\phi)^3 a^2/45\phi^2$. Cette relation est néanmoins discutable, ainsi que l'hypothèse d'écoulement homogène qui est souvent effectuée. On pourra consulter l'article \url{http://pubs.rsc.org/en/content/articlelanding/2014/sm/c3sm52528g}.} existent pour relier la perméabilité aux caractéristiques géométriques du milieu poreux.

\textbf{Références :} 
\cite{rheophysique} Chapitre 3 - 7.9.

\paragraph{Récupération du pétrole :}

Un des moteurs majeurs de la recherche actuelle sur les écoulements dans les milieux poreux est la récupération assistée de pétrole\footnote{Enhanced Oil Recovery en anglais.}, ou récupération tertiaire. Le pétrole est emprisonné dans des roches poreuses et son extraction se fait en trois étapes. La récupération primaire se produit simplement en forant un puits jusqu'aux roches : le pétrole jaillit alors spontanément jusqu'à équilibrage de la pression. Il est ainsi possible de récupérer environ $10\%$ du pétrole. Ensuite, on injecte de l'eau sous pression dans la roche afin de poursuivre l'extraction : il s'agit de la récupération secondaire permettant une extraction de $30\%$ de la ressource disponible. Cependant, l'eau n'envahit pas uniformément la roche. \'Etant moins visqueuse que le pétrole, le front d'invasion se déstabilise et forme des doigts\footnote{Il s'agit de l'instabilité de digitation visqueuse de Saffman-Taylor, que nous étudierons au paragraphe \ref{subsec:saffman}.}, finissant par percoler à travers le poreux, et la tension de surface eau/pétrole étant élevée, le pétrole reste piégé à l'extérieur du canal d'écoulement de l'eau. Il faut alors passer à la récupération tertiaire, en injectant un fluide, idéalement aussi visqueux que le pétrole et ayant une faible tension de surface.

\section{Compétition entre gravité et mouillage}
\label{sec:gravite}
De nombreux phénomènes mettent en jeu une compétition entre la capillarité et la pesanteur dont nous ne nous sommes pas préoccupés jusqu'ici. Les effets de la gravitation étant volumiques, on s'attend à ce qu'ils deviennent dominants par rapport à la capillarité à de grandes échelles de longueur. Nous nous intéressons à cette compétition dans ce paragraphe.

\subsection{Gouttes dans le champ de pesanteur}
\label{subsec:gouttespes}

\subsubsection{Longueur capillaire et nombre de Bond}
\label{subsubsec:gouttespes_longueur_cap}

Considérons une goutte posée sur un substrat, dans une situation de mouillage partiel. Les effets surfaciques tendent à lui donner une forme de calotte sphérique, de rayon $R$ imposé par les tensions superficielles entre les phases en contact. Le gain en énergie surfacique est d'ordre $\delta E_\text{s} \sim \gamma R^2$\footnote{On a utilisé la relation de Young-Dupré.}. Néanmoins, adopter une telle forme élève le centre de gravité de la goutte, ce qui représente un coût en énergie de pesanteur $\delta E_\text{p} \sim \rho V_\text{goutte} g R \sim \rho g R^4$. On définit le nombre de Bond comme le rapport de ces deux contributions\footnote{Ou de façon équivalente comme le rapport de la pression hydrostatique sur la pression de Laplace.} :
\begin{equation}
\mathrm{Bo} = \frac{\rho g R^2}{\gamma} = \left(\frac{R}{\ell_\text{c}}\right)^2.
\end{equation}
\noindent Le nombre de Bond fait apparaître une longueur caractéritique $\ell_\text{c} =\sqrt{\gamma/\rho g}$ appelée longueur capillaire. Il s'agit de l'échelle de transition entre les régimes dominés par la capillarité et par la gravité, comme nous l'avons déjà vu au paragraphe sur les ondes de surface.

Pour l'eau à température ambiante, la longueur capillaire vaut $\ell_\text{c} \simeq \SI{2.7}{\milli\meter}$.

\textbf{Références :} 
\cite{gouttes} Chapitre 2 - 1. ; \cite{GHP} Chapitre 1 - 4.4.

\subsubsection{Forme de gouttes larges}
\label{subsubsec:gouttespes_gouttes_larges}

Des gouttes de taille petite devant la longueur capillaire $\ell_\text{c}$ ont une forme de calotte sphérique, de façon à minimiser la surface de l'interface liquide-gaz, et nous avons vu que leur courbure est déterminée par la compétition entre les différentes tensions au niveau de la ligne triple.

\begin{figure}[htb]
\centering
\includegraphics{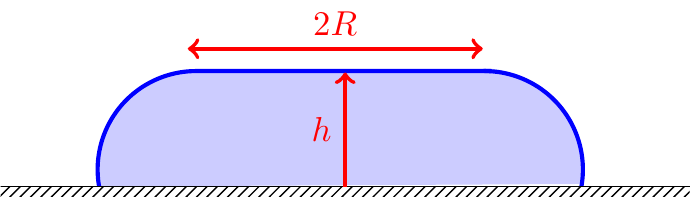}
\caption{Une goutte large devant la longueur capillaire adopte dans le champ de pesanteur une forme de galette aplatie, de rayon $R$ et d'épaisseur $h$. Nous avons ici exagéré la dimension des zones recourbées en périphérie de la goutte. \label{fig:forme_grosse_goutte}}
\end{figure}

Des gouttes de grande taille devant la longueur capillaire ont une forme de galette aplatie au centre, d'épaisseur $h$ sensiblement constante, comme représenté figure \ref{fig:forme_grosse_goutte}. Essayons d'estimer cette épaisseur, dans la limite de gouttes de rayon $R$ grand devant $\ell_\text{c}$. Dans ce cas, le volume du ménisque est négligeable. Le volume de la goutte vaut alors $V \simeq \pi R^2 h$ et la surface de ses deux interfaces, $A \simeq \pi R^2$. On considère que la goutte est à volume constant : le potentiel adapté est alors $F^* = U - T_0S$ où $T_0$ est la température du gaz environnant. Sa différentielle pour une transformation à volume constant s'écrit\footnote{On n'oubliera pas que dans l'énergie de pesanteur, c'est l'élévation du centre de masse de la goutte qui compte, d'où le facteur $2$.}
\begin{align}
\mathrm{d} F^* &= (T-T_0) \mathrm{d}S + (\gamma + \gamma_\text{SL} -\gamma_\text{SG}) \mathrm{d}A + \rho V g \frac{\mathrm{d} h}{2} \nonumber \\
			   &= (T-T_0) \mathrm{d}S + (\gamma + \gamma_\text{SL} -\gamma_\text{SG}) \cdot 2\pi R \mathrm{d} R + \rho \pi R^2 h g \frac{\mathrm{d} h}{2}.
\end{align}
\noindent La contrainte de volume constant impose $2\mathrm{d}R/R=-\mathrm{d}h/h$ donc
\begin{equation}
\mathrm{d} F^* = (T-T_0) \mathrm{d}S + \pi R^2 \left[ \frac{(\gamma_\text{SG} - \gamma - \gamma_\text{SL})}{h} + \frac{\rho g h}{2} \right] \mathrm{d} h.
\end{equation}

La minimisation du potentiel nous donne d'une part, sans surprise, que la température d'équilibre de la goutte est $T_0$, et d'autre part, que l'épaisseur d'équilibre est donnée par
\begin{equation}
h_\text{eq} = \sqrt{\frac{2(\gamma_\text{SG}-\gamma-\gamma_\text{SL})}{\rho g}}.
\end{equation}
\noindent En utilisant la relation de Young-Dupré et en se souvenant que $1-\cos{\theta} = 2 \sin{(\theta / 2)}^2$, on obtient en introduisant la longueur capillaire l'expression de l'épaisseur d'équilibre
\begin{equation}
h_\text{eq} = 2 \ell_\text{c} \sin{\frac{\theta}{2}}.
\label{eq:hauteur_goutte}
\end{equation}

\noindent Comme attendu, $h$ tend vers zéro dans une situation de mouillage total ($\theta =0$) : dans ce cas, un film mince apparaît dont l'étude nécessite une prise en compte plus fine des interactions entre le fluide et la paroi.

\textbf{Références :} 
\cite{gouttes} Chapitre 2 - 2.3.

\subsubsection{Tensiométrie par analyse de forme}
\label{subsubsec:gouttespes_tensiometrie}

\paragraph{Méthode de la goutte sessile :}

En observant la forme d'une grosse goutte posée sur une surface, il est possible de mesurer la tension de surface liquide/gaz $\gamma$ en analysant sa forme. Si l'on mesure l'angle de contact $\theta$ et la hauteur de la goutte en son centre $h$, la relation \eqref{eq:hauteur_goutte} permet d'obtenir la tension de surface.

\paragraph{Méthode de la goutte pendante :}

\begin{figure}[!htb]
\centering
\includegraphics{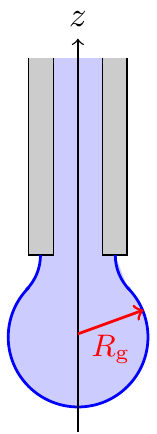}
\caption{Forme d'une goutte pendante au bout d'un capillaire. L'analyse de sa forme permet de mesurer la tension de surface. \label{fig:goutte_pendante}}
\end{figure}

On laisse pendre une goutte à partir d'un fin tube capillaire comme illustré figure \ref{fig:goutte_pendante}. La forme théorique $r(z)$ prise par la goutte est obtenue en minimisant l'énergie totale de la goutte. Elle satisfait à l'équation\footnote{On l'obtient de façon similaire à ce que l'on a fait au paragraphe \ref{subsec:gen_forme}, mais en ajoutant une énergie de pesanteur et sans contrainte de volume.}
\begin{equation}
\gamma \left[ - \frac{r''}{(1+r'^2)^{3/2}} + \frac{1}{r\sqrt{1+r'^2}} \right] = \rho g z.
\end{equation}

Cette équation peut être résolue numériquement, puis adaptée à la forme observée expérimentalement en prenant la tension de surface comme paramètre d'ajustement.

\paragraph{Stalagmométrie :}

Une goutte pendante se détache quand son poids excède la force capillaire qui la retient au niveau de la ligne triple, c'est-à-dire $2\pi R \gamma = \rho g V_\text{g}$ où $V_\text{g}$ est le volume de la goutte. Quand elle chute, la goutte reprend une forme sphérique de rayon $R_\text{g}$. Si $R$ est le rayon intérieur du tube capillaire, on obtient ainsi la loi de Tate
\begin{equation}
R_\text{g} = \left( \frac{3}{2} \ell_\text{c}^2 R \right)^{1/3}.
\end{equation}

Cependant, lors du décrochage, la goutte s'étire et un pincement apparaît : le fluide au-dessus du pincement reste attaché au capillaire, et seule une fraction $\alpha V_\text{g}$ de la goutte pendante choit. La valeur de $\alpha$ se situe généralement autour de~$0.6$ et dépend du rapport $R/R_\text{g}$. La loi de Tate devient
\begin{equation}
R_\text{g} = \left( \frac{3}{2 \alpha(R/R_\text{g})} \ell_\text{c}^2 R \right)^{1/3}.
\end{equation}

Le coefficient $\alpha$ est tabulé dans certaines conditions, permettant ainsi la mesure de la tension de surface à partir du poids d'une goutte tombante. Cela rend discutable la méthode pour la détermination absolue d'une tension de surface. Elle peut néanmoins être employée pour une mesure comparative si l'on dispose d'un fluide étalon, de tension de surface connue.

\textbf{Références :} 
\cite{gouttes} Chapitre 2 - 6.1.1.

\subsection{Forme d'un ménisque}
\label{subsec:menisque}

\begin{figure}[htb]
\centering
\includegraphics{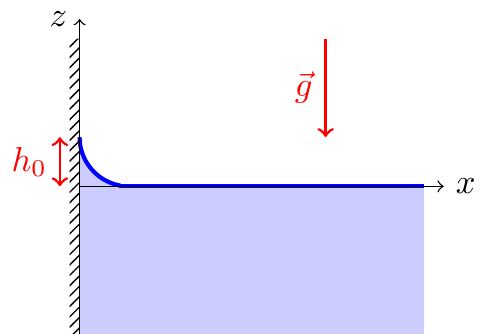}
\caption{Du fait de la compétition entre les différentes tensions de surface, la ligne triple d'un fluide plutôt mouillant (respectivement plutôt non mouillant) s'élève (respectivement s'abaisse) d'une hauteur $h_0$ le long d'une paroi solide.\label{fig:forme_menisque}}
\end{figure}

On s'intéresse à la forme que prend l'interface liquide-gaz près d'une paroi solide mouillée par le liquide. D'une part, le mouillage favorable tend à faire monter le liquide le long de la paroi, mais cela augmente l'énergie de pesanteur du système. Raisonnons en termes de pression. À l'équilibre, le fluide au repos et il ne doit donc pas y avoir de gradient de pression selon $x$, donc $P(x,0) = P(x\rightarrow \infty,0) = P_0$. La relation de l'hydrostatique nous donne $P(x,0) = P(x,z_0(x)) + \rho g z_0(x)$, et nous utilisons la loi de Laplace pour exprimer $P(x,z_0(x))$. Le rayon de courbure dans la direction $y$ est infini, et dans la direction $x$, en notant $z_0(x)$ l'altitude de la surface libre, il vaut
\begin{equation}
R(x) = \frac{(1+z_0'^2)^{3/2}}{z_0''}.
\end{equation}
\noindent L'équilibre des pressions donne donc finalement : 
\begin{equation}
\rho g z_0 = \gamma \frac{z_0''}{(1+z_0')^{3/2}} \quad \text{soit :} \quad  z_0 = \ell_\text{c}^2 \frac{z_0''}{(1+z_0')^{3/2}}.
\end{equation}
\noindent En multipliant par $z_0'$ de part et d'autre et en intégrant, on obtient :
\begin{equation}
z_0^2 = -\frac{2\ell_\text{c}^2}{\sqrt{1+z_0'^2}} + K
\end{equation}
\noindent où $K$ désigne une constante.

Loin de la paroi, l'interface redevient plane, donc $z_0(x\rightarrow \infty) = 0$ et $z_0'(x\rightarrow \infty)=0$. Ainsi, $K = 2\ell_\text{c}^2$. Le profil du ménisque est donc donné par l'équation différentielle non linéaire
\begin{equation}
z_0^2 = 2\ell_\text{c}^2 \left( 1 - \frac{1}{\sqrt{1+z_0'^2}} \right).
\label{eq:ED_menisque}
\end{equation}

La loi de Young Dupré impose l'angle de contact $\theta_\text{E}$ à la paroi, donc $z_0'(0) = - 1/\tan{\theta_\text{E}}$. On en tire donc la hauteur d'ascension le long de la paroi
\begin{equation}
h_0^2 = 2\ell_\text{c}^2 (1-\sin{\theta_\text{E}}).
\end{equation}

On peut également obtenir l'expression du profil du ménisque loin de la paroi, c'est-à-dire pour $z_0 \ll \ell_\text{c}$. Réécrivons l'équation \eqref{eq:ED_menisque} :
\begin{equation}
1+ z_0'^2 = \left( 1 - \frac{z_0^2}{2\ell_\text{c}^2}\right)^2
\end{equation}
\noindent En développant au deuxième ordre en $z_0/\ell_\text{c}$, et en considérant que la hauteur de l'interface diminue en s'éloignant de la paroi, on obtient
\begin{equation}
z_0' + \frac{z_0}{\ell_\text{c}} \simeq 0
\end{equation}
\noindent qui admet une solution exponentiellement décroissante $z_0(x) \simeq \exp{(-x/\ell_\text{c})}$. Les perturbations de la surface décroissent donc exponentiellement à grande distance.

Mentionnons enfin le cas de mouillage total où $\theta_\text{E} = 0$. Cette condition paraît impossible à satisfaire géométriquement, tout en gardant une hauteur d'ascension finie. Dans ce cas, au delà d'une hauteur de l'ordre de $h_0$, un film nanométrique de fluide s'élève pour recouvrir la paroi. La description de ce film nécessite la prise en compte des détails microscopiques des interactions entre le fluide et la paroi\footnote{Le lecteur intéressé pourra se reporter au chapitre 4 de \cite{gouttes}.}.

\textbf{Références :} 
\cite{gouttes} Chapitre 2 - 3.1. et 3.2. ; \cite{GHP} Chapitre 1 - 4.4.

\subsection{Ascension capillaire}
\label{subsec:jurin}

\begin{figure}[htb]
\centering
\includegraphics{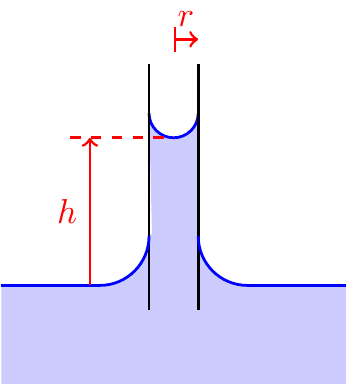}
\caption{Ascension d'un fluide plutôt mouillant dans un tube capillaire de rayon $r$. Du fait de la compétition entre les différentes tensions de surface, le fluide s'élève à une hauteur $h$ au-dessus de la surface libre. \label{fig:ascension_capillaire}}
\end{figure}

Nous avons vu au paragraphe \ref{subsubsec:poreux_washburn} sur la loi de Washburn que le mouillage permettait à un liquide d'envahir un tube capillaire. Nous allons maintenant considérer l'effet de la pesanteur sur l'ascension capillaire. Intéressons-nous donc au liquide à l'intérieur d'un tube cylindrique de rayon $r$ plongeant dans un réservoir de liquide à température $T_0$ et à pression $P_0$. Le potentiel thermodynamique adapté est $G^*$, et nous considérons que l'équilibre thermique et l'équilibre mécanique au niveau du ménisque\footnote{C'est-à-dire que le ménisque a sa forme d'équilibre et que la loi de Young-Dupré est vérifié.} sont réalisés. Dans ces conditions, la différentielle du potentiel s'exprime, si $h$ est la hauteur dont le fluide s'est élevé\footnote{Plus précisément, $h$ désigne la hauteur du bas du ménisque par rapport au niveau du fluide dans le réservoir, loin du tube capillaire.},
\begin{equation}
\mathrm{d} G^* = \rho g \pi r^2 h \frac{\mathrm{d}h}{2} + (\gamma_\text{SL}-\gamma_\text{SG})\cdot 2\pi r \mathrm{d}h.
\end{equation}
\noindent Le premier terme traduit l'augmentation d'énergie potentielle de pesanteur du fluide\footnote{Le facteur $2$ vient du fait que l'on regarde l'élévation du centre de gravité de la colonne de fluide, et nous supposons le réservoir de fluide suffisament large pour que l'abaissement de son niveau soit négligeable.} et le second, le changement d'énergie surfacique lié au mouillage. Notons que nous avons ici négligé la contribution du ménisque au poids de la colonne : cette approximation est valable si la hauteur de montée est grande devant la hauteur du ménisque. Nous avons vu au paragraphe \ref{subsec:menisque} que cette dernière est de l'ordre de la longueur capillaire : il faut donc avoir $r \leq \ell_\text{c}$. Il s'agit de la définition d'un tube capillaire. Cette condition est en fait très peu restrictive dans la mesure où le ménisque  reste confiné à proximité de la paroi : son volume n'augmente donc quasiment pas quand $r>\ell_\text{c}$.

La minimisation du potentiel à l'équilibre fournit la hauteur maximale d'ascension du fluide
\begin{equation}
h_\text{J} = \frac{2(\gamma_\text{SG} - \gamma_\text{SL})}{\rho g r} = \frac{2 \gamma \cos{\theta}}{\rho g r}.
\end{equation}
\noindent Il s'agit de la loi de Jurin. On constate que plus le tube capillaire est large, plus la hauteur d'ascension est faible.

La mesure de hauteur d'ascension capillaire est une manière de mesurer la tension de surface. Cependant, l'angle de contact est mal connu pour les raisons évoquées au paragraphe \ref{subsubsec:gentriple_youngdupre} : on le suppose généralement faible pour ne pas se préoccuper du terme en $\cos{\theta}$, mais cela entache la mesure d'une erreur systématique. De plus, l'hystérésis de l'angle de contact, lié à l'ancrage de la ligne triple aux défauts du capillaire, peut causer une montée plus faible que celle prévue par la loi de Jurin. Pour limiter cet effet, il convient de nettoyer soigneusement les capillaires, et de travailler en laissant redescendre le ménisque : de la sorte, la pesanteur aide la ligne triple à se décrocher des défauts.

Dans un capillaire de quelques centaines de microns, la hauteur d'ascension de l'eau est de l'ordre de quelques centimètres. Dès lors, l'ascension capillaire permet d'expliquer la montée de la sève dans de petites plantes. Elle ne suffit néanmoins pas, comme on l'entend parfois dire, pour expliquer la montée de la sève dans les arbres : pour cette dernière, un mécanisme supplémentaire intervient, l'évapotranspiration\footnote{Rapidement, l'évaporation de l'eau au niveau des feuilles, en sortie de vaisseaux capillaires, induit une forte courbure des ménisques, ce qui met le fluide en forte dépression et engendre ainsi un écoulement ascendant.}.

En parlant d'ascension, nous évoquons implicitement le cas d'un liquide plutôt mouillant, pour lequel $\theta_\text{E} \leq \pi/2$ et $h_\text{J}>0$. Il ne s'agit pas néanmoins d'une hypothèse dans nos calculs, qui s'appliquent aussi au cas d'un liquide plutôt non-mouillant : dans ce cas, $h_\text{J}<0$ et le ménisque s'abaisse dans le capillaire. Notons enfin que la remarque de la fin paragraphe \ref{subsec:menisque} sur le cas de mouillage total $\theta_\text{E}=0$ reste valable.

\textbf{Références :} 
\cite{gouttes} Chapitre 2 - 4. ; \cite{GHP} Chapitre 1 - 4.4.

\section{Hydrodynamique aux petites échelles}
\label{sec:hydro}
La tension de surface influence de façon importante l'hydrodynamique des films minces, ce qui induit des écoulements et des instabilités des interfaces : l'étude de ces effets fait l'objet de ce paragraphe.

\subsection{\'Ecoulements quasi-parallèles}
\label{subsec:parallele}

Nous allons ici nous intéresser à des écoulements de fluides dont les lignes de courant sont quasiment parallèles, comme cela se produit dans les écoulements confinés entre deux parois proches, ou dans un film mince.

\subsubsection{Approximation de lubrification}
\label{subsubsec:parallele_lubrification}

\begin{figure}[htb]
\centering
\includegraphics{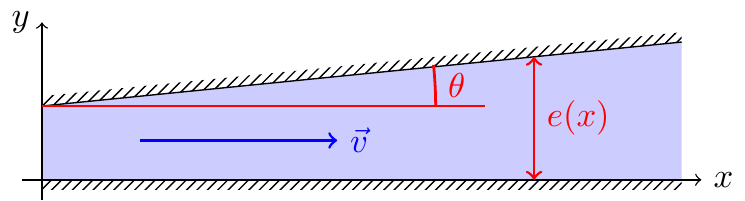}
\caption{\'Ecoulement quasi-parallèle entre deux plaques planes formant un petit angle $\theta$.\label{fig:lubrification}}
\end{figure}

Considérons le cas particulier, schématisé figure \ref{fig:lubrification}, d'un écoulement principalement selon la direction $\vec{e}_x$, entre deux parois planes\footnote{L'approximation de lubrification reste utilisable pour des parois quelconques tant que la variation de leur distance est lente à l'échelle de l'écoulement. Le cas de parois planes permet d'utiliser l'angle qu'elles forment comme petit paramètre, ce qui est commode.}. La paroi inférieure définit le plan $(xz)$, et la paroi supérieure forme un angle $\theta$ avec l'autre, de sorte que la distance séparant les plans dans la direction $\vec{e}_y$ s'écrit $e(x) = e_0 + x \sin{\theta}$.

La première hypothèse de l'approximation de lubrification consiste à supposer que les variations des grandeurs dans la direction orthogonale à l'écoulement moyen sont lentes (écoulement quasi-parallèle), soit ici
\begin{equation}
\frac{\partial e}{\partial x} = \sin{\theta} \ll 1 \quad \text{soit :} \quad \sin{\theta} \sim \theta \ll 1.
\end{equation} 
\noindent L'angle $\theta$ nous servira donc de petit paramètre dans la suite.

L'écoulement est supposé incompressible, donc $\overrightarrow{\nabla} \cdot \vec{v} = 0$. Notons $U$ la vitesse caractéristique dans la direction de l'écoulement moyen (c'est-à-dire selon $\vec{e}_x$) et $V$, celle selon la direction transverse (c'est-à-dire selon $\vec{e}_y$). En ordres de grandeur, nous avons donc
\begin{equation}
\frac{V}{U} \sim \frac{\delta e}{\delta x} = \theta \ll 1.
\end{equation}
\noindent  Ainsi, la vitesse transverse est du premier ordre par rapport à la vitesse moyenne car pour un écoulement quasi-parallèle, les lignes de vitesse suivent la paroi.

L'écoulement est solution de l'équation de Navier-Stokes
\begin{equation}
\left(\frac{\partial}{\partial t} + \vec{v}\cdot \overrightarrow{\nabla}\right) \vec{v} = -\frac{\overrightarrow{\nabla} p}{\rho} + \nu \Delta \vec{v}
\end{equation}
\noindent où $p$ désigne la pression corrigée de son éventuelle composante hydrostatique.

La taille typique de variation des propriétés de l'écoulement selon la direction transverse est donnée par l'épaisseur de la paroi, soit en ordres de grandeur $\partial_y \sim 1/e_0$. Estimons alors les ordres de grandeur des dérivées premières des composantes de la vitesse :
\begin{equation}
\frac{\partial v_x}{\partial y} \sim \frac{U}{e_0} \quad \text{et} \quad \frac{\partial v_y}{\partial y} \sim \frac{V}{e_0} \sim \frac{U}{e_0} \theta 
\end{equation}
\noindent d'où l'on déduit
\begin{equation}
\frac{\partial v_x}{\partial x} = - \frac{\partial v_y}{\partial y}  \sim \frac{U}{e_0} \theta .
\end{equation}

Nous pouvons en tirer les dérivées secondes suivantes
\begin{equation}
\frac{\partial^2 v_x}{\partial y^2} \sim \frac{U}{e_0^2} \quad \text{et} \quad \frac{\partial^2 v_x}{\partial x \partial y} = \frac{\partial^2 v_y}{\partial y^2} \sim \frac{U}{e_0^2} \theta .
\end{equation}

Pour poursuivre, nous devons nous doter d'une grandeur typique $L$ de variation dans la direction de l'écoulement moyen, par exemple la longueur sur laquelle se fait l'écoulement. Dans ce cas, on peut estimer les dérivées par rapport à $x$ par $\partial_x \sim 1/L$. Dès lors,
\begin{equation}
\frac{\partial v_y}{\partial x} \sim \frac{U}{L} \theta \; \text{,} \quad \frac{\partial^2 v_x}{\partial x^2} = - \frac{\partial^2 v_y}{\partial x \partial y} \sim  \frac{U}{e_0 L} \theta \quad \text{et} \quad \frac{\partial^2 v_y}{\partial x^2} \sim \frac{U}{L^2} \theta.
\end{equation}

La seconde hypothèse de l'approximation de lubrification consiste à négliger les termes non linéaires d'accélération convective devant les termes visqueux. En projection dans la direction de l'écoulement moyen, nous avons
\begin{equation}
v_x \frac{\partial v_x}{\partial x} \sim v_y \frac{\partial v_x}{\partial y} \sim \frac{U^2}{e_0}{\theta} \quad \text{et} \quad \nu \frac{\partial^2 v_x}{\partial x^2} \sim \frac{U\nu}{e_0 L}\theta \ll \nu \frac{\partial^2 v_x}{\partial y^2} \sim \frac{U\nu}{e_0^2}
\end{equation}
\noindent donc le terme visqueux domine si $\mathrm{Re} = Ue_0 / \nu \ll 1/\theta$. L'approximation de lubrification comporte donc une hypothèse de faible nombre de Reynolds, mais moins restrictive que la formulation en géométrie quelconque $\mathrm{Re} \ll 1$. La projection de l'équation de Navier-Stokes sur la direction de l'écoulement moyen se résume donc à l'équation de Stokes stationnaire
\begin{equation}
\frac{1}{\rho} \frac{\partial p}{\partial x} = \nu \frac{\partial^2 v_x}{\partial y^2}.
\label{eq:stokes_x}
\end{equation}

Préoccupons nous maintenant de la projection dans la direction transverse. Nous avons
\begin{equation}
v_x \frac{\partial v_y}{\partial x} \sim \frac{U^2}{L} \theta \; \text{,} \quad v_y \frac{\partial v_y}{\partial y} \sim \frac{U^2}{e_0}{\theta^2} \quad \text{et} \quad \nu \frac{\partial^2 v_y}{\partial x^2} \sim \frac{U \nu}{L^2}\theta \ll \nu \frac{\partial^2 v_y}{\partial y^2} \sim \frac{U \nu}{e_0^2} \theta
\end{equation} 
\noindent donc le gradient de pression $\partial_y p$ selon $y$ est d'ordre $\theta$ à comparer au gradient $\partial_x p$ selon $x$, donné par \eqref{eq:stokes_x}, qui n'est pas infinitésimal. Nous allons donc le supposer nul\footnote{Cela me semble être une escroquerie : il n'y a aucune raison de comparer ces deux gradients puisqu'ils ne sont pas en compétition dans les équations du système, et nous sommes qui plus est restés à l'ordre 1 jusque là. Je ne doute néanmoins pas de la pertinence de l'approximation de lubrification, mais je n'ai pas lu d'explication qui me semble satisfaisante sur ce sujet.}.

Revenons enfin sur l'hypothèse de stationnarité : elle est valable dans la limite où le terme $\partial_t \vec{v}$ de l'équation de Navier-Stokes est négligeable devant le terme visqueux $\nu \Delta \vec{v}$. En notant $T$ le temps caractéristique de l'écoulement, cette hypothèse est valable si
\begin{equation}
\frac{U}{T} \ll \frac{\nu U}{e_0^2} \quad \text{soit} \quad \frac{e_0^2}{\nu} \ll T.
\end{equation}
\noindent L'hypothèse de stationnarité revient donc à supposer que la diffusion de la quantité de mouvement sur l'épaisseur $e_0$ est rapide par rapport à l'écoulement.

Récapitulons enfin ce que nous avons vu. L'approximation de lubrification est constituée de trois hypothèses : 
\begin{itemize}
\item écoulement quasi-parallèle : les variations de direction de la vitesse sont lentes à l'échelle de l'écoulement, c'est-à-dire $\theta = \partial_x e \ll 1$.
\item écoulement visqueux : le nombre de Reynolds est faible, soit $\mathrm{Re} \ll 1/\theta$.
\item écoulement stationnaire : la diffusion est rapide sur l'épaisseur de fluide, soit $T \gg e_0^2 / \nu$.
\end{itemize}
\noindent L'écoulement est alors régi par les équations
\begin{align}
\frac{1}{\rho} \frac{\partial p}{\partial x} &= \nu \frac{\partial^2 v_x}{\partial y^2} \label{eq:lubrification_Stokes} \\ 
\frac{\partial p}{\partial y} &= 0. \label{eq:lubrification_nogradient}
\end{align}

\textbf{Références :} 
\cite{GHP} Chapitre 8 - 1.1. à 1.4. ; \cite{rheophysique} Chapitre 3 - 7.5.

\subsubsection{\'Equation de Reynolds pour un film mince}
\label{subsubsec:parallele_reynolds}

Considérons maintenant le cas d'un écoulement mince à surface libre au dessus d'une paroi plane immobile, en conservant les notations du paragraphe précédent. La continuité des contraintes tangentielles pour un fluide visqueux, ainsi que l'impénétrabilité de la paroi nous donnent les conditions aux limites suivantes :
\begin{equation}
\vec{v}(x,y=0) = \vec{0} \quad \text{et} \quad \frac{\partial v_x}{\partial y} (x,y=e(x,t)) = 0.
\label{eq:CL_film}
\end{equation}
\noindent Nous pouvons alors obtenir le profil de l'écoulement. L'équation \eqref{eq:lubrification_nogradient} nous informe que la pression est constante sur l'épaisseur du fluide, donc l'équation de Stokes stationnaire \eqref{eq:lubrification_Stokes} s'intègre en
\begin{equation}
v_x (x,y) = \frac{1}{2\eta} \left(\frac{\partial p}{\partial x}\right) y^2 + Ay + B
\end{equation}
\noindent et la prise en compte des conditions aux limites \eqref{eq:CL_film} nous permet d'aboutir à l'expression
\begin{equation}
v_x(x,y) = \frac{1}{2\eta} \left(\frac{\partial p}{\partial x}\right)[y^2 - 2e(x,t)y].
\end{equation}

Calculons alors le débit volumique de l'écoulement :
\begin{equation}
Q(x,t) = \int_0^{e(x,t)} v_x(x,y) \mathrm{d}y = -\frac{1}{3\eta} \left(\frac{\partial p}{\partial x}\right) e^3(x,t).
\end{equation}

Or, l'écoulement est incompressible, donc le débit se conserve : un bilan de matière sur une tranche fine du film de fluide nous donne $\partial_t e + \partial_x Q = 0$ soit
\begin{equation}
\frac{\partial e}{\partial t} = \frac{1}{3\eta} \frac{\partial}{\partial x}\left[\frac{\partial p}{\partial x} e^3(x,t)\right].
\label{eq:reynolds}
\end{equation}
\noindent Il s'agit d'un cas particulier de l'équation de Reynolds\footnote{L'équation de Reynolds correspond généralement au cas de l'écoulement de lubrification entre deux parois mobiles. Elle permet alors de calculer les forces exercées par le film de lubrification sur les parois.}.

\textbf{Références :} 
\cite{GHP} Chapitre 8 - 1.6. ; \cite{rheophysique} Chapitre 3 - 7.5.

\subsection{Instabilité de Rayleigh-Taylor}
\label{subsec:taylor}

\subsubsection{Analyse qualitative}
\label{subsubsec:taylor_qualit}

Nous allons nous intéresser à l'instabilité d'un film mince infini suspendu : on constate qu'il se déforme en une série de goutelettes, organisées de façon assez régulière si la surface est lisse. C'est ce que l'on observe pour les goutelettes de condensation sur la paroi intérieure haute d'un réfrigérateur, sur la plaque recouvrant une casserole, ou encore dans une bouteille d'eau.

Cette instabilité résulte de la compétition entre la tension de surface, qui tend à aplanir le film, et la gravité qui tend à abaisser son centre de gravité\footnote{J'espère que vous voyez la longueur capillaire approcher à grands pas.}, mais à volume imposé.

\begin{figure}[htb]
\centering
\includegraphics{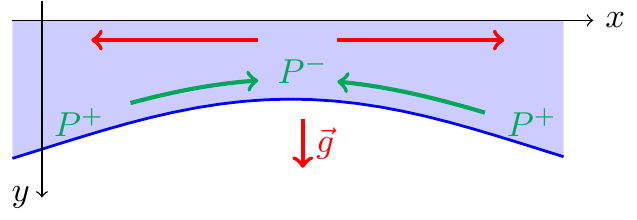}
\caption{Un film de liquide suspendu à une paroi se déstabilise pour former un ensemble de goutelettes. L'instabilité résulte de la compétition entre la pesanteur, qui tend à abaisser le centre de gravité du film en induisant un écoulement vers les zones les plus basses, et la tension superficielle, qui tend à minimiser la courbure de l'interface en ramenant du fluide vers les régions concaves.\label{fig:rayleigh_taylor_analyse}}
\end{figure}

Plus précisément, considérons que le film s'amincisse localement suite à une perturbation, comme représenté figure \ref{fig:rayleigh_taylor_analyse}. La courbure occasionnée induit une légère dépression derrière le creux du fait de la tension de surface : on a donc génération d'un gradient de pression, induisant à son tour un écoulement en direction du creux. Au contraire, la pesanteur induit un écoulement vers les zones qui se sont abaissées, c'est-à-dire vers l'extérieur du creux.

\subsubsection{Recherche des modes instables}
\label{subsubsec:taylor_modes_instables}

Pour simplifier le formalisme, considérons une situation bidimensionnelle\footnote{Ou de façon équivalente, cherchons une déformation en tole ondulée. Dans ce cas, l'énergie $\Delta E$ considérée ensuite est une énergie linéique.}, dans le plan $(xy)$. Nous allons maintenant effectuer une analyse en modes normaux, c'est-à-dire que nous allons étudier l'évolution temporelle de perturbations de la forme $e(x,t)=e_0 + \delta e(t) \cos{(qx)}$, avec $\delta e (t) \ll e_0$. Cherchons tout d'abord quels modes $q$ sont instables, et pour ce faire, évaluons la variation d'énergie associée à la perturbation sur une longueur d'onde~$\lambda = 2\pi/q$. Cette énergie contient un terme volumique dû à la pesanteur et un terme surfacique, et s'écrit
\begin{equation}
\Delta E = E(\delta e) - E(\delta e = 0) = -\int_0^\lambda \frac{\rho g}{2}(e(x,t)^2-e_0^2)\mathrm{d}x + \int_0^\lambda \gamma (\mathrm{d}s - \mathrm{d}x).
\end{equation}

On a $\mathrm{d}s = \sqrt{(\mathrm{d}e)^2 + (\mathrm{d}x)^2}$ donc pour de petites déformations, $\mathrm{d}s \simeq \mathrm{d}x [1+(\mathrm{d}e/\mathrm{d}x)^2/2]$ et ainsi,
\begin{equation}
\Delta E = \frac{\lambda \gamma (\delta e)^2}{4 \ell_\text{c}^2} [ (q \ell_\text{c})^2 -1] + \mathcal{O}(\delta e^3).
\end{equation}
\noindent Ainsi, $\Delta E < 0$ pour $q<1/\ell_\text{c}$ : tous les modes de longueur d'onde supérieure à $2\pi \ell_\text{c}$ sont donc instables, c'est-à-dire correspondent à une perturbation d'amplitude exponentiellement croissante avec le temps.

\subsubsection{Taux de croissance de l'instabilité}
\label{subsubsec:taylor_dynamique}

Bien qu'il existe une infinité de modes instables, seul l'un d'entre eux est expérimentalement observé : la sélection est faite par la cinétique de croissance. Le mode observé est celui dont le taux de croissance est le plus élevé.

Afin d'étudier la dynamique de l'instabilité, nous utilisons l'équation de Reynolds\footnote{On n'oubliera pas, pour orienter les axes, que dans notre étude, l'axe $y$ pointait vers la surface libre : il est donc descendant ici.} \eqref{eq:reynolds} obtenue au paragraphe \ref{subsubsec:parallele_reynolds}. Le gradient de pression le long de l'interface est ici égal au gradient de la pression de Laplace $\nabla P = \partial_x(-\gamma \partial^2_x e)$\footnote{Attention au signe ! Avec nos notations, la courbure est $C = - \partial^2_x e$.}. On obtient donc\footnote{On n'aura pas oublié que dans \eqref{eq:reynolds}, $p=P-\rho g y$ avec nos conventions d'orientation. Dans l'approximation de lubrification, le gradient de pression dans la direction transverse à l'écoulement est supposé nul, donc on peut écrire $p=P - \rho g e$.}
\begin{flalign}
&& \frac{\partial e}{\partial t} &= \frac{1}{3\eta}\frac{\partial}{\partial x}\left[\left(-\rho g \frac{\partial e}{\partial x} - \gamma \frac{\partial^3 e}{\partial x^3}\right) e^3 \right] \nonumber && \\
\text{d'où} && \frac{\mathrm{d} \delta e}{\mathrm{d} t} \cos{(qx)} = \frac{1}{3\eta}&\frac{\partial}{\partial x}\left[\left( \rho g q \delta e - \gamma q^3 \delta e\right) \sin{(qx)} (e_0 + \delta e \cos{(qx)})^3 \right] \nonumber && \\
\text{soit} &&  \frac{\mathrm{d} \delta e}{\mathrm{d} t} &= \frac{\rho g e_0^3}{3\eta} q^2 [1-(q\ell_\text{c})^2] \delta e
\end{flalign}
\noindent en restant au premier ordre en perturbation.

\begin{figure}[htb]
\centering
\includegraphics{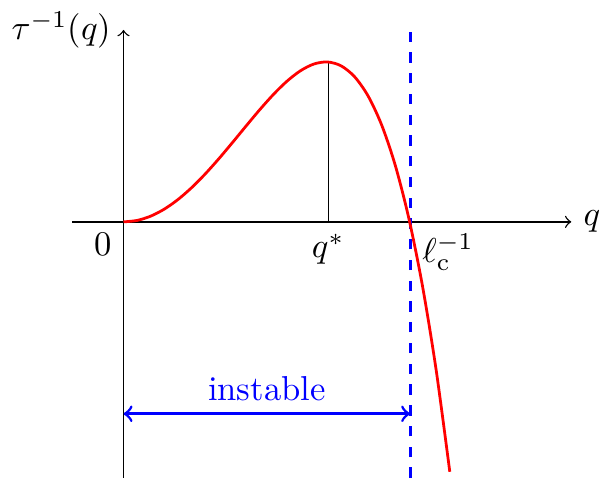}
\caption{Taux de croissance $\tau^{-1}(q)$ de l'instabilité de Rayleigh-Taylor. Les modes de vecteurs d'onde inférieurs à $\ell_\text{c}$ sont instables, et le mode de vecteur d'onde $q^*$ est cinétiquement sélectionné. \label{fig:courbe_rayleigh_taylor}}
\end{figure}

La perturbation $\delta e$ évolue donc exponentiellement selon $\delta e (t) = \delta e (0) \mathrm{e}^{t/\tau(q)}$ avec un temps caractéristique\footnote{Le cas traité ici peut aisément se transposer au problème de la relaxation d'un film mince de liquide posé sur une surface : le signe de la contribution de la pesanteur est simplement inversé. La perturbation est alors toujours exponentiellement amortie et le temps caractéristique est similaire, à un signe près.} 
\begin{equation}
\tau(q) = \frac{3\eta}{\rho g e_0^3 q^2} \frac{1}{1-(q \ell_\text{c})^2}
\end{equation}
\noindent tracé figure \ref{fig:courbe_rayleigh_taylor}. Nous retrouvons bien que les modes pour lesquels $q \ell_\text{c} < 1$ sont instables.

En outre, le temps de croissance $\tau(q)$ passe par un minimum pour $\tau^*$ en $q^*$ donnés par
\begin{equation}
\tau^* = \frac{12 \ell_\text{c}^2 \eta}{\rho g e_0^3} \quad \text{et} \quad q^* = \frac{2\pi}{\lambda^*} = \frac{1}{\sqrt{2} \, \ell_\text{c}}.
\end{equation}
\noindent Cela définit le mode le plus instable, cinétiquement sélectionné. La croissance exponentielle est, comme souvent dans les études d'instabilités, saturée lorsque la surface devient trop déformée (non linéarité géométrique) ou lorsque le film reliant les goutelettes devient trop fin (il convient alors de prendre en compte plus finement les interactions microscopiques entre le fluide et la paroi solide).

\textbf{Références :} 
\cite{gouttes} Chapitre 5 - 2.3. ; \cite{GHP} Chapitre 1 - 4.6.

\subsection{Instabilité de Rayleigh-Plateau}
\label{subsec:plateau}

\subsubsection{Analyse qualitative}
\label{subsubsec:plateau_qualit}

Intéressons-nous maitenant à une seconde instabilité, d'origine purement capillaire cette fois-ci : l'instabilité de Rayleigh-Plateau. Un jet de liquide peut se déstabiliser pour former des goutelettes : on peut l'observer sur le jet d'un robinet, s'il est assez long et à débit suffisamment faible, ou encore avec les perles de rosée sur les toiles d'araignées.

Schématiquement, il s'agit de comparer la surface d'un cylindre $A_\text{c} = 2 \pi R h$, de rayon $R$ et de hauteur $h$, et d'une assemblée de $N$ gouttes sphériques de rayon $r$, $A_\text{g} = 4 \pi r^2 N$. La conservation du volume impose $\pi R^2 h =4 N \pi r^3 /3$, donc $A_\text{c} > A_\text{g}$ si $r > 3R/2$.

Avant de quantifier plus précisément cette instabilité, assurons-nous que la pesanteur n'intervient pas dans le problème en estimant le nombre de Bond. La contribution de l'énergie de pesanteur est négligeable tant que $\mathrm{Bo}=(R/\ell_\text{c})^2 \ll 1$ : c'est très correct pour la formation de perles de rosées ou pour de fins filets d'eau.

\subsubsection{Analyse en modes normaux}
\label{subsubsec:plateau_analyse}

\begin{figure}[htb]
\centering
\includegraphics{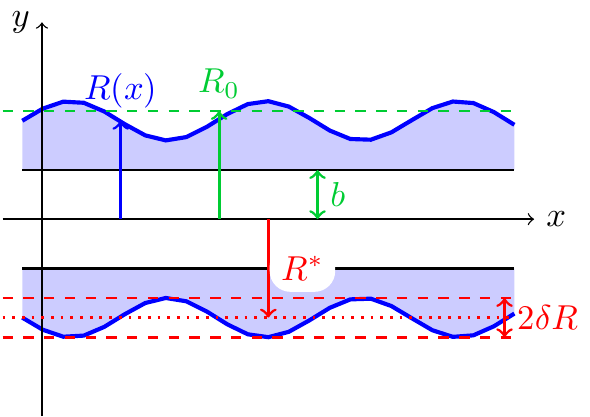}
\caption{Un film de liquide de rayon $R_0$ est déposé sur un cylindre de rayon $b$. Une perturbation axisymétrique de l'interface de longueur d'onde suffisante peut croître : le cylindre de fluide se déstabilise alors pour former un chapelet de goutelettes. \label{fig:rayleigh_plateau}}
\end{figure}

Considérons un film d'eau de rayon $R_0$ autour d'un fil cylindrique de rayon $b$\footnote{L'épaisseur du film est donc $e_0=R_0-b$.} et d'axe $\vec{e}_x$. On étudie l'évolution temporelle d'une perturbation axisymétrique de la forme $R(x) = R^* + \delta R(t) \cos{(qx)}$, avec $\delta R \ll R^*$, représentée figure \ref{fig:rayleigh_plateau}. En raison de la contrainte de conservation du volume\footnote{Nous pourrions, plutôt que de l'imposer ici, la rajouter dans l'expression de l'énergie avec un multiplicateur de Lagrange.}, $R^* \neq R_0$. Plus précisément, nous avons sur une longueur d'onde
\begin{equation}
\int_0^\lambda \pi R^2(x) \mathrm{d}x - \pi b^2 \lambda = \pi (R_0^2 -b^2) \lambda
\end{equation}
\noindent dont nous tirons
\begin{equation}
R^* \simeq R_0- \frac{\delta R^2}{4R_0} + \mathcal{O}(\delta R^3).
\end{equation}

La variation énergétique associée à cette perturbation s'écrit, pour une longueur d'onde, 
\begin{equation}
\Delta E = \gamma \left[ \int_0^\lambda 2\pi R(x) \mathrm{d}s - 2\pi R_0 \lambda \right]. 
\end{equation}

Une fois de plus, pour de petites déformations, on a $\mathrm{d}s \simeq \mathrm{d}x [1+R'(x)^2/2]$ donc
\begin{align}
\Delta E &\simeq 2\pi \gamma \left[ \int_0^\lambda (R^* + \delta R \cos{(qx)}) \left(1+\frac{q^2 (\delta R)^2}{2} \sin{(qx)}^2 \right)\mathrm{d}x -R_0 \lambda \right] \nonumber \\
		 &= 2\pi \gamma \left[ (R^* - R_0) \lambda + \frac{R^* (\delta R)^2 q^2 \lambda}{4} \right] + \mathcal{O}(\delta R^3) \nonumber \\
		 &= \frac{\pi \lambda \gamma}{4R_0} (\delta R)^2 [(qR_0)^2-1] + \mathcal{O}(\delta R^3).
\end{align}
\noindent Ainsi, les modes satisfaisant $qR_0 < 1$, donc de longueurs d'onde supérieures à $2\pi R_0$ sont instables. À nouveau, malgré l'existence d'une infinité de modes instables, seul se développe le plus rapide. Pour le déterminer, nous nous plaçons dans l'approximation de lubrification en supposant en particulier l'épaisseur $e=R-b$ du film faible devant le rayon $b$ du cylindre. Le gradient de pression responsable de l'écoulement est le gradient de pression de Laplace. La courbure du fluide perturbé est donnée par\footnote{La surface est courbée dans les dimensions : dans le plan de la figure, la courbure est celle que nous avions obtenue au paragraphe précédent, et dans le plan orthogonal, la section est circulaire donc le rayon de courbure est $R(x)$.}
\begin{equation}
C(x) = \frac{1}{R(x)} - \frac{\mathrm{d}^2R}{\mathrm{d}x^2} = \frac{1}{b+e(x)} - \frac{\mathrm{d}^2e}{\mathrm{d}x^2}
\end{equation}

\noindent Dès lors, l'équation de Reynolds \eqref{eq:reynolds} s'écrit 
\begin{equation}
\frac{\mathrm{d} \delta e}{\mathrm{d} t} = \frac{1}{3\eta} \frac{\mathrm{d}}{\mathrm{d} x} \left[\gamma \left(\frac{-1}{(b+e)^2} \frac{\mathrm{d}e}{\mathrm{d}x} - \frac{\mathrm{d}^3e}{\mathrm{d}x^3} \right) e^3\right].
\end{equation}
\noindent En écrivant $e(x,t)=(R^*-b) + \delta e (t) \cos{(qx)}$ et en restant à l'ordre $1$ dans le développement en $\delta e$, on obtient\footnote{Un peu laborieusement, j'en conviens.}
\begin{equation}
\frac{\mathrm{d} \delta e}{\mathrm{d} t} = \frac{\gamma e_0^3}{3\eta b^2} q^2[1-(qb)^2] \delta e.
\end{equation}

À nouveau, la perturbation évolue exponentiellement avec un temps caractéristique
\begin{equation}
\tau(q) = \frac{3\eta b^2}{\gamma e_0^3} \frac{1}{q^2[1-(qb)^2]}.
\end{equation}

On retrouve le critère d'instabilité : il y a croissance de la perturbation si $qb <1$\footnote{On rappelle que l'on a choisi $R-b \ll b$ donc $R\simeq b$ : on retrouve bien le même critère.}. En outre le mode le plus rapide correspond à
\begin{equation}
\lambda^* = \frac{2\pi}{q^*} = 2\pi\sqrt{2}b \quad \text{et} \quad \tau^* = \frac{12\eta b^4}{\gamma e_0^3}.
\end{equation}

\textbf{Références :} 
\cite{gouttes} Chapitre 5 - 2.4. ; \cite{GHP} Chapitre 8 - 3.2. ; \cite{fluides} - Partie 3, 5.

\subsection{Instabilité de Saffman-Taylor}
\label{subsec:saffman}

Jusque là, nous nous sommes intéressés à des instabilités dont la capillarité est l'un des moteurs. Elle intervient de façon plus générale dans les instabilités impliquant des déformations d'une interface : la tension de surface doit être prise en compte dans les conditions aux limites sur la pression, et elle introduit des termes non-linéaires qui saturent l'instabilité quand elle se développe. On peut citer par exemple l'instabilité de Kelvin-Helmholtz, apparition de vagues à l'interface entre deux fluides de viscosités différentes et poussés à des vitesses différentes, ou l'instabilité de Faraday, apparition de vagues à la surface d'un fluide que l'on fait vibrer verticalement.

Intéressons-nous plus précisément à l'instabilité de digitation visqueuse de Saffman-Taylor. Lorsqu'un fluide est poussé dans un autre fluide plus visqueux, le front tend à se déstabiliser en doigts de différentes tailles. Nous nous contenterons d'une analyse qualitative de cette instabilité et d'une estimation dimensionnelle de la longueur d'onde critique.

On considère une cellule de Hele-Shaw, c'est-à-dire un canal de section rectangulaire, de largeur $b$ et d'épaisseur $e \ll b$, initialement rempli d'un fluide (2) de viscosité $\eta$ et dans lequels on injecte un fluide (1) de viscosité négligeable. L'injection est réalisée en maintenant une différence de pression entre l'entrée (à la pression $P_1$) et la sortie (à la pression $P_\text{atm}<P_1$) de la cellule. L'écoulement obtenu est alors de type Poiseuille, bidimensionnel dans le plan $(xy)$ où $\vec{e}_x$ désigne la direction d'injection du fluide, et satisfait à la loi de Darcy\footnote{Selon les références, la loi de Darcy peut faire référence soit à cette relation locale, soit à la relation intégrée sur la section du canal proposée au paragraphe \ref{subsubsec:poreux_rqs}. Dans le cas d'un écoulement de Poiseuille, comme considéré ici, ces relations peuvent être démontrées. Dans le cas des écoulements en milieux poreux, où l'écoulement est éventuellement inhomogène, on utilise plutôt la loi intégrée, qui est alors empirique.}
\begin{equation}
\vec{v} = -\frac{b^2}{12\eta} \overrightarrow{\nabla}P.
\end{equation}

L'interface est initialement plane et nous considérons en une perturbation, de sorte qu'un point $A$ de l'interface se retrouve en avance par rapport à un point $B$ proche. La viscosité du fluide (1) étant négligeable devant celle du fluide (2), la chute de pression due à la dissipation visqueuse se produit principalement dans le fluide (2). La pression en amont de l'interface peut donc être considérée comme homogène et égale à $P_1$.

\begin{figure}[htb]
\centering
\includegraphics{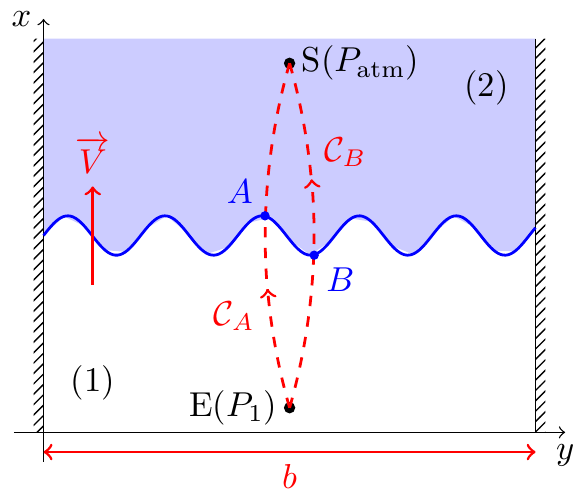}
\caption{Dans une cellule de Hele-Shaw de largeur $b$, un fluide (1) injecté au point E pousse un fluide (2) plus visqueux, ressortant par le point S. La pression est imposée en E et en S. La loi de Darcy implique que la vitesse le long du chemin $\mathcal{C}_A$ est plus élevée que celle le long du chemin $\mathcal{C}_B$, ce qui mène à une croissance de la perturbation en l'absence de tension de surface. \label{fig:saffman}}
\end{figure}

Dans un premier temps, si l'on ne considère pas la tension de surface, la pression est également constante et égale à $P_1$ juste en aval de l'interface. Considérons alors deux lignes de courant $\mathcal{C}_A$ et $\mathcal{C}_B$, reliant l'entrée de la cellule et la sortie de la cellule, l'une passant par $A$ et l'autre par $B$, et représentées figure \ref{fig:saffman}. Le point $B$ étant en arrière, $\mathcal{C}_B$ parcourt une plus grande distance dans le fluide (2) que $\mathcal{C}_A$ : la pression en sortie étant égale à $P_\text{atm}$ pour les deux lignes, le gradient de pression le long de $\mathcal{C}_A$ est donc nécessairement supérieur à celui le long de $\mathcal{C}_B$. Dès lors, la loi de Darcy impose que la vitesse soit plus élevée le long de $\mathcal{C}_A$, entraînant une croissance de la perturbation. Ce mécanisme est déstabilisant.

Si l'on prend en compte la tension de surface, il y a une discontinuité de pression due à la courbure de l'interface : la pression n'est plus constante juste en aval de l'interface, ce qui tend à atténuer l'instabilité. La tension de surface a un effet stabilisant sur l'interface plane.

L'instabilité va donc se développer si le gradient de pression associé à la loi de Darcy domine le gradient de pression de Laplace. Notons $V$ la vitesse de l'interface, qui donne l'ordre de grandeur de la vitesse de l'écoulement. Pour une déformation sinusoïdale de l'interface, de longueur d'onde $\lambda$, le gradient de pression lié à la loi de Darcy peut être estimé par $(\nabla P)_\text{Darcy} \sim \eta V /b^2$ alors que le gradient de pression de Laplace est donné par $(\nabla P)_\text{Laplace} \sim \gamma / \lambda^2$, $\lambda$ donnant à la fois un ordre de grandeur de la courbure et de la distance typique de variation de pression. Dès lors, l'interface se déstabilise quand
\begin{equation}
(\nabla P)_\text{Darcy} > (\nabla P)_\text{Laplace} \quad \text{soit} \quad \lambda > b \sqrt{\frac{\gamma}{\eta V}} \equiv \frac{b}{\sqrt{\mathrm{Ca}}}.
\end{equation}

Nous avons fait ici apparaître le nombre capillaire 
\begin{equation}
\mathrm{Ca} = \frac{\eta V}{\gamma}
\end{equation}
\noindent qui compare l'influence de la dissipation visqueuse à la tension de surface. Ce nombre sans dimension est pertinent dès que l'on s'intéresse aux propriétés interfaciales d'un système où une vitesse d'écoulement est imposée. C'est par exemple le cas pour l'étude de l'entraînement d'un film fluide par tirage d'une plaque (films de Landau Levich) ou pour l'étude des angles de mouillage dynamiques\footnote{Pour plus de détails, on consultera \cite{gouttes} paragraphe 5.3. et \cite{GHP} pararaphe 8.2.2.}. Le nombre capillaire intervient aussi en rhéologie de fluides contenant des bulles ou des gouttes (comme les émulsions) afin de déterminer si les inclusions se déforment lors de l'écoulement.

Un calcul plus précis et plus général permet d'obtenir la relation de dispersion reliant le taux de croissance au vecteur d'onde de la perturbation, comme nous l'avons fait pour l'étude des instabilités de Rayleigh-Taylor et Rayleigh-Plateau. Il est en outre possible\footnote{Mais il est nécessaire de disposer de temps, de motivation, et de bonnes aptitudes au calcul, ce qui est le cas de l'auteur de \cite{rheophysique}, où l'on trouvera des détails au paragraphe 3.7.10.} de déterminer la forme du doigt principal, une fois l'instabilité développée. On trouve que sa largeur diminue quand la vitesse d'injection du fluide augmente : c'est un problème dans le cadre de l'extraction secondaire de pétrole, discutée au paragraphe \ref{subsubsec:poreux_rqs}. Non seulement, la formation d'un doigt empêche la récupération de l'intégralité du pétrole, mais en plus, augmenter la vitesse d'injection, et donc le débit, diminue la fraction récupérée.

\textbf{Références :} 
\cite{rheophysique} Chapitre 3 - 7.10.

\subsection{Effet Marangoni}
\label{subsec:marangoni}

Nous avons vu que l'existence de gradients de pression de Laplace pouvaient induire un écoulement : c'est également le cas si des gradients de tension de surface existent à la surface d'un fluide. Un écoulement se produit alors en direction des zones de forte tension de surface. Il s'agit de l'effet Marangoni. Le traitement rigoureux des écoulements engendrés par cet effet peut par exemple se faire dans le cadre de l'approximation de lubrification. Nous nous contenterons ici de quelques remarques qualitatives.

De tels gradients de pression de Laplace peuvent être causés par un gradient thermique, la tension de surface dépendant de la température. On parle alors d'effets thermocapillaires. Un cas classique d'effet thermocapillaire est l'instabilité d'une couche de fluide à surface libre chauffée par le bas. On observe l'apparition de cellules de convection s'organisant en un réseau hexagonal : il s'agit de l'instabilité de Bénard-Marangoni.

Ils peuvent également être créés en présence de tensioactifs répartis de façon inhomogène à la surface du fluide. C'est ce qui explique la montée d'un film liquide d'une solution contenant un alcool\footnote{En tant que fonctionnaires de l'\'{E}tat, éthiques et responsables, nous ne citerons pas le vin ou les diverses autres liqueurs permettant d'observer cet effet.} le long des parois d'un récipient. Dans le ménisque, l'évaporation de l'alcool est plus rapide que dans le volume. Or, l'alcool abaisse la tension de surface de l'eau : le haut du ménisque a donc une tension superficielle plus importante, et le ménisque monte le long de paroi, laissant un film mince derrière lui. Le film se finit, dans sa partie supérieure, par un bourrelet qui finit par se déstabiliser sous l'influence de la pesanteur, et par former des gouttes qui redescendent (les larmes).

Deux expériences simples permettent de mettre cet effet en évidence. Dans un cristallisoir d'eau (propre) à la surface de laquelle on a répandu du poivre, on plonge un couteau dont la pointe a été légèrement enduite de savon. On observe que le poivre s'éloigne du couteau. En effet, les molécules de savon, tensioactives, se répandent à partir de la pointe en abaissant la tension de surface localement : elles induisent ainsi un écoulement vers les zones de forte tension de surface, qui éloigne le poivre du couteau. Cependant, l'effet ne peut se répéter un grand nombre de fois : quand la surface est recouverte de molécules de tensioactifs, il n'y a plus de modification de la tension de surface. De la même façon, on peut faire un \og bateau \fg{} autopropulsé avec une aiguille dont on a plongé une extrémité dans du savon : l'écoulement généré par la dissolution des tensioactifs induit un mouvement de l'aiguille tant que la surface de l'eau n'est pas saturée.

\textbf{Références :} 
\cite{gouttes} Chapitre 10 - 1.1. ; \cite{GHP} Chapitre 8 - 2.4. ; \cite{fluides} - Partie 3, 6.

\section{Aspects microscopiques}
\label{sec:micro}
Dans ce paragraphe, nous nous penchons sur quelques aspects microscopiques liés aux interfaces, plus anecdotiques du point de vue de la leçon, mais qui peuvent être réutilisés dans d'autres leçons, ou faire l'objet de problèmes aux écrits.

\subsection{Modèle de Ginzburg-Landau}

\subsubsection{Modèle de gaz sur réseau}

On considère un modèle de gaz sur réseau où $N$ particules se répartissent sur $M$ sites, en contact avec un thermostat à température $T_0$ et un réservoir de particules de potentiel chimique $\mu_0$. Pour chaque site $i$, on note $n_i$ l'occupation du site, valant $1$ si le site est occupé et $0$ s'il est vacant. On interdit l'occupation multiple d'un site et on ne considère que des interactions entre plus proches voisins, avec une énergie $-\epsilon$.

L'énergie du système s'écrit alors
\begin{equation}
E = - \epsilon \sum_{(i,j)} n_i n_j
\label{eq:energie_gaz_reseau}
\end{equation} 
\noindent où la première somme porte sur les couples de plus proches voisins et où $\mu$ désigne le potentiel chimique du système.

Si le système comporte $N$ particules, le nombre de configurations accessibles au système s'écrit $\Omega =\dbinom{M}{N}$. L'entropie du  système à l'équilibre est donnée par la relation de Boltzmann
\begin{equation}
S = k_\text{B} \ln{\Omega} = k_\text{B} \ln{\left[\frac{M!}{N!(M-N)!}\right]}.
\end{equation}

À la limite thermodynamique, on considère $M$, $N$ et $(M-N)$ grands en conservant la densité moyenne $n = N/M$\footnote{Nous ne traitons pas les choses de façon très rigoureuse ici, car nous mélangeons une approche de physique statistique avec de la thermodynamique. La formule de Boltzmann pour l'entropie est valable dans l'ensemble microcanonique, et nous travaillons ici dans l'ensemble grand canonique. Passant \textit{in fine} à la limite thermodynamique, les résultats resteront valables, mais on peut les obtenir rigoureusement en calculant la fonction de partition grand canonique du système comme fait dans \cite{barrat_hansen}, paragraphe 4.1.} constante et supposée homogène. La formule de Stirling nous permet alors d'estimer l'entropie par
\begin{equation}
S = - M k_\text{B} [n \ln{n} + (1-n) \ln{(1-n)}].
\end{equation}

Pour calculer l'énergie du système, nous faisons une approximation de champ moyen. \'Ecrivons $n_i = n + \delta n_i$ et supposons $\delta n_i \ll n$ : à l'ordre 1, l'énergie s'écrit
\begin{equation}
E \simeq - \epsilon \sum_{(i,j)} (n^2 + \delta n_i n + \delta n_j n).
\end{equation}

L'énergie moyenne du système est alors donnée par
\begin{equation}
\langle E \rangle = - \epsilon \sum_{(i,j)} (n^2 + \langle\delta n_i \rangle n + \langle \delta n_j \rangle n)  = - \frac{\epsilon M z}{2}n^2 \equiv \mu_\text{CM} M n
\end{equation}
\noindent où $z$ est la coordinence du réseau. Les interactions agissent donc comme un potentiel extérieur homogène $\mu_\text{CM} = -\epsilon z n /2$ appliqué en chaque site et dû au champ moyen.

En champ moyen, le potentiel thermodynamique adapté par site s'écrit donc
\begin{equation}
\omega^*(\langle E \rangle,n;T_0,\mu_0) = \frac{\langle E \rangle -T_0 S -\mu_0 N}{M} = -\frac{\epsilon z}{2} n^2 + k_\text{B} T_0[n\ln{n} + (1-n) \ln{(1-n)}] - \mu_0 n.
\label{eq:potentiel_gazreseau}
\end{equation}

\noindent Ce potentiel décrit une transition de phase avec point critique. Les densités d'équilibre satisfont à 
\begin{equation}
\frac{\partial \omega^*}{\partial n} = 0 \quad \text{soit} \quad 2n - 1 = \tanh{\left(\frac{\epsilon z n + \mu_0}{2 k_\text{B} T_0} \right)}.
\label{eq:autocoherence}
\end{equation}
\noindent Il s'agit d'une relation dite d'autocohérence.

\begin{figure}[htb]
\centering
\includegraphics{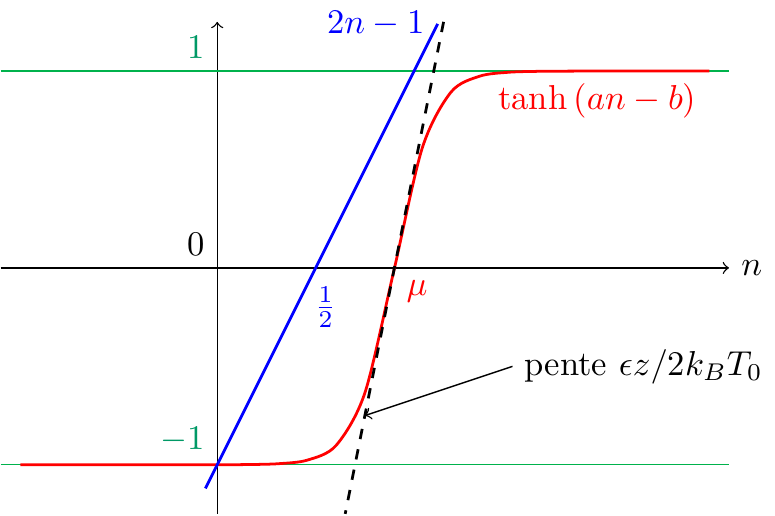}
\caption{Représentation des fonctions impliquées dans l'équation d'autocohérence \eqref{eq:autocoherence}. Les intersections des deux courbes correspondent aux extrema du grand potentiel. \label{fig:autocoherence}}
\end{figure}

On peut discuter graphiquement le comportement du système : les extrema du potentiel correspondent aux intersections des courbes tracées figure \ref{fig:autocoherence}, en gardant en mémoire le fait que $n$ doit rester comprise entre $0$ et $1$ par définition.

Plusieurs situations se présentent alors.
\renewcommand{\labelitemii}{$\diamond$}
\begin{itemize}
\item Si $\mu > 0$ ou $\mu < -\epsilon z$, il n'existe qu'une densité d'équilibre, respectivement supérieure (phase liquide) ou inférieure (phase gazeuse) à $1/2$. 
\item Pour $\mu$ compris entre ces deux valeurs et différent de $-\epsilon z / 2$, il y a à nouveau deux possibilités :
   \begin{itemize} 
   \item si la température est suffisamment élevée, il n'y a qu'une seule phase d'équilibre, liquide si $\mu > -\epsilon z / 2$ ou gazeuse sinon.
   \item si la température est suffisamment basse, le potentiel admet deux minima et un maximum. Les deux minima ne sont pas de même profondeur : si $\mu > -\epsilon z / 2$, la phase d'équilibre est le liquide mais une phase gazeuse peut être métastable, et inversement dans le cas contraire.
   \end{itemize}
\item Le cas $\mu = -\epsilon z / 2 \equiv \mu_\text{c}$ est particulier. Si $T > \epsilon z / 4 k_\text{B} \equiv T_\text{c}$, il n'existe qu'une phase d'équilibre correspondant à $n = 1/2 \equiv n_\text{c}$. Sinon, le grand potentiel admet deux minima $n=n_\text{c} \pm \eta$, symétriques par rapport à $n_\text{c}$ et de profondeurs égales. Il y a alors coexistence entre les phases liquide et gaz. Ces conditions correspondent au point critique, où la transition devient de second ordre\footnote{On peut s'assurer que $\partial^2_n \omega^* (n_\text{c},\mu_\text{c},T_\text{c}) = 0$ et $\partial^3_n \omega^* (n_\text{c},\mu_\text{c},T_\text{c}) = 0$. Au point critique, $n_\text{c}$ est une racine triple.}.
\end{itemize} 

\begin{figure}[htb]
\centering
\includegraphics{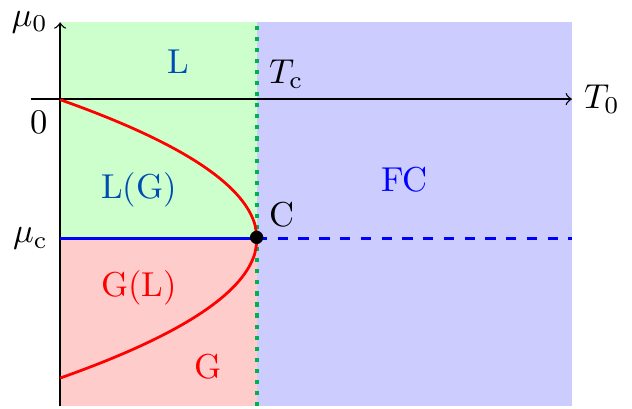}
\caption{Diagramme de phase du modèle de gaz sur réseau. C : point critique, G : zone de stabilité exclusive du gaz, G(L) : zone de stabilité du gaz avec liquide métastable, L : zone de stabilité exclusive du liquide, L(G) : zone de stabilité du liquide avec gaz métastable, FC : fluide critique. \label{fig:landau_diagramme}}
\end{figure}

Le diagramme de phase du gaz sur réseau\footnote{\samepage On remarquera la troublante ressemblance de ce diagramme avec celui du modèle d'Ising en présence d'un champ extérieur. Cela n'a rien de surprenant car en champ moyen, le modèle d'Ising en champ extérieur dans l'ensemble canonique est équivalent au modèle de gaz sur réseau dans l'ensemble grand canonique. Le passage de l'un à l'autre se fait en remplaçant les nombres d'occupation $n_i \in \{0,1\}$ par $S_i=2n_i - 1 \in \{-1,1\}$.} est résumé figure \ref{fig:landau_diagramme}.

\subsubsection{Théorie de Landau de la transition liquide-gaz}

\paragraph{Généralités sur la théorie de Landau :}

La théorie de Landau vise à donner une description qualitative la plus simple possible des transitions de phase. Pour ce faire, on effectue un développement du potentiel thermodynamique~$\Psi$ adapté à la situation en puissances du paramètre d'ordre~$\phi$ autour du point critique\footnote{Pas trop près cependant, la théorie de Landau reposant sur des arguments de champ moyen qui ne sont plus valables à proximité du point critique. Le lecteur intéressé se penchera sur le critère dit de Ginzburg, qui quantifie la pertinence du champ moyen au voisinage du point critique.}
\begin{equation}
\Psi(T,\phi) = \Psi_0(T) + \sum_{k=1}^{+\infty} \frac{a_k (T)}{k} \phi^k.
\end{equation}

L'ordre $1$ peut être absorbé si au lieu du paramètre d'ordre, on utilise son écart au point critique $\eta = \phi-\phi_\text{c}$. Il faut ensuite limiter au maximum le développement : pour ce faire, on élimine les termes ne respectant pas les symétries de la phase la plus symétrique et on s'arrête à l'ordre le plus bas permettant de rendre compte des propriétés de la transition étudiée.  Le terme d'ordre le plus haut doit être pair et avec un coefficient positif, de façon à ce que $\Psi \rightarrow +\infty$ pour $\phi \rightarrow \pm \infty$. On considère un maximum de coefficients $a_k$ comme indépendants de $T$, et on prend une dépendance linéaire $\alpha_k \cdot (T-T_\text{c})$ pour les coefficients restants.

On peut alors rechercher les valeurs du paramètre d'ordre minimisant le potentiel selon la température considérée et obtenir les valeurs des divers exposants critiques.

\paragraph{Cas du modèle de gaz sur réseau :}

Reprenons le potentiel \eqref{eq:potentiel_gazreseau} obtenu pour le modèle de gaz sur réseau et développons-le autour du point critique ($n_\text{c} = 1/2$, $\mu_\text{c} = \epsilon z / 2$, $k_\text{B} T_\text{c} = \epsilon z / 4$), la densité étant ici le paramètre d'ordre. En poussant le développement à l'ordre $4$, on obtient
\begin{flalign}
&& \omega^*(n=n_\text{c} + \delta n;T_0=T_\text{c} + \delta T &; \mu_0=\mu_\text{c} + \delta \mu) = \omega^*_\text{c} - \delta \mu \delta n + 2k_\text{B} \delta T (\delta n)^2 + 32 k_\text{B} T_0 (\delta n)^4 + \mathcal{O}(\delta n^6) \nonumber &&\\
\text{soit} && \omega^*(n,T_0, \mu) &\simeq -\delta \mu (n-n_\text{c}) + \frac{\alpha_2 (T-T_\text{c})}{2} (n-n_\text{c})^2 + \frac{\alpha_4}{4} (n-n_\text{c})^4 
\end{flalign}
\noindent où $\alpha_2 = 4$ et $\alpha_4 = 128 k_\text{B} T_0 \simeq 128 k_\text{B} T_\text{c}$.

\begin{figure}[htb]
\centering
\includegraphics{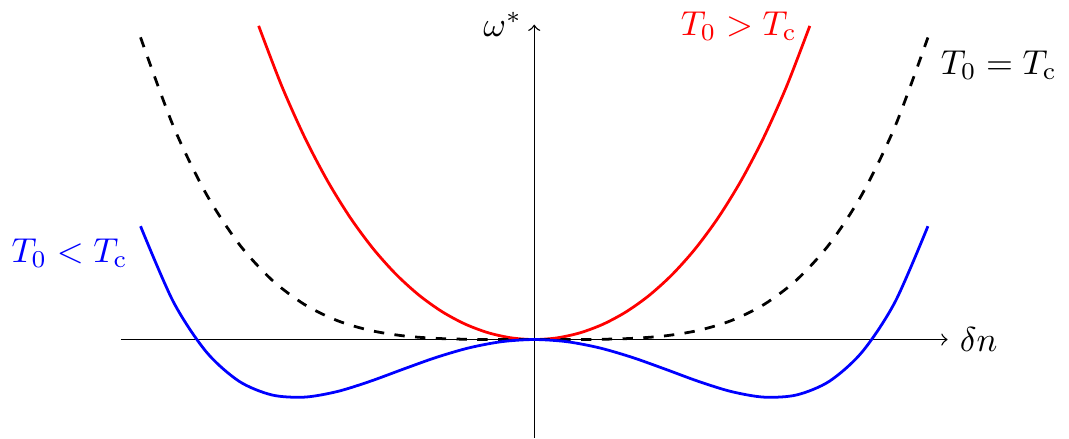}
\caption{Grand potentiel $f^*$ en fonction de $\delta n = n - n_\text{c}$, écart du paramètre d'ordre à sa valeur au point critique, pour différentes températures : en rouge, $T_0>T_\text{c}$, en pointillés noirs, $T_0=T_\text{c}$, et en bleu, $T_0<T_\text{c}$ \label{fig:landau_potentiel}.}
\end{figure}

En se plaçant en champ nul, c'est-à-dire en prenant $\delta \mu = 0$, on peut calculer les densités minimisant le potentiel. Comme illustré figure \ref{fig:landau_potentiel}, pour $T_0>T_\text{c}$, le grand potentiel admet un unique minimum en $n=n_\text{c}$ alors que pour $T_0<T_\text{c}$, $n=n_\text{c}$ devient un maximum, et le grand potentiel admet deux minima symétriques
\begin{equation}
n_\text{eq}^\pm = n_\text{c} \pm \sqrt{\frac{\alpha_2}{\alpha_4} (T_\text{c}-T_0)}.
\end{equation}

Nous pouvons en outre calculer la susceptibilité au point critique. En champ non-nul, c'est-à-dire pour $\mu_0 \neq \mu_\text{c}$, les extrema de $\omega^*$ vérifient
\begin{equation}
\alpha_2 \cdot (T_0-T_\text{c}) \delta n + \alpha_4 (\delta n)^3 = \delta \mu.
\end{equation}
\noindent En différenciant cette expression par rapport à $\mu$, on obtient la susceptibilité du système au point critique
\begin{equation}
\chi = \left(\frac{\partial n}{\partial \mu}\right)_{\mu=\mu_\text{c}} = \left\{\begin{array}{c}
\frac{1}{\alpha_2(T_0-T_\text{c})} \quad \text{pour $T_0>T_\text{c}$} \\
\frac{1}{2\alpha_2(T_\text{c}-T_0)} \quad \text{pour $T_0<T_\text{c}$}. \end{array} \right. 
\end{equation}

\paragraph{Transition liquide-gaz :}

Nous avons jusque là illustré la démarche à l'envers, en partant d'un modèle microscopique pour remonter au développement autour du point critique. La théorie de Landau peut se construire à partir des observations expérimentales, ce que nous allons maintenant faire pour le cas de la véritable transition liquide gaz. Nous nous plaçons par commodité dans l'ensemble canonique, c'est-à-dire que nous supposons maintenant fixé le nombre de particules du système.

Nous n'avons pas ici a priori de symétrie particulière permettant d'éliminer un ordre dans le développement\footnote{On peut légitimement se demander pouquoi tous les termes impairs dans le développement de Landau du modèle gaz sur réseau sont nuls.}. En développant autour de $n_\text{c}$, nous avons
\begin{equation}
f^*(n,T_0) \simeq f^*(n_\text{c},T_0) + \frac{a_2(T)}{2} (n -n_\text{c})^2 + \frac{a_3(T)}{3} (n -n_\text{c})^3 + \frac{a_4(T)}{4} (n -n_\text{c})^4.
\end{equation}

Nous négligeons la dépendance en température de $a_4(T)$ en le supposant égal à sa valeur en $T_\text{c}$. Ce coefficient doit être positif pour que l'énergie soit positive pour $n \rightarrow \pm \infty$. Nous prenons $a_2(T) = \alpha_2 \cdot (T-T_\text{c})$, de façon à avoir une transition entre l'existence d'un minimum unique en $T>T_\text{c}$ à deux minima pour $T<T_\text{c}$. D'autre part, pour avoir un point critique, la dérivée de $f^*$ doit admettre une racine triple  en $T=T_\text{c}$ : nous prenons donc $a_3(T) = \alpha_3 \cdot (T-T_\text{c})$.

Munis de ce développement, il est possible d'étudier la transition liquide-gaz à proximité de son point critique. Les résultats sont finalement proches de ceux que nous avons obtenus précédemment : le diagramme de phase et les exposants critiques sont similaires\footnote{Le terme cubique introduit ici n'est pas pertinent du point de vue du groupe de renormalisation.}. Le principal effet du terme cubique ajouté est d'introduire une dissymétrie entre les densités des phases liquide et gazeuse par rapport à la densité critique.

En revanche, si l'on considère que le coefficient $a_3(T)$ est indépendant de la température, le comportement est drastiquement modifié et on obtient une transition du premier ordre, sans point critique.

\textbf{Références :} \cite{barrat_hansen} Chapitre 2 - 4.2. ; \cite{chaikin} Chapitre 4 - 1., 2. et 4.

\subsubsection{Profil d'interface et tension de surface}

\paragraph{Théorie de Ginzburg-Landau :}

La théorie de Landau présentée au paragraphe précédent permet la description de l'énergie d'une phase homogène. Nous souhaiterions la généraliser au cas d'un système inhomogène afin d'étudier l'interface entre les phases liquide et gaz. Considérons un système de densité $\rho(\vec{r})$ inhomogène.

Nous faisons une approximation de densité locale\footnote{La même que pour les théories de fonctionnelles de la densité.} et considérons que l'énergie de la phase est donnée par l'intégrale de l'énergie libre volumique de la théorie de Landau :
\begin{equation}
F[\rho(\vec{r})] = \int f_\text{Landau}(\rho(\vec{r})) \mathrm{d}\vec{r}.
\end{equation}
\noindent En nous plaçant à la coexistence entre phases, c'est-à-dire à $\mu_0= \mu_\text{c}$, on peut écrire $f_\text{Landau}(\rho)  = C(\rho-\rho_\ell)^2(\rho-\rho_\text{g})^2/2$ et on peut utiliser l'énergie libre s'identifie au grand potentiel.

Cependant, une telle fonctionnelle va favoriser une démixion complète du système, ce qui est peu réaliste : de tels gradients de concentration seront lissés par la diffusion. Il nous faut donc pénaliser les variations brusques de densité, tout en conservant le caractère scalaire de l'énergie libre. Pour cela, nous rajoutons à l'énergie volumique un terme proportionnel à la norme au carré du gradient de densité\footnote{Cette justification phénoménologique du terme en gradient carré est assez dans l'esprit initial de la théorie de Landau. Il peut néanmoins être obtenu rigoureusement pour un modèle microscopique donné, comme effectué dans \cite{barrat_hansen} ou \cite{stat_interfaces}. Par exemple, pour le modèle de gaz sur réseau, on peut réécrire $2 n_i n_j=(n_i-n_j)^2-n_i^2-n_j^2$ dans l'énergie \eqref{eq:energie_gaz_reseau}. En passant à la limite continue, la somme sur $(n_i-n_j)^2$ fait apparaître le terme en gradient carré.} et obtenons l'énergie de Landau-Ginzburg 
\begin{equation}
\mathcal{F}_\text{LG}[\rho(\vec{r})] = \int \left[f_\text{Landau}(\rho(\vec{r})) + \frac{b}{2}\|\overrightarrow{\nabla} \rho\|^2\right] \mathrm{d}\vec{r} \equiv \int f_\text{LG}(\rho(\vec{r})) \mathrm{d}\vec{r}
\end{equation}
\noindent où le coefficient $b$ dépend a priori de $\rho$ et $T$. Par simplicité, nous le supposerons indépendant de $\rho$.

Nous cherchons alors le profil de concentration qui minimise cette fonctionnelle et satisfait donc à l'équation d'Euler-Lagrange\footnote{Notons que nous ne sommes pas préoccupés des conditions aux limites, ce qui est mal car si on en autorise des variations, l'équation d'Euler-Lagrange contient des termes supplémentaires. Mais écrire des conditions aux limites en 3D est fastidieux, on attendra la suite pour s'en préoccuper.} :
\begin{equation}
\frac{\delta \mathcal{F}_\text{LG}}{\delta \rho(\vec{r})} = 0 \quad \text{soit} \quad \frac{\mathrm d}{\mathrm{d} \vec{r}} \cdot \left[\frac{\partial f_\text{LG}}{\partial \overrightarrow{\nabla}\rho} \right] = \frac{\partial f_\text{LG}}{\partial \rho}.
\label{eq:ginzburg_fctelle}
\end{equation}
\noindent En utilisant la forme proposée de la fonctionnelle de Landau-Ginzburg, nous obtenons
\begin{equation}
b \Delta \rho = \frac{\mathrm{d} f_\text{Landau}}{\mathrm{d} \rho}.
\label{eq:ginzburg_profil}
\end{equation}

Considérons maintenant une interface plane, parallèle au plan $(xy)$, de sorte que la densité ne dépende que de la coordonnée transverse $z$. On recherche une interface connectant une phase liquide pure située en $z \rightarrow -\infty$ ($\rho \rightarrow \rho_\ell$, $\rho' \rightarrow 0$) et une phase gazeuse pure située en $z \rightarrow +\infty$ ($\rho \rightarrow \rho_\text{g}$, $\rho' \rightarrow 0$). De la sorte, l'intégrale première de l'équation \eqref{eq:ginzburg_profil} s'écrit
\begin{equation}
\frac{b}{2} \left(\frac{\mathrm{d} \rho}{\mathrm{d}z} \right)^2 = f_\text{Landau}(\rho).
\label{eq:ginzburg_profil_1D}
\end{equation}
\noindent En notant $A$ l'aire de l'interface dans le plan $(xy)$, le minimum de la fonctionnelle de Landau-Ginzburg \eqref{eq:ginzburg_fctelle} s'exprime selon
\begin{equation}
\mathcal{F}_\text{LG}^\text{min} = A \int_{-\infty}^{+\infty} b \left(\frac{\mathrm{d} \rho}{\mathrm{d}z}\right)^2 \mathrm{d}z = A \int_{\rho_\ell}^{\rho_\text{g}} b \frac{\mathrm{d}\rho}{\mathrm{d} z} \mathrm{d} \rho = A \int_{\rho_\ell}^{\rho_\text{g}} \sqrt{2 b f_\text{Landau}(\rho)} \mathrm{d}\rho.
\end{equation}
\noindent En utilisant l'expression proposée de l'énergie libre de Landau à l'équilibre entre phases, on peut donc en déduire la tension de surface
\begin{equation}
\gamma = \frac{\mathcal{F}_\text{LG}^\text{min}}{A} = \sqrt{bC} (\rho_\text{g} - \rho_\ell)^3.
\end{equation}
\noindent Puisque $\rho_\text{g} - \rho_\ell \sim (T_\text{c} - T)^{1/2}$ près du point critique, nous obtenons le comportement critique de la tension superficielle en champ moyen $\gamma \sim (T_\text{c} - T)^{3/2}$. On notera en particulier que la tension de surface s'annule au point critique, ce qui paraît raisonnable au vu des propriétés du fluide supercritique, pour lequel les états gazeux et fluides sont indistincts.

\begin{figure}[htb]
\centering
\includegraphics{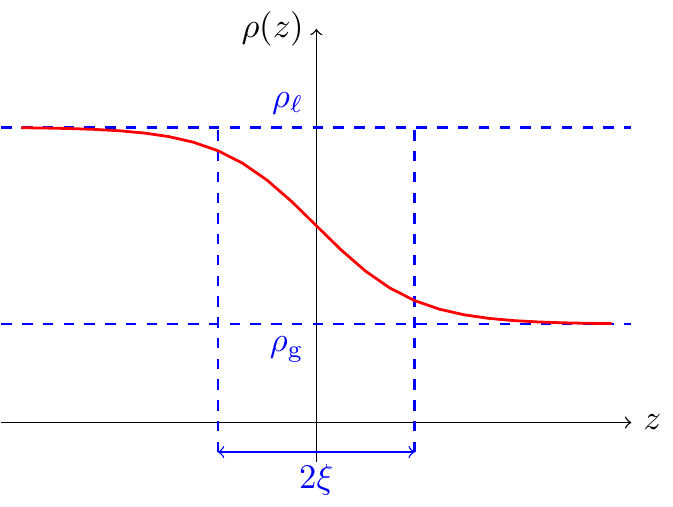}
\caption{Profil d'interface liquide-gaz prédit par la théorie de Landau-Ginzburg à l'équilibre liquide-vapeur. \label{fig:profil_ginzburg}}
\end{figure}

Nous pouvons poursuivre le calcul un peu plus loin en résolvant l'équation \eqref{eq:ginzburg_profil_1D}. En choisissant l'origine des axes de sorte que $\rho(0) = (\rho_\ell+\rho_\text{g})/2$, on obtient\footnote{Après un peu de calcul il est vrai, mais sans astuce géniale. Il s'agit d'une équation différentielle non linéaire à variables séparables, il  faut simplement faire attention aux signes.}
\begin{equation}
\rho (z) =  \frac{\rho_\ell+\rho_\text{g}}{2} + \frac{\rho_\ell-\rho_\text{g}}{2} \tanh{\left(-\frac{z}{2\xi}\right)}
\end{equation}
\noindent où $\xi = (m/C)^{1/2}(\rho_\ell+\rho_\text{g})^{-1}$ représente la largeur caractéristique de l'interface. On note que cette largeur diverge au point critique, ce qui est à nouveau raisonnable. Ce profil est représenté figure \ref{fig:profil_ginzburg}.

\textbf{Références :} 
\cite{barrat_hansen} Chapitre 7 - 3. ; \cite{stat_interfaces} Chapitre 2 - 3.

\subsection{Tensioactifs}

\subsubsection{Molécules amphiphiles et micelles}

\begin{figure}[htb]
\centering
\includegraphics{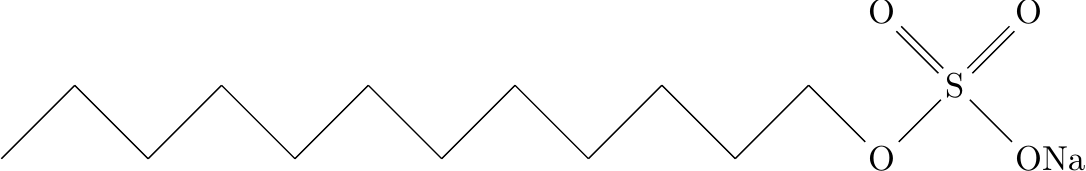}
\caption{Formule topologique du laurylsulfate de sodium ou SDS (Sodium DodecylSulfate en anglais). Il s'agit d'un tensioactif cationique couramment utilisé de formule brute $\mathrm{C_{12}H_{25}NaSO_4}$. On en identifie aisément la queue apolaire hydrophobe (chaîne carbonée) et la tête polaire hydrophile (groupement sulfate). \label{fig:SDS}}
\end{figure}

Comme nous l'avons déjà mentionné à plusieurs reprises, la présence d'impuretés modifie la tension de surface, l'élevant ou l'abaissant, ce dernier cas étant le plus fréquent. Nous allons considérer ici le cas des tensioactifs, encore appelés amphiphiles ou surfactants\footnote{Ce dernier terme étant un anglicisme, venant de \textit{Surf}ace \textit{act}ing \textit{a}ge\textit{nts}.}. Il s'agit de molécules possédant une zone polaire hydrophile (souvent une tête ionique) et une zone apolaire hydrophobe (souvent une queue formée d'une chaîne carbonée plus ou moins longue) comme représenté figure \ref{fig:SDS}.

Dans de l'eau pure, les tensioactifs se disposent de façon à ce que leur queue hydrophobe ne soit pas en contact avec de l'eau : ils se placent donc préférentiellement à l'interface et abaissent ainsi la tension superficielle. Schématiquement, l'interface liquide-gaz est remplacée par une double interface liquide-tête hydrophile et gaz-queue hydrophobe, moins défavorable. Au-delà d'une certaine concentration, appelée concentration micellaire critique (CMC), la surface est saturée et les tensioactifs restent en solution : ajouter des tensioactifs ne modifie alors plus la tension de surface. Afin de minimiser les contacts entre l'eau et les queues hydrophobes, les tensioactifs s'agrègent alors en micelles, queues vers l'intérieur et tête vers l'extérieur. Il convient néanmoins de garder à l'esprit que la micelle est un objet dynamique, où les tensioactifs s'échangent constamment avec ceux qui sont libres dans la solution environnante.

Si l'eau contient des impuretés (graisse ou bulles), les tensioactifs vont venir se placer aux interfaces et les stabiliser. De la sorte, on peut ralentir le vieillissement de mousses ou d'émulsions en limitant le processus de coalescence, fusion de deux gouttes ou bulles qui entrent en contact. C'est par exemple le rôle de la moutarde pour stabiliser une mayonnaise. C'est aussi le principe de la détergence : les savons sont composés de molécules amphiphiles, qui viennent entourer les graisses et former des micelles qui peuvent être évacuées par de l'eau.

\textbf{Références :} 
\cite{gouttes} Chapitre 8 - 1.1. et 1.3. ; \cite{israelachvili} Chapitre 19, 1.

\subsubsection{Agrégation micellaire}

\paragraph{Forme des micelles :}

Il existe un grand nombre de phases micellaires\footnote{Ces phases possèdent souvent des propriétés de cristaux liquides, dont les transitions sont contrôlables par la concentration. On parle de cristaux liquides lyotropes.}. La forme d'une micelle peut être prédite empiriquement par des critères géométriques : notons $a$ l'aire occupée par une tête polaire, $\ell$ la longueur maximale d'étirement d'une chaîne et $v$ le volume occupé par une molécule tensioactive. Ces différentes valeurs peuvent être obtenues par des simulations numériques, pour une molécule donnée. On définit le paramètre d'empilement par
\begin{equation}
\phi = \frac{v}{a\ell}.
\end{equation}

La forme permettant le plus grand recouvrement entre chaînes ainsi qu'une entropie de mélange maximale est la sphère. Une sphère de rayon $R$ contient un nombre $N = 4\pi R^2 / a = 4\pi R^3/3v$ de molécules : ainsi, $R = 3v/a$. Pour que la micelle puisse exister, il faut que $R<\ell$ donc des tensioactifs ne forment des micelles sphériques seulement si $\phi < 1/3$.

Si cette condition n'est pas satisfaite, les molécules tendent à s'organiser en cylindres. On désigne ces solutions par l'appellation de polymères vivants\footnote{Ou vers micelles selon un bon mot de de Gennes, ou encore \textit{wormlike micelles} pour les anglophones, qui n'ont pas compris la blague.}, car les cylindres sont perpétuellement en train de se couper et de se reformer. Un cylindre de rayon $R$ contient $n = 2\pi R / a = \pi R^2 / v$ molécules par unité de longueur, donc $R=2v/a$. La condition $R< \ell$ indique que des micelles cylindriques ne peuvent se former que si $\phi < 1/2$.

Si ce critère géométrique n'est pas satisfait non plus, les micelles formeront des lamelles, pouvant éventuellement se replier sur elles-mêmes pour former des vésicules ou des tubes. Pour une lamelle d'épaisseur $d$, le nombre de molécules par unité de surface est $\nu = 2/a=d/v$ soit $d=2v/a$. Les lamelles ne peuvent se former que si $\phi < 1$.

\begin{figure}[htb]
\centering
\includegraphics{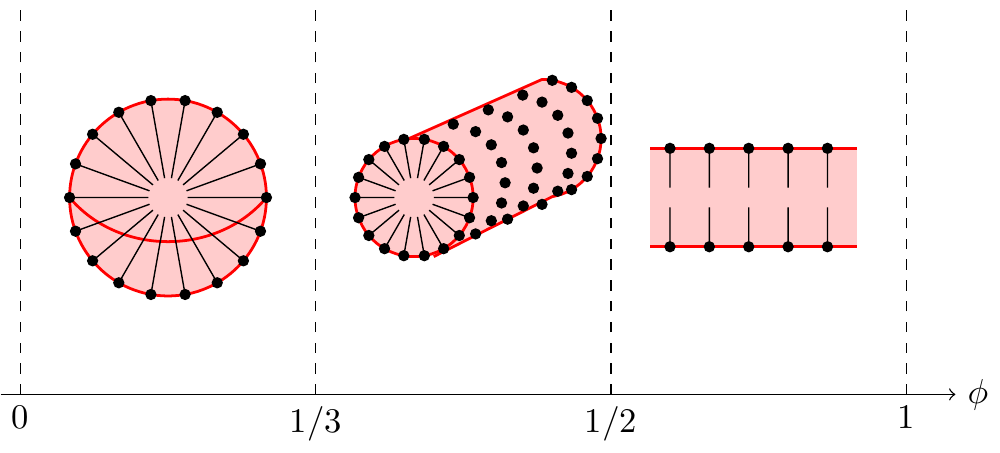}
\caption{Diagramme de forme des micelles en fonction du paramètre d'empilement $\phi=v/a\ell$. Les micelles sont successivement sphériques, cylindriques et lamellaires, puis d'autres phases plus exotiques peuvent apparaître pour $\phi \geq 1$. La représentation est schématique : il ne faut pas oublier que les tensioactifs s'échangent perpétuellement entre la micelle et la solution environnante, et qu'ils ne sont pas organisés régulièrement. \label{fig:micelles}}
\end{figure}

Dans le cas où $\phi=1$, les tensioactifs s'organisent en lamelles planes (phases éponges), et si $\phi >1$, on obtient des structures inverses, où l'eau est emprisonnée dans les vésicules. Le diagramme\footnote{Diagramme de phase simplifié, où ne sont pas incluses les phases plus exotiques observées pour $\phi \geq 1$. Notons de plus que tout cela est très empirique.} de phases en fonction du paramètre d'empilement est récapitulé figure \ref{fig:micelles}.

\paragraph{Concentration micellaire critique :}

Considérons une solution de tensioactifs. Notons $X_i = N_i / N_\text{tot} \leq 1$ la concentration numérique des molécules présentes dans des micelles à $i$ molécules ($i=1$ correspondant aux tensioactifs libres) : la concentration totale en tensioactifs est donc $c=\sum_i X_i$. La solution est maintenue à température $T$ et pression $P$.

En solution diluée, le potentiel chimique d'un agrégat de taille $N$ s'écrit
\begin{equation}
\mu_{\text{ag},N} = \psi_N(T,P) + k_\text{B} T \log{c_{\text{ag},N}} = \psi_N(T,P) + k_\text{B} T \log{\frac{X_N}{N}}
\end{equation}
\noindent où $\psi_N(T,P)$ est le potentiel chimique d'agrégats de taille $N$ en solution infiniment diluée. Le potentiel chimique d'une molécule présente dans un tel agrégat est donc donné par
\begin{equation}
\mu_N = \frac{\mu_{\text{ag},N}}{N} = \mu^0_N(T,P) + \frac{k_\text{B} T}{N} \log{\frac{X_N}{N}}
\end{equation}
\noindent où $\mu^0_N(T,P)$ est le potentiel chimique standard pour une molécule dans un agrégat unique de taille $N$.

À l'équilibre thermodynamique, l'égalité des potentiels chimiques des molécules nous permet d'écrire
\begin{equation}
X_N = N \left[X_1 \exp{\left(\frac{\mu^0_1-\mu^0_N}{k_\text{B}T}\right)} \right]^N \equiv N \left[X_1 \mathrm{e}^{\alpha_N} \right]^N.
\end{equation}
\noindent Le paramètre $\alpha_N$ représente l'énergie de liaison entre deux molécules dans un monomère de taille N en unités de $k_\text{B}T$.

Nous pouvons donner une estimation de $\mu^0_N$, qui représente le coût énergétique pour ajouter une molécule à un agrégat à $N$ molécules. Il contient d'une part une contribution surfacique $\gamma a$ liée à l'agrandissement de l'interface, et une contribution électrostatique répulsive provenant de l'interaction entre la molécule ajoutée et les autres. On peut estimer cette dernière contribution par un terme de la forme $\kappa / a$. On a donc
\begin{equation}
\mu^0_N(a) \sim \gamma a + \frac{\kappa}{a}
\end{equation}
\noindent qui passe par un minimum pour $a=a_0 \equiv (\kappa/\gamma)^{1/2}$. Cela correspond, pour une forme de micelle donnée, à un certain nombre $M$ de surfactant. L'énergie de liaison $\alpha_N=(\mu^0_1-\mu^0_N)/k_\text{B}T$ passe donc par un maximum pour une taille de micelle $M$ et nous noterons $\alpha=\alpha_M$. 

Considérons alors la limite de basse concentration où $X_1 \mathrm{e}^{\alpha}  \ll 1$ : dans ce cas $X_N \ll \cdots \ll X_2 \ll X_1 \ll 1$. Dans cette limite, toutes les molécules sont sous forme de monomères et $X_1 \simeq c$. Alors, on peut écrire $X_N \simeq N (c\mathrm{e}^\alpha)^N$ : si l'on augmente $c$, il finit par être impossible de conserver $X_N < 1$. On définit la concentration micellaire critique\footnote{Cette définition ne coïncide donc pas vraiment avec la saturation de la surface : une fois atteinte, les tensioactifs commencent à être présents dans le volume, et il faut alors que leur concentration soit suffisante pour former des micelles. Cependant, les effets sur la tension de surface sont quant à eux liés à la saturation.} par $c^* = \mathrm{e}^{-\alpha}$ : pour $c$ approchant $c^*$, des agrégats commencent à se former. Au-delà de la concentration micellaire critique, $X_1 \simeq c^*$ et les micelles commencent à se former.

\paragraph{Distribution en taille de micelles sphériques :}

Détaillons le cas de micelles sphériques de rayon $R$. Le volume de la micelle est alors $N v = 4\pi R^3/3$ et sa surface, $Na = 4\pi R^2$. On peut donc écrire $Na^3=36\pi v^2$ d'où l'on tire le nombre optimal de molécules dans une micelle $M=36 \pi v^2 / a_0^3$.

Développons alors le potentiel $\mu^0_N$ autour de $M$ : on obtient après un peu de calcul
\begin{equation}
\mu^0_N - \mu^0_M = \Lambda (N-M)^2 \quad \text{où} \quad \left\{\begin{array}{l}
\mu^0_M = 2\sqrt{\gamma\kappa} \\
\Lambda = \displaystyle{\frac{\kappa^{7/2}}{9(36\pi)^2 \gamma^{5/2} v^4}} \end{array}\right.
\end{equation}
\noindent d'où nous tirons aisément, en supposant $M$ suffisamment proche de $N$ et de 1, $\mu^0_1-\mu^0_N = \Lambda (1-N)(2M-N-1)$.

\'Etudions la distribution des concentrations des différentes micelles. En posant $\Delta N = N-M$, on peut écrire
\begin{equation}
X_N = N \left[\frac{X_M}{M} \exp{\left(-\frac{M\Lambda (\Delta N)^2}{k_\text{B} T}\right)} \right]^{N/M}.
\end{equation}
\noindent Développons alors autour du maximum $N=M$ : on a
\begin{equation}
X_N \sim \exp{\left[- \frac{(\Delta N)^2}{2\sigma^2} \right]} \quad \text{avec} \quad \sigma = \sqrt{\frac{k_\text{B}T}{2 M \Lambda}} \simeq \sqrt{\frac{M}{2\alpha}}.
\end{equation}
Il s'agit d'un distribution gaussienne, centrée sur $M$ indépendant de la concentration, et de largeur $\sigma$ qui augmente avec $c$. On peut mener une étude similaire pour d'autres formes de micelles.

\textbf{Références :} 
\cite{israelachvili} Chapitre 19, 2. et 5.

\subsection{Mouillages spéciaux}

Nous avons vu au paragraphe \ref{subsubsec:gentriple_youngdupre} que la présence de rugosité ou d'impuretés chimiques pouvait modifier l'angle de contact. Nous proposons ici des modèles permettant d'en estimer la valeur.

\subsubsection{Modèle de Wenzel}
\label{subsubsec:mouillages_wenzel}

Nous considérons une goutte posée sur une surface chimiquement homogène présentant une rugosité à une échelle petite devant la goutte. On note $r$ la rugosité de la surface, c'est-à-dire le rapport de son aire réelle sur son aire apparente : pour une surface lisse $r=1$ alors que pour une surface quelconque, $r>1$. Comme nous l'avons vu, la goutte adopte une forme de calotte sphérique et on note $x$ le rayon de l'interface solide-liquide. 

\begin{figure}[htb]
\centering
\includegraphics{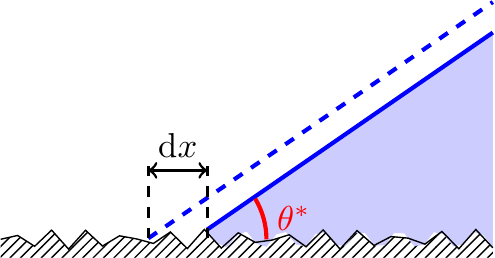}
\caption{Une goutte déposée sur une surface rugueuse présente un angle apparent $\theta^*$ différent de celui prévu par la relation de Young-Dupré. \label{fig:wenzel}}
\end{figure}

Le modèle de Wenzel suppose la loi de Young-Dupré valable localement, et cherche la valeur de l'angle de contact apparent $\theta^*$ au niveau de la ligne triple. Comme illustré figure \ref{fig:wenzel}, si l'on déplace la ligne triple de $\mathrm{d}x$, une surface de solide, rugueuse, est recouverte de liquide, et l'aire de contact liquide-gaz augmente. L'énergie de surface varie donc de
\begin{equation}
\mathrm{d}E_\text{s} = (\gamma_\text{SL} - \gamma_\text{SG}) \mathrm{d}A_\text{SL} + \gamma \mathrm{d}A_\text{LG} = (\gamma_\text{SL} - \gamma_\text{SG}) \cdot (2\pi x r \mathrm{d}x) + \gamma \cdot (2\pi x \cos{\theta^*} \mathrm{d}x).
\end{equation}

En utilisant la loi de Young-Dupré, la condition d'équilibre $\mathrm{d}E_\text{s}=0$ fournit la relation de Wenzel
\begin{equation}
\cos{\theta^*} = r \cos{\theta_\text{E}}.
\end{equation}

Cette relation montre que la rugosité amplifie les effets de mouillage : elle abaisse l'angle de contact d'un fluide plutôt mouillant ($\theta^*<\theta_\text{E}$) et augmente celui d'un fluide plutôt non-mouillant ($\theta^*>\theta_\text{E}$). Elle implique également la possibilité, pour une rugosité suffisante, d'avoir une transition vers un mouillage total ou nul selon le fluide : cette prédiction n'est cependant pas vérifiée expérimentalement. En effet, pour un fluide mouillant, le liquide va envahir spontanément les anfractuosités en amont de la ligne triple, alors que pour un fluide non mouillant, des bulles de gaz vont rester piégées dans les anfractuosités sous la goutte : dans les deux cas, on ne peut plus considérer la surface solide comme chimiquement homogène.

\textbf{Références :}
\cite{gouttes} Chapitre 9 - 2.1.2.

\subsubsection{Modèle de Cassie-Baxter}

Nous nous intéressons maintenant aux effets d'une hétérogénéité chimique de la surface. Considérons une goutte de liquide déposée sur une surface possédant deux constituants occupant des fractions de surface respectives $f_1$ et $f_2$ ($f_1+f_2=1$). On note $\theta_1$ et $\theta_2$ les angles de contacts du fluide sur une surface homogène respectivement composée de l'espèce $1$ et $2$.

Nous supposons la surface hétérogène à l'échelle de la goutte, de sorte que les fractions que nous  avons définies soient représentatives des surfaces rencontrées par la goutte, et nous considérons à nouveau que la loi de Young-Dupré est valable localement. En conservant les notations du paragraphe \ref{subsubsec:mouillages_wenzel}, un raisonnement similaire mène à
\begin{equation}
\mathrm{d}E_\text{s} = f_1 (\gamma_\text{SL,1} - \gamma_\text{SG,1}) \cdot (2\pi x r \mathrm{d}x) + f_2 (\gamma_\text{SL,2} - \gamma_\text{SG,2}) \cdot (2\pi x r \mathrm{d}x) + \gamma \cdot (2\pi x \cos{\theta^*} \mathrm{d}x)
\end{equation}

En utilisant la loi de Young-Dupré, on obtient dès lors la relation de Cassie-Baxter
\begin{equation}
\cos{\theta^*}=f_1\cos{\theta_1} + f_2\cos{\theta_2}.
\end{equation}

L'angle apparent obtenu est compris entre $\theta_1$ et $\theta_2$. Pour aller plus loin, nous pourrions combiner les points de vue de Wenzel et Cassie-Baxter, afin de tenir compte d'une rugosité et d'une hétérogénéité chimique conjointe\footnote{Le lecteur intéressé se reportera à \cite{gouttes}, paragraphe 9.2.2.}.

\textbf{Références :}
\cite{gouttes} Chapitre 9 - 2.1.3.

\newpage

\bibliographystyle{unsrt}

\begin{thebibliography}{10}

\bibitem{gouttes}
P-G. de~Gennes, F. Brochard-Wyart et D. Quéré,
\newblock {\em Gouttes, bulles, perles et ondes}.
\newblock Belin - \'Echelles, 2002.

\bibitem{diu_thermo}
B. Diu, C. Guthmann, D. Lederer et B. Roulet,
\newblock {\em Thermodynamique}.
\newblock Hermann, 2007.

\bibitem{GHP}
J.P. Hulin, L.~Petit et \'E.~Guyon.,
\newblock {\em Hydrodynamique physique, 3e édition}.
\newblock EDP Sciences, 2012.

\bibitem{handbook}
W.M. Haynes,
\newblock {\em Handbook of chemistry and physics - 95th edition}.
\newblock CRC, 2014-2015.

\bibitem{fluides}
J.P. Hulin, L.~Petit et \'E.~Guyon.,
\newblock {\em Ce que disent les fluides}.
\newblock Belin - Pour la science, 2005.

\bibitem{barrat_hansen}
J.L. Barrat et J.P. Hansen,
\newblock {\em Basic concepts for simple and complex liquids}.
\newblock Cambridge University Press, 2003.

\bibitem{portelli}
J. Barthes et B. Portelli,
\newblock {\em La physique par la pratique}.
\newblock H et K, 2005.

\bibitem{andreotti}
O. Pouliquen, B. Andreotti et Y. Forterre,
\newblock {\em Les milieux granulaires, entre fluide et solide}.
\newblock EDP Sciences, 2011.

\bibitem{israelachvili}
J.N. Israelachvili,
\newblock {\em Intermolecular and Surface Forces, third edition}.
\newblock Elsevier, 2011.

\bibitem{rheophysique}
P. Oswald,
\newblock {\em Rhéophysique}.
\newblock Belin - \'Echelles, 2005.

\bibitem{chaikin}
P.M. Chaikin et T.C. Lubensky,
\newblock {\em Principles of condensed matter physics}.
\newblock Cambridge University Press, 1995.

\bibitem{stat_interfaces}
S.A. Safran,
\newblock {\em Statistical Thermodynamics of Surfaces, Interfaces and
  Membranes}.
\newblock Westview, 2003.

\bibitem{Marchand_2011}
A. Marchand, J.H. Weijs, J.H. Snoeijer et B. Andreotti,
\newblock {\em Why is surface tension a force parallel to the interface?}.
\newblock Am.J.Phys., \textbf{79}, 999 (2011).

\end{thebibliography}

\end{document}